\documentclass[aps,prd,amssymb,nofootinbib,floatfix,a4paper,eqsecnum,twocolumn]{revtex4}

\usepackage{mathrsfs}
\usepackage{latexsym,amsmath,amsfonts,amssymb}
\usepackage{bm}
\usepackage{graphicx}

\allowdisplaybreaks

\newcommand{\beq}{\begin{equation}}
\newcommand{\eeq}{\end{equation}}
\newcommand{\bea}{\begin{eqnarray}}
\newcommand{\eea}{\end{eqnarray}}

\newcommand{\e}{\widehat{\mathcal E}_{\rm eff}}

\begin{document}

\title{Binary dynamics at the fifth and fifth-and-a-half post-Newtonian orders}

\author{Donato Bini$^{1,2}$, Thibault Damour$^3$, Andrea Geralico$^1$}
  \affiliation{
$^1$Istituto per le Applicazioni del Calcolo ``M. Picone,'' CNR, I-00185 Rome, Italy\\
$^2$INFN, Sezione di Roma Tre, I-00146 Rome, Italy\\
$^3$Institut des Hautes \'Etudes Scientifiques, 91440 Bures-sur-Yvette , France.
}

\date{\today}

\begin{abstract}
Using the new methodology  introduced in a recent Letter [Phys.\ Rev.\ Lett.\  {\bf 123}, 231104 (2019)], 
we present the details of the computation
of the conservative dynamics of gravitationally interacting binary systems at the fifth post-Newtonian (5PN) level,
together with its extension at the fifth-and-a-half post-Newtonian (5.5PN) level.  We present also the sixth post-Newtonian (6PN)
contribution to the third-post-Minkowskian (3PM) dynamics.
Our strategy combines 
several theoretical formalisms: post-Newtonian, post-Minkowskian, multipolar-post-Minkowskian, gravitational self-force,
effective one-body, and Delaunay averaging.   We determine the full functional structure of the 5PN Hamiltonian (which 
involves 95 non-zero numerical coefficients), except for  two undetermined 
coefficients proportional to the cube of the symmetric mass ratio, and to the fifth and sixth power of the gravitational constant, $G$. 
We present  not only the 5PN-accurate, 3PM contribution to the scattering angle, but also
 its 6PN-accurate generalization.   Both results agree with the corresponding truncations of the recent 3PM
result of Bern et al. [Phys.\ Rev.\ Lett.\  {\bf 122},  201603 (2019)].
 We also compute the 5PN-accurate, fourth-post-Minkowskian (4PM) contribution to the scattering angle,
 including its nonlocal contribution, thereby offering checks for future  4PM calculations.
We point out a remarkable hidden simplicity of the gauge-invariant functional relation between the radial
action and the effective-one-body energy and angular momentum.
\end{abstract}

\maketitle

\section{Introduction}

The main tool used up to now  for the theoretical description of the general relativistic dynamics of a two-body system is the post-Newtonian 
(PN) formalism \cite{Schafer:2018kuf,Blanchet:2013haa}. It encodes the corrections to the Newtonian Hamiltonian due to the 
weak-field,  slow-motion, and small-retardation interaction between the bodies, expressed as a power series in inverse powers of the speed of light $c$. The PN knowledge of the conservative dynamics must then be completed by an analytical description of the gravitational-wave
emission and back-reaction. The main tool currently used for the latter task is the 
(PN-matched~\cite{Blanchet:1987wq,Blanchet:1989ki,Poujade:2001ie}) multipolar-post-Minkowskian (MPM) formalism~\cite{Blanchet:1985sp}.

The present status of PN knowledge is the fourth post-Newtonian (4PN) accuracy, corresponding to $O(1/c^8)$ fractional
corrections to the Newtonian Hamiltonian.
A conceptually (and technically) important new feature of the 4PN Hamiltonian is the presence of a non-local-in-time interaction due
to tail-transported large-time-separation correlations \cite{Blanchet:1987wq}. The current direct perturbative
computations  of the 4PN-level reduced action~\cite{Damour:2014jta,Jaranowski:2015lha,Bernard:2015njp,Damour:2016abl,Marchand:2017pir,Foffa:2019rdf,Foffa:2019yfl,Blumlein:2020pog} have succeeded in tackling this time-non-locality issue in various ways.
However, this variety of approaches, which included discrepant intermediate results~\cite{Bernard:2015njp} before complete agreement was reached, shows that straightforward perturbative PN computations have reached their limit of 
easily verifiable reliability This clearly implies that any $n$-PN computation, with $n \geq 5$, is significantly
 more challenging than  lower-order ones. Let us note in this respect that the recent 5PN-level works~\cite{Foffa:2019hrb,Blumlein:2019zku}
 based on using the standard PN expansion have  computed only the  small,
and non gauge-invariant, subset of ``static''  contributions to the 5PN Hamiltonian. 

The present status of complete MPM knowledge of the  gravitational-wave emission is the third-and-a-half post-Newtonian (3.5PN) 
level (see Ref. \cite{Blanchet:2013haa} for a review). The MPM formalism  led to the  
discovery of (tail-transported) nonlocal dynamical correlations at the 4PN level \cite{Blanchet:1987wq} (later discussed within a different perspective in 
Refs.~\cite{Foffa:2011np,Ross:2012fc,Galley:2015kus}). When projected on the {\it conservative} (time-symmetric)
dynamics the 4PN tail effects lead to a nonlocal action~\cite{Damour:2014jta,Jaranowski:2015lha,Bernard:2015njp,Damour:2016abl,Marchand:2017pir,Foffa:2019rdf,Foffa:2019yfl}. Here, we shall make use of the 5PN-accurate generalization
of the latter tail-related action, first obtained by using results of the MPM formalism in section IXA of ~\cite{Damour:2015isa}, and 
recently discussed within a different perspective in  Ref.~\cite{Foffa:2019eeb} (see also \cite{Blanchet:2019rjs}). Note that the
MPM formalism is used here both to discuss tail-transported correlations and to control the needed PN-corrected multipole
moments.

In view of this situation, we have recently introduced \cite{Bini:2019nra} a new strategy for computing
 the conservative two-body dynamics to higher PN orders. This strategy combines information from different formalisms besides the 
 PN and MPM ones, namely:  
 gravitational Self Force (SF) theory (see, e.g., Ref. \cite{Barack:2018yvs} for a recent review),  post-Minkowskian (PM) 
 theory (see, e.g., Ref.~\cite{Damour:2017zjx,Bern:2019nnu,Antonelli:2019ytb,Bern:2019crd} for latest achievements),  
 effective one-body (EOB) theory \cite{Buonanno:1998gg,Damour:2000we}, and Delaunay averaging \cite{BrouwerClemence}.
The SF formalism has previously allowed the computation of several gauge-invariant quantities (redshift factor, gyroscope precession angle, etc.) at very high PN orders, but its validity is limited to  small values of the mass ratio between the bodies (and to the first order up to now). SF computations do not distinguish local from nonlocal parts of the various quantities, and give results that include both parts.
The PM formalism is a weak-field expansion in powers of the gravitational constant $G$, which does not make any slow-motion assumption. 
An explicit spacetime metric associated with a two-body system was computed at the second postMinkowskian (2PM) 
approximation in the 80s \cite{Bel:1981be}. The corresponding 2PM-accurate equations of motion (and scattering angle) were
computed at the time \cite{Westpfahl:1979gu,Bel:1981be,Westpfahl:1985}. A corresponding 2PM-accurate Hamiltonian was
computed recently \cite{Damour:2017zjx} (see also \cite{Cheung:2018wkq}). A recent breakthrough work of 
Bern et al.  \cite{Bern:2019nnu,Bern:2019crd} has deduced a 3PM-accurate ($O(G^3)$) scattering angle (and Hamiltonian) from a two-loop quantum
scattering amplitude computation. No other complete 3PM calculation exists at present. As we explain below, one
consequence of our new strategy is to allow for a 3PM-complete computation of the scattering angle at the PN accuracy
at which we implement our method. We give here the details of our 5PN-accurate implementation, include its 5.5PN generalization,
and will also
mention the result of a recent 6PN extension of our method \cite{BDG6PN}. Our results provide a 6PN-level
 confirmation of the $O(G^3)$ scattering angle of Refs. \cite{Bern:2019nnu,Bern:2019crd}. A similar confirmation was
 independently recently obtained, within a different approach, in Ref. \cite{Blumlein:2020znm}.

Combining PN, SF, and PM information is efficiently done within the EOB formalism,  which condenses any available 
 analytical information (including nonlocal information) into a few gauge-fixed potentials. See, for example, the EOB
 formulation of the full (nonlocal) 4PN dynamics in Ref. \cite{Damour:2015isa}. We shall use below the EOB formalism
 as a convenient common language for extracting and comparing the gauge-invariant information contained in
 various other formalisms.

Here we detail the application of our new strategy to the 5PN level. 
Essentially, we complete the 5PN-accurate (tail-related) nonlocal part of the
action by constructing a complementary 5PN-accurate local Hamiltonian. The latter local Hamiltonian is obtained, modulo two undetermined
 coefficients, by combining the result of a new SF computation to sixth order in eccentricity with a general
 result  within EOB-PM theory concerning the mass-ratio dependence of the scattering angle \cite{Damour:2019lcq}.
 The transcription of the SF result into dynamical information is obtained by combining the first law of binary dynamics
 ~\cite{LeTiec:2011ab,Barausse:2011dq,Tiec:2015cxa} with the EOB formalism.
 
In principle, our method can be extended to higher PN orders. We have recently been able to extend it to the next two
PN levels, namely the 5.5PN and 6PN levels. We present below our computation of the 5.5PN Hamiltonian. Our results
extend previous studies of 5.5PN effects \cite{Shah:2013uya,Bini:2013rfa,Blanchet:2013txa}, and do not rely on
SF computations but on the 5.5PN conservative action obtained in  Ref. \cite{Damour:2015isa}. We leave to a future
publication the details of the extension of our strategy to the 6PN level, and only cite its consequences at the $G^3$ order.

Note that, at each PN order, our strategy leaves undetermined a relatively small number of
coefficients multiplying the cube of the symmetric mass ratio $\nu$ (defined below). [On the other hand,
we can determine many other coefficients entering the Hamiltonian multiplied by higher powers of $\nu$.]
 Computing these missing coefficients presents a challenge that must be tackled by a complementary
 method. However, we wish to stress that our present 5PN-accurate results (as well as their
 5.5PN and 6PN extensions) are complete at the 3PM and 4PM levels.
 In other words, all the terms $O(G^3)$ and $O(G^4)$ in the  Hamiltonian are fully derived by our method at
 the PN accuracy of its implementation.
 It is this property which allows us to probe the recent 3PM result of Refs. \cite{Bern:2019nnu,Bern:2019crd}
 at the  6PN level, and to make predictions about the 4PM dynamics. 

We denote the masses of the two bodies as $m_1$ and $m_2$.  We then define:
the reduced mass of the system $\mu\equiv m_1 m_2/(m_1+m_2)$, the total mass $M=m_1+m_2$,
and the symmetric mass ratio
\beq
\nu=\frac{m_1m_2}{(m_1+m_2)^2}\,.
\eeq
We use a mostly plus signature. Depending on the context we shall sometimes keep all $G$'s and $c$'s,
and sometimes set them (especially $c$) to one. Beware also that it is often convenient to work with 
dimensionless rescaled quantities, such as radial distance, momenta, Hamiltonian, orbital frequency, etc.

To help the reader to follow the logic of our strategy, let us sketch the plan of our paper:
Working in harmonic coordinates, we first compute the 5PN-accurate nonlocal part of the action\footnote{It will be
convenient to introduce some additional flexibility in the definition of the nonlocal action. For simplicity, we do not mention 
this technical detail here.}. We then consider
an ellipticlike bound state motion and take the (Delaunay) time
 average of the associated nonlocal (harmonic-coordinates)  Hamiltonian 
 \beq
 \langle\delta H_{\rm  nonloc}^{\rm 4PN+5PN, h}\rangle
 = \frac1{\oint dt_h} \oint \delta H_{\rm  nonloc}^{\rm 4PN+5PN, h}(t_h) dt_h \,.
 \eeq  
 In this way we get a gauge invariant function of two orbital parameters. We use here as orbital parameters some 
 harmonic-coordinates semi-latus rectum $a_r^h$ and eccentricity $e_t^h$, but these are known functions of the 
 energy and angular momentum. 

Next, parametrizing with unknown coefficients the nonlocal part of the Hamiltonian expressed in EOB-coordinates (labelled with ``e'',
instead of ``h''), we compute the corresponding Delaunay average
\beq
 \langle\delta H_{\rm eob, nonloc}^{\rm 4PN+5PN, e}\rangle
 = \frac1{\oint dt_e} \oint \delta H_{\rm eob, nonloc}^{\rm 4PN+5PN, e}(t_e) dt_e \,.
 \eeq  
Identifying the two Delaunay averages (when using the 1PN-accurate relation between the harmonic-coordinates orbital parameters 
$a_r^h, e_t^h$ and the corresponding EOB parameters $a_r^e, e_t^e$) then determines  the unknown coefficients used to parametrize 
the 5PN nonlocal part of the EOB Hamiltonian.

Having in hands the latter 5PN-accurate nonlocal part of the EOB Hamiltonian, we then determine the complementary
5PN-accurate local part of the EOB Hamiltonian. 
This is done by using SF information about small-eccentricity ellipticlike motions.
 Namely, we first compute the averaged redshift factor \cite{Barack:2011ed} to the sixth order in eccentricity.
 We had to generalize to the sixth order
previous results that extended only to the fourth order in eccentricity \cite{Bini:2015bfb,Bini:2016qtx}.  To relieve the tedium, we
relegated some of our derivations and results to Appendices. We notably list in Appendix \ref{app_self_force} the result of our 
SF computation of the (averaged) redshift factor along eccentric orbits in the Schwarzschild spacetime (accurate to the 9.5PN level), 
and its conversion  into the EOB potential $q_6$ through the first law of eccentric binaries \cite{Tiec:2015cxa}.

This determines the sum of the local and the nonlocal
EOB Hamiltonian, but only at the second order in the symmetric mass ratio $\nu$. [Here, we are talking about
the unrescaled Hamiltonian, such that the test-particle Hamiltonian is $O(\nu)$.] Subtracting the above-determined
nonlocal EOB Hamiltonian determines the local part of the EOB Hamiltonian up to $O(\nu^2)$ included (corresponding
to an $O(\nu)$ knowledge of the potentials entering the effective EOB Hamiltonian).

Ref. \cite{Damour:2019lcq} has recently uncovered a simple property of the $\nu$-dependence of the scattering angle 
for hyperbolic encounters. This property plays a crucial role in allowing us to  complete the previously discussed
$O(\nu^2)$ SF-based knowledge of the Hamiltonian, and to determine most of the $O(\nu^{n \geq 3})$
contributions to the Hamiltonian. In order to use the result of Ref. \cite{Damour:2019lcq} (which concerns the
structure of the total scattering angle $\chi^{\rm tot} = \chi^{\rm loc} + \chi^{\rm nonloc}$) two separate steps
are needed. On the one hand, we need to compute the nonlocal contribution $\chi^{\rm nonloc}$ to the scattering angle by 
generalizing the technique used at 4PN in \cite{Bini:2017wfr}. On the other hand, it is convenient, in order to separately
compute the local contribution  $\chi^{\rm loc}$ to the scattering angle, to convert the local EOB Hamiltonian, so far
obtained in the standard $p_r$-gauge \cite{Damour:2000we}, into the so-called energy-gauge \cite{Damour:2017zjx}.
Indeed, the latter gauge  is more convenient for discussing hyperboliclike scattering motions. The computation of
 the total scattering angle $\chi^{\rm tot} = \chi^{\rm loc} + \chi^{\rm nonloc}$, together with 
  the knowledge of the exact 2PM EOB Hamiltonian, then allows us to 
fix most of the parametrizing coefficients of the EOB potentials (actually all coefficients with two exceptions only: $\bar d_5^{\nu^2}$ and $a_6^{\nu^2}$, i.e., the $O(\nu^2)$ coefficients of the local potentials $\bar D$ and $A$ at 5PN).

Besides the results just summarized (which constitute the core of the present work), let us highlight other new results obtained
below as  by-products of our computations:

\begin{enumerate}
\item We have evaluated the averaged value of the 5.5PN Hamiltonian. It is entirely
given by the (scale-independent) second-order-in-tail nonlocal Hamiltonian $H_{{\rm tail}^2}$, from which we have computed the half-PN-order coefficients $A_{6.5}$, $\bar D_{5.5}$, $q_{4,4.5}$, $q_{6,3.5}$, and $q_{8,2.5}$. The last one,  $q_{8,2.5}$ is new and a prediction for future SF calculations (see Section VI).
\item We have shown how to use an (inverse) Abel transform  to compute in closed-form the standard $p_r$-gauge 
version of the 2PM  energy-gauge EOB potential $q_{2\rm EG}$ (see Appendix \ref{Abel_tr}).
\item We have explicitly computed the local contribution to the 5PN radial action, as well as the
corresponding local Delaunay Hamiltonian ({\it i.e.}, the local Hamiltonian expressed in terms of action variables).
We find that the radial action has a remarkably simple structure.
See Section \ref{delaunay}.
\end{enumerate}

\section{The 5PN-accurate nonlocal  action and its associated Hamiltonian} 

The complete, reduced two-body conservative action ($S_{\rm tot}$)  can be decomposed, at any given PN accuracy,
by using the  PN-matched~\cite{Blanchet:1987wq,Blanchet:1989ki,Poujade:2001ie}  
multipolar-post-Minkowskian (MPM) formalism~\cite{Blanchet:1985sp},
in two separate pieces: a nonlocal-in-time part ($S_{\rm nonloc}$) 
and a local-in-time part ($S_{\rm loc}$) ,
\beq
\label{Sdecomp}
S_{\rm tot}^{\leq n \rm PN}= S_{\rm loc, f}^{\leq n \rm PN}+S_{\rm nonloc, f}^{\leq n \rm PN}\,.
\eeq
Here each action piece is a time-symmetric functional of the worldlines of the two bodies, say $x_1(s_1)$ and $x_2(s_2)$.
The original total action $S_{\rm tot}[x_1(s_1),x_2(s_2)]$ (before approximating it at some PN accuracy)  is
defined as a PM-expanded Fokker action~\cite{Damour:1995kt}. The PN-truncated
 nonlocal action $S_{\rm nonloc, f}^{\leq n \rm PN}$ (which starts at the 4PN level~\cite{Blanchet:1987wq,Damour:2014jta})
 is defined by using the MPM formalism. Its 5PN-accurate value was first obtained in Section IXA of Ref. ~\cite{Damour:2015isa}
 (based on the effective action used in Ref.~\cite{Damour2010}).
 It was recently derived in a different (though related) way in Ref.~\cite{Foffa:2019eeb}. [See also 
 Refs.~\cite{Blanchet:2010zd,LeTiec:2011ab} for the related 5PN logarithmic terms, and Ref. \cite{Blanchet:2019rjs}
 for higher-order tail-related logarithms.]. From Eq. (9.12) of ~\cite{Damour:2015isa}, it reads
\begin{eqnarray} 
\label{Snonloc}
S_{\rm nonloc, f}^{4+5 \rm PN}[x_1(s_1), x_2(s_2)]&=& \frac{G{\cal M}}{c^3} \int dt {\rm Pf}_{2 r_{12}^f(t)/c}\times  \nonumber\\
&& \int  \frac{ dt'}{|t-t'|} {\cal F}_{1 \rm PN}^{\rm split}(t,t')\,.
\end{eqnarray} 
Here, ${\cal M}$ denotes the total ADM conserved mass-energy of the binary system, while ${\cal F}_{1 \rm PN}^{\rm split}(t,t')$ is
the time-split version of the fractionally 1PN-accurate gravitational-wave energy flux emitted by the system, namely 
\begin{eqnarray}
\label{flux1PNdef}
 &&{\cal F}_{1 \rm PN}^{\rm split}(t,t')= \frac{G}{c^5} \left( \frac15 I_{ab}^{\rm (3)}(t) I_{ab}^{\rm (3)}(t') \right. \nonumber\\
 &&\left.+  
  \frac1{189 c^2} I_{abc}^{\rm (4)}(t) I_{abc}^{\rm (4)}(t') +\frac{16}{45 c^2} J_{ab}^{\rm (3)}(t) J_{ab}^{\rm (3)}(t') \right),
 \end{eqnarray}
where the superscript in parenthesis  denotes  repeated time-derivatives. The specific choice of the time scale $2 r_{12}^f(t)/c$ entering
the {\it partie finie} (Pf) operation  used in the definition of the nonlocal action,
Eq. \eqref{Snonloc} (whose integral over $t'$ is  logarithmically divergent when $t' \to t$) will be discussed below.

The quantities $I_{ab}$, $I_{abc}$, $J_{ab}$ entering Eq. \eqref{flux1PNdef} are the MPM-derived Blanchet-Damour (1PN-accurate)  mass and spin multipole moments defined by suitable integrals over the stress-energy tensor of the source~\cite{Blanchet:1989ki}. Their (center-of-mass, harmonic coordinates) expressions for a binary system read (see Eqs. (3.32) and (3.33) of Ref.~\cite{Blanchet:1989cu})  
\begin{eqnarray}
I_{ij}
&=& \mu r_{\langle i j \rangle} \left[1+\frac{29}{42c^2}(1-3\nu)v^2-\frac{(5-8\nu)}{7c^2} \frac{GM}{r} \right]\nonumber\\
&+&\mu \frac{1-3\nu}{21c^2}\left[-12 ({\mathbf v}\cdot {\mathbf r})r_{\langle i}v_{j \rangle}+11 r^2 v_{\langle ij \rangle}  \right]\,,
\nonumber\\
I_{ijk}&=& \mu \sqrt{1-4\nu} r_{\langle ijk \rangle}\,,\nonumber\\
J_{ij}&=& \mu \sqrt{1-4\nu}\epsilon_{kl \langle i}r_{j\rangle k} v_l\,,
\end{eqnarray}
with $M\sqrt{1-4\nu}=m_2-m_1$ and
\begin{eqnarray}
r_{\langle ijk \rangle}&=& r_{ijk}-\frac{3}5 r^2 \delta_{(ij}r_{k)}\,,\nonumber\\
\epsilon_{kl \langle i}r_{j\rangle k} v_l&=&({\mathbf r}\times {\mathbf v})_{(i}  r_{j)}\,,
\end{eqnarray}
where the standard notations  $A^{ijk\ldots}=A^i A^j A^k\ldots$ for tensorial products, $S^{(ij)}=\frac12(S^{ij}+S^{ji})$ for the symmetric part of a tensor, and $S^{\langle ij\rangle}$ for the symmetric and trace-free part of a tensor have been used.

As stated above, Eq. \eqref{Snonloc} defines (for any choice of  $r_{12}^f$) an explicit functional of the two worldlines, and subtracting it from the (in principle PM-computable) total action $S_{\rm tot}$ defines the corresponding local-in-time contribution 
$S_{\rm loc, f}^{\leq 5 \rm PN}$ to the two-body dynamics. There is some {\it flexibility} in the choice
of the time-scale $2 r_{12}^f/c$ entering the {\it partie finie} (Pf) operation used in Eq. \eqref{Snonloc}. Let us first
 point out that the meaning here of $S_{\rm loc}$ (and its corresponding $H_{\rm loc}$)  
differs from the one in Refs. \cite{Damour:2015isa,Bini:2017wfr},  where the time-scale entering the partie-finie defining $S_{\rm nonloc}$
was taken to be a fixed scale $2 s/c$. The explicit results of  Ref. \cite{Damour:2015isa} show that, with such a choice,
the 4PN-accurate local Hamiltonian $H_{\rm loc}$ then includes several terms proportional to the logarithm $\ln (r_{12}/s)$.
Choosing as length scale $s$ the radial distance $r_{12}$ between the two bodies has therefore the technical advantage
of simplifying the local part of the Hamiltonian by removing all logarithms from it. At the 4PN level, a Newtonian-accurate definition
of the radial distance $r_{12}$ is adequate. However, as we are now working at a higher PN accuracy we need to define the
time-scale $2 r_{12}^f/c$ with at least 1PN fractional accuracy. Let us emphasize that the choice of any precise definition
of $r_{12}^f$ is purely conventional, and will affect in no way the end results of our methodology. Indeed, the total
action $S_{\rm loc, f}^{\leq n \rm PN}+S_{\rm nonloc, f}^{\leq n \rm PN}$ will always be defined so as to be independent
of the flexibility in the definition of $r_{12}^f$. Only the separation between $S_{\rm loc, f}$ and $S_{\rm nonloc, f}$
depends on this flexibility. 
Though it would be perfectly acceptable (and would lead, when consistently used, to the same final results)
to use everywhere the harmonic-coordinate radial distance  $r_{12}^h$ as length scale,
we shall show here that there are some technical advantages to employing a more
general scale of the general form
\beq
r_{12}^f(t) = f(t) \, r_{12}^h(t) \; , 
\eeq
where $f(t)=1+O(\frac1{c^2})$ is a combination of dynamical variables of the type 
\beq \label{f}
f(t) =1+ c_1 \left( \frac{ p_r}{\mu c} \right)^2 + c_2\left( \frac{{\bf p}}{\mu c} \right)^2  + c_3 \frac{GM}{ r c^2} + \ldots
\eeq
A convenient criterion for choosing the (5PN-level) {\it flexibility} parameter $f(t)$ will be discussed below. It will imply,
in particular, the fact that the dimensionless coefficients  $c_1, c_2, c_3, \ldots$ entering the definition of $f(t)$ are
proportional to the symmetric mass ratio $\nu$.

It is convenient to rewrite $S_{\rm nonloc}^{4+5 \rm PN}$ as 
\beq
S_{\rm nonloc, f}^{4+5 \rm PN}= - \int dt\, \delta H_{\rm  nonloc, f}^{4+5 \rm PN}(t)\,,
\eeq
and to rewrite $\delta H_{\rm  nonloc, f}^{4+5 \rm PN}(t)$ as
\beq \label{deltaHnlf}
\delta H_{\rm  nonloc, f}^{4+5 \rm PN}(t) = \delta H_{\rm  nonloc, h}^{4+5 \rm PN} + \Delta^{\rm f-h} H(t)\,,
\eeq
where 
\begin{eqnarray} 
\label{delta_H_nonloc}
\delta H_{\rm nonloc, h}^{4+5 \rm PN}(t)&=&
-\frac{G^2 H_{}}{c^{5}}{\rm Pf}_{2s/c}\int \frac{d\tau}{|\tau|}{\mathcal F}^{\rm split}_{\rm 1PN}(t,t+\tau)\nonumber\\
&+&2\frac{G^2 H_{}}{c^{5}}{\mathcal F}^{\rm split}_{\rm 1PN}(t,t)  \ln \left( \frac{r_{12}^h(t)}{s}\right)
\,,
\end{eqnarray} 
and
\beq \label{DHfh}
\Delta^{\rm f-h} H(t)= + 2\frac{G^2 H_{}}{c^{5}}{\mathcal F}^{\rm split}_{\rm 1PN}(t,t)  \ln \left( f(t)\right)\,.
\eeq
Here ${\mathcal F}^{\rm split}(t,t) \equiv {\mathcal F}^{GW}(t)$ is the instantaneous gravitational-wave energy flux, so that
the flexibility term $\Delta^{\rm f-h} H(t)$ is a purely local additional contribution to $\delta H_{\rm  nonloc, h}$.
In the following, we will keep indicating by a label $f$ or $h$ nonlocal (or local) contributions that depend on choosing
as partie-finie scale $2 r_{12}^f/c$ or $2 r_{12}^h/c$.

We now compute the time average of $\delta H_{\rm real, nonloc, h}^{4+5 \rm PN}(t)$ along an elliptic-like bound-state motion,
using its well known (harmonic coordinates) 1PN-accurate quasi-Keplerian parametrization \cite{DD1}, i.e.,
\beq
\langle\delta H_{\rm nonloc, h}^{4+5 \rm PN} \rangle\equiv  \frac1{\oint dt} \oint \delta H_{\rm  nonloc, h}^{4+5 \rm PN}(t) dt\,.
\eeq
The quasi-Keplerian parametrization of the orbit (needed at the 1PN level of accuracy for the purposes of the present paper) is summarized in Table \ref{table_relations}. The functional relations shown there are also valid at 1PN in Arnowitt-Deser-Misner (ADM) coordinates 
(which start differing only at 2PN from harmonic coordinates) and in EOB coordinates (with different numerical values of the orbital elements
 $a_r$ and $e_t$).  See \cite{Damour:1988mr} for the 2PN generalization of the
quasi-Keplerian parametrization.

The temporal average is conveniently transformed as an integral over the azimuthal angle, namely
\begin{eqnarray}
\langle \delta H_{\rm }^{h}\rangle&=&\frac{n}{2\pi}\int_0^{2\pi/n}  \delta H_{\rm }^{h} dt\nonumber\\
&=&\frac{n}{2\pi}\int_0^{2\pi K}  \frac{\delta H_{\rm }^{h}}{\dot \phi^h} d\phi^h\nonumber\\
&=&\frac{nK}{2\pi}\int_0^{2\pi }  \frac{\delta H_{\rm }^{h}}{\dot \phi^h}\bigg|_{\phi^h=K\bar \phi^h} d\bar \phi^h\,,
\end{eqnarray} 
where $\bar \phi^h \equiv\phi^h/K$ (see the caption of Table \ref{table_relations} for the definition of the various orbital parameters).
The result of the average is a gauge-invariant function (say $F$) of a set of independent orbital parameters. The latter are chosen here to be the (harmonic-coordinates) semi-latus rectum $a_r^h$ and eccentricity $e_t^h$, so that
\beq
\langle \delta H_{\rm }^{h}\rangle \equiv \langle\delta H_{\rm  nonloc}^{4+5 \rm PN} \rangle= F^h(a_r^h,e_t^h)
 \,,
\eeq
with
\begin{eqnarray}
F^h(a_r^h,e_t^h)&=&\frac{\nu^2}{(a_r^h)^5}\left[{\mathcal A}^{\rm 4PN}(e_t^h)+{\mathcal B}^{\rm 4PN}(e_t^h)\ln a_r^h\right]\nonumber\\
&&
+\frac{\nu^2}{(a_r^h)^6}\left[{\mathcal A}^{\rm 5PN}(e_t^h)+{\mathcal B}^{\rm 5PN}(e_t^h)\ln a_r^h\right]
\,.\nonumber\\
\end{eqnarray} 
The  expansion in powers of $e_t^h$ of the coefficients parametrizing $F^h(a_r^h,e_t^h)$ are listed in Table \ref{table_F_h} 
up to the order $O((e_t^h)^{10})$ included. [We use here  $G=1=c$.]

Let us also mention that the average of the $f$-related contribution (with the parametrization \eqref{f}), Eq. \eqref{DHfh}, 
to the nonlocal Hamiltonian,  starts at the 5PN order, and reads
\begin{eqnarray}
&& \frac1{\oint dt} \oint \Delta^{\rm f-h} H(t) dt =\frac{\nu^2}{a_r^6} 
\left[
\frac{64}{5}  (c_2+c_3) \right. \nonumber\\
&&\qquad +\frac{8}{15} (229 c_3+12 c_1+301 c_2) e_t^2\nonumber\\
&&\qquad
+\frac{2}{5} (1183 c_3+1761c_2+97 c_1) e_t^4\nonumber\\
&&\qquad
+\frac{14}{15} (134 c_1+1341 c_3+2137 c_2) e_t^6\nonumber\\
&&\qquad+\frac{7}{4} (1529 c_3+171 c_1+2543 c_2) e_t^8 \nonumber\\
&&\qquad\left. +\frac{231}{80} (208 c_1+2961 c_2+1729 c_3) e_t^{10}
\right]\nonumber\\
&&\qquad+O(e_t^{12})\,.
\end{eqnarray} 

The functional dependence on the orbital parameters  $a_r^h$ and  $e_t^h$ could be replaced by a dependence
on the gauge-invariant energy and angular momentum (see below).


\begin{table*}
\caption{\label{table_F_h} Coefficients of the averaged nonlocal Hamiltonian (with scale $2 r_{12}^h/c$) in harmonic coordinates}
\begin{ruledtabular}
\begin{tabular}{ll }
Coefficient & Expression \\
\hline
${\mathcal A}^{\rm 4PN}(e_t^h)$ & $\frac{128}{5}\ln(2)+\frac{64}{5}\gamma
+\left(\frac{1256}{15}\gamma-\frac{176}{5}+\frac{729}{5}\ln(3)+\frac{296}{15}\ln(2)\right)(e_t^h)^2$\\
& $+\left(\frac{29966}{15}\ln(2)-\frac{2681}{15}-\frac{13851}{20}\ln(3)+242\gamma\right)(e_t^h)^4$\\
& $+\left(\frac{1953125}{576}\ln(5)+\frac{1526}{3}\gamma+\frac{419661}{320}\ln(3)-\frac{90017}{180}-\frac{116722}{15}\ln(2)\right)(e_t^h)^6$\\
& $+\left(\frac{3605}{4}\gamma-\frac{83984375}{4608}\ln(5)-\frac{306433}{288}+\frac{5381201}{180}\ln(2)+\frac{26915409}{2560}\ln(3)\right)(e_t^h)^8$\\
& $+\left(-\frac{4697998651}{54000}\ln(2)-\frac{138733913079}{2048000}\ln(3)+\frac{678223072849}{18432000}\ln(7)-\frac{18541327}{9600}\right.$\\
& $\left.
+\frac{18736328125}{442368}\ln(5)+\frac{114807}{80}\gamma\right)(e_t^h)^{10}$\\
\hline
${\mathcal B}^{\rm 4PN}(e_t^h)$ & $-\frac{32}{5}-\frac{628}{15}(e_t^h)^2-121(e_t^h)^4-\frac{763}{3}(e_t^h)^6-\frac{3605}{8}(e_t^h)^8-\frac{114807}{160}(e_t^h)^{10}$\\
\hline
${\mathcal A}^{\rm 5PN}(e_t^h)$ & $\left(-\frac{11708}{105}-\frac{112}{5}\nu\right)\gamma+\left(-\frac{25276}{105}+\frac{912}{35}\nu\right)\ln(2)+
\left(-\frac{486}{7}\nu+\frac{243}{14}\right)\ln(3)+\frac{32}{5}\nu-\frac{96}{5}$\\
& $+\left[\left(\frac{19024}{35}-\frac{57284}{15}\nu\right)\ln(2)+\left(\frac{94041}{70}\nu-\frac{27702}{35}\right)\ln(3)+\left(-\frac{10672}{35}-\frac{1364}{3}\nu\right)\gamma
-\frac{5441}{35}+\frac{4672}{21}\nu\right](e_t^h)^2$\\
& $+\left[\left(-\frac{599911}{105}+\frac{3476231}{105}\nu\right)\ln(2)+\left(-\frac{303993}{896}-\frac{6268671}{1120}\nu\right)\ln(3)
+\left(-\frac{9765625}{672}\nu+\frac{9765625}{2688}\right)\ln(5)\right.$\\
& $\left.
+\left(\frac{14003}{35}-2241\nu\right)\gamma+\frac{59756}{35}\nu-\frac{1160639}{840}\right] (e_t^h)^4$\\
& $+\left[\left(\frac{16502161}{945}-\frac{73289299}{315}\nu\right)\ln(2)+\left(-\frac{70956243}{896}\nu+\frac{23153769}{896}\right)\ln(3)\right.$\\
& $\left.
+\left(-\frac{434140625}{24192}+\frac{162109375}{1152}\nu\right)\ln(5)
+\left(-\frac{20027}{3}\nu+\frac{17797}{5}\right)\gamma+\frac{474653}{72}\nu-\frac{15761437}{2520}\right] (e_t^h)^6$\\
& $+\left[\left(-\frac{604746629}{10080}+\frac{4166409179}{3780}\nu\right)\ln(2)+\left(-\frac{18879921207}{163840}+\frac{237686858217}{286720}\nu\right)\ln(3)\right.$\\
& $\left.
+\left(-\frac{843248046875}{1548288}\nu-\frac{25425078125}{6193152}\right)\ln(5)+\left(\frac{96889010407}{884736}-\frac{96889010407}{221184}\nu\right)\ln(7)\right.$\\
& $\left.
+\left(\frac{356481}{32}-\frac{61285}{4}\nu\right)\gamma-\frac{168508293}{8960}+\frac{2591779}{144}\nu\right](e_t^h)^8$\\
& $+\left[\left(\frac{189246461867}{126000}-\frac{990620463289}{108000}\nu\right)\ln(2)+\left(\frac{2103914638719}{14336000}-\frac{95555936957967}{28672000}\nu\right)\ln(3)\right.$\\
& $\left.
+\left(\frac{68322265625}{1032192}+\frac{2431021484375}{2064384}\nu\right)\ln(5)+
\left(\frac{455281459902493}{110592000}\nu-\frac{11520188079967}{18432000}\right)\ln(7)\right.$\\
& $\left.
+\left(\frac{2035869}{80}-\frac{4794867}{160}\nu\right)\gamma+\frac{255777929}{6400}\nu-\frac{1492974817}{33600}\right](e_t^h)^{10}$\\
\hline
${\mathcal B}^{\rm 5PN}(e_t^h)$ & $\frac{5854}{105}+\frac{56}{5}\nu
+\left(\frac{682}{3}\nu+\frac{5336}{35}\right)(e_t^h)^2
+\left(-\frac{14003}{70}+\frac{2241}{2}\nu\right) (e_t^h)^4
+\left(\frac{20027}{6}\nu-\frac{17797}{10}\right)(e_t^h)^6$\\
& $+\left(\frac{61285}{8}\nu-\frac{356481}{64}\right)(e_t^h)^8
+\left(\frac{4794867}{320}\nu-\frac{2035869}{160}\right) (e_t^h)^{10}$\\
\hline
\end{tabular}
\end{ruledtabular}
\end{table*}


\begin{table}
\caption{\label{table_relations} Quasi-Keplerian representation of the 1PN motion (valid both in harmonic and EOB coordinates with $e_t=e_t^{\rm coord}$, etc). Here $n=2\pi/P$  (with $P$ the radial period) and $K=\Phi/2\pi$ (periastron advance) are gauge invariant, 
while the various eccentricity parameters, $e_t$, $e_r$ and $e_\phi$, and the semi-latus rectum, $a_r$,  are coordinate-dependent;  $u$ 
denotes here the eccentric anomaly (not to be confused with the inverse radial variable largely used in the rest of the paper). We use here
mass-rescaled  radial variables: $r \equiv r^{\rm phys}/(G M)$, $a_r \equiv a_r^{\rm phys}/(G M)$.}
\begin{ruledtabular}
\begin{tabular}{l }
\hline
$ nt=\ell=u-e_t\sin u$ \\
$r=a_r(1-e_r\cos u)$\\
$\dot r =\frac{n a_r e_r \sin u}{1-e_t \cos u}$\\
$\phi-\phi_P=2K\arctan \left[\left( \frac{1+e_\phi}{1-e_\phi} \right)^{1/2}\tan \frac{u}{2}  \right]$\\
$\dot \phi =n K \frac{\sqrt{1-e_\phi^2}}{(1-e_t \cos u)(1-e_\phi \cos u)}$\\
$\tan \frac{u}{2}=\sqrt{\frac{1-e_\phi}{1+e_\phi}}\tan \frac{\bar \phi}{2}\,,\qquad \bar \phi\equiv\frac{\phi}{K}$\\ 
$\sin u=\frac{\sqrt{1-e_\phi^2} \sin \bar \phi}{1+e_\phi \cos \bar \phi}\,,\qquad 
\cos u =\frac{\cos \bar \phi+e_\phi}{1+e_\phi \cos \bar \phi}$
\end{tabular}
\end{ruledtabular}
\end{table}

Using the known transformation between harmonic and EOB coordinates \cite{Buonanno:1998gg,Damour:2000we},
the relations between the harmonic-coordinates orbital parameters $(a_r^h,e_t^h)$ and the corresponding EOB-coordinate orbital parameters $(a_r^e,e_t^e)$, read (with $\eta\equiv1/c$)
\beq
a^h_r= a_r^e-\eta^2\,,\qquad e_t^h=e_t^e\left( 1+\frac{\nu}{a_r^e}\eta^2 \right)\,.
\eeq
 The invariant function $F^h$ can then be reexpressed in terms of the corresponding EOB parameters
\beq \label{Delaunaynonloc}
F^h(a_r^h,e_t^h)=\tilde F^h(a_r^e,e_t^e)\,.
\eeq
We list the relations between the various harmonic-coordinate orbital parameters and the corresponding EOB-coordinates ones as functions of (rescaled) energy ($\bar E \equiv (H_{\rm }-M c^2)/\mu$) and angular momentum ($j \equiv J/(G m_1 m_2)$) in Table \ref{table_h_and_EOB}.\footnote{
Throughout the paper we use several energy-related variables,  $\bar E$, $\tilde E$, $\bar E_{\rm eff}$, etc., which are defined in the
section where they appear.
}


\begin{table*}
\caption{\label{table_h_and_EOB}1PN quasi-Keplerian orbital parameters as functions of 
$\bar E\equiv (H-M)/\mu$ and $j \equiv J/(G M \mu)$}
\begin{ruledtabular}
\begin{tabular}{lll}
Orbital parameter, $X$ & harmonic, $X^h$ & EOB, $X^e$ \\
\hline
$a_r$ & $\frac{1}{(-2\bar E)}\left[1+\eta^2\frac{7-\nu}{2}\bar E  \right]$ & $\frac{1}{(-2\bar E)}\left(1-\frac{\nu-3}{2}\bar E \eta^2 \right)$\\
$n $ & $(-2\bar E)^{3/2}\left[1+\eta^2\frac{15-\nu}{4}\bar E  \right] $& $(-2\bar E)^{3/2}\left(1+\frac{15-\nu}{4}\bar E \eta^2 \right)$\\
$e_t $& $\left\{1+ 2\bar Ej^2 \left[1+\left(\frac{17}{2}-\frac{7}{2}\nu \right) \bar E \eta^2 +\frac{2-2\nu}{j^2}\eta^2  \right]  \right\}^{1/2} $ & $\left\{ 1+2\bar E j^2 \left[1+ \left(\frac{\nu+17}{2}\bar E+\frac{2}{j^2}\right)\eta^2  \right] \right\}^{1/2} $\\
$e_r$ &$\left\{1+ 2\bar Ej^2 \left[1+\left(-\frac{15}{2}+\frac{5}{2}\nu \right) \bar E \eta^2 +\frac{-6+\nu}{j^2}\eta^2  \right]   \right\}^{1/2}$& $\left\{ 1+2\bar E j^2 \left[1+ \left(\frac{\nu-7}{2}\bar E-\frac{4}{j^2}\right)\eta^2  \right] \right\}^{1/2}$\\
$e_\phi $ &$ \left\{1+ 2\bar Ej^2 \left[1+\left(-\frac{15}{2}+\frac{1}{2}\nu \right) \bar E \eta^2 -\frac{6}{j^2}\eta^2  \right]  \right\}^{1/2}$& $\left\{ 1+2\bar E j^2 \left[ 1+ \left(\frac{\nu-15}{2}\bar E-\frac{6}{j^2}\right)\eta^2 \right]\right\}^{1/2}$ \\
$\delta_r=\frac{e_r}{e_t}-1$ & $(3\nu-8)\bar E \eta^2 $ & $-6\bar E \eta^2$\\
$\delta_\phi=\frac{e_\phi}{e_t}-1$ &$(2\nu-8)\bar E\eta^2 $ & $-8\bar E \eta^2$ \\
$K$ & $1+\frac{3}{j^2}\eta^2$& $1+\frac{3}{j^2}\eta^2$\\
$k$ &$\frac{3}{a_r(1-e_t^2)}$ & $\frac{3}{a_r(1-e_t^2)}$\\
$\bar E$ &$ -\frac{1}{2a_r}-\frac1{2a_r^2}\left(-\frac74 +\frac{\nu}{4}\right)\eta^2$ & $-\frac{1}{2a_r}-\frac1{2a_r^2}\left(-\frac34+\frac{\nu}{4}\right)\eta^2$\\
$j^2$ &$a_r(1-e_t^2)\left(1+\frac{4+2(\nu-3)e_t^2}{j^2}\eta^2 \right)$ & $a_r(1-e_t^2)\left(1-\frac{5e_t^2-3}{j^2}\eta^2 \right)$\\
\end{tabular}
\end{ruledtabular}
\end{table*}
 
Let us  now use the (Delaunay-averaged) 4+5PN information about the nonlocal action contained in Eq. \eqref{Delaunaynonloc}
to compute  a corresponding (squared, rescaled) effective nonlocal 4+5PN-accurate EOB Hamiltonian, say
\beq
\delta [\hat H_{\rm eff, nonloc, h}^{\rm 4PN+5PN}]^2=\delta  \hat H_{\rm eff, h}^2\,.
\eeq
We recall the (universal) EOB link between the usual Hamiltonian and the effective one: 
\beq
H_{\rm eob} = M c^2 \sqrt{1 + 2 \nu(\hat H_{\rm eff}-1)}\,.
\eeq

This is achieved (as at the 4PN level \cite{Damour:2015isa}) by parametrizing a general squared effective EOB
Hamiltonian (in standard $p_r$-gauge \cite{Damour:2000we}) in terms of PN-expanded EOB potentials $A(u)$, $\bar D(u)$ and $Q(u,p_r)$
(where $u \equiv GM/(c^2 r^{\rm phys})=\eta^2/r$, $p_r=\eta p_r^{\rm phys}/\mu$ and $p_\phi=j/\eta$): 
\begin{eqnarray}
\delta  \hat H_{\rm eff}^2(u,p_r,p_\phi)&=&(1+2p_r^2+p_\phi^2 u^2)\delta A(u)\nonumber\\
&&+ (1-4u) p_r^2 \delta \bar D(u)\nonumber\\
&&+(1-2u)\delta Q(u,p_r)\,.
\end{eqnarray}
Here the notation $\delta$ refers to the looked-for additional 4+5PN nonlocal contribution, and we have
written the right-hand side at the needed 1PN fractional accuracy. The general parametrization of 
$\delta A(u)$, $\delta \bar D(u)$ and $\delta Q(u,p_r)$ read
\begin{eqnarray} \label{ADQnonloc}
\delta A&=&a_5^{\rm nonloc}u^5+a_6^{\rm nonloc}u^6\,,\nonumber\\
\delta \bar D &=& \bar d_4^{\rm nonloc}u^4+\bar d_5^{\rm nonloc}u^5\,,\nonumber\\
\delta Q&=&p_r^4 (q_{43}^{\rm nonloc}u^3+q_{44}^{\rm nonloc}u^4)\nonumber\\
&& +p_r^6 (q_{62}^{\rm nonloc}u^2+q_{63}^{\rm nonloc}u^3)\nonumber\\
&& +p_r^8 (q_{81}^{\rm nonloc}u+q_{82}^{\rm nonloc}u^2) + \ldots\,.
\end{eqnarray}
Here $a_5^{\rm nonloc}=a_5^{\rm nl,c}+a_5^{\rm nl,\ln{}}\ln(u)$, etc., are {\it a priori}  unknown (logarithmically varying)
coefficients parametrizing the 4+5PN nonlocal EOB Hamiltonian. Here, we have not indicated any additional label $f$ or $h$,
because the form of Eqs. \eqref{ADQnonloc} is valid for both cases. It is only when computing specific values of the 
$\geq 5$PN EOB parameters $a_6^{\rm nonloc}, \bar d_5^{\rm nonloc}, q_{44}^{\rm nonloc}, \ldots$ that we will need to
specify whether they correspond to the unflexed $h$ case, or to some specified flexed case $f$. [$a_5^{\rm nonloc}$ belongs
to the 4PN approximation and does not depend on the choice of $f=1+O(1/c^2)$.]

When doing explicit computations one needs to
 truncate the $p_r$ expansion of $\delta Q$ to a finite order. The $n$-th order in $p_r^2$ in $\delta Q$ corresponds to the 
$n$-th order in $e^2$ when correspondingly  computing the redshift $\delta z_1$, as we shall do below. 

Converting $\delta  H_{\rm eff}^2$ into the usual Hamiltonian $\delta  H_{\rm eob}$ is straightforward,
\beq
\delta H_{\rm eob}^{4+5 \rm PN}=\frac{\mu M}{2H_{\rm eob} \hat H_{\rm eff}} \delta \hat H_{\rm eff}^2\,,
\eeq
as well as taking the Delaunay time average over the orbital motion. As before, the latter average is conveniently done 
 in terms of an integral over the EOB azimuthal angle by using Hamilton's equations to express $dt^e$ in terms of $d\phi^e$ along the orbit, i.e.,
\beq
dt^e=\frac{H_{\rm eob} \hat H_{\rm eff}}{Au^2j}d\phi^e\,,
\eeq
so that
\begin{eqnarray}
&&\langle \delta H_{\rm eob}^{4+5 \rm PN}\rangle= \frac{\mu M}{2H_{\rm eob}  \hat H_{\rm eff}} 
 \frac{nK}{2\pi} \frac{H_{\rm eob} \hat H_{\rm eff}}{j} \int \frac{\delta  \hat H_{\rm eff}^2 }{ Au^2}d\bar \phi^e\nonumber\\
&&=  
 \frac{\nu M^2 nK}{4\pi j}   \int \frac{\delta  \hat H_{\rm eff}^2 }{ Au^2}d\bar \phi^e\,,
\end{eqnarray}
where the 1PN-accurate expression for $K$ is given in Table \ref{table_h_and_EOB}. 

Let us now focus on the $r_{12}^h$-version of the nonlocal action (considered both in harmonic and EOB coordinates).
In close correspondence to what we have done before in harmonic coordinates, the time average of the
above parametrized EOB nonlocal Hamiltonian  provides a function of the EOB orbital parameters,
say 
\beq
F_{\rm eob \, nonloc, h}^{4+5\rm PN}(a_r^e,e_t^e; a_5^{\rm nonloc}, a_6^{\rm nonloc, h}, \ldots)\,.
\eeq
This invariant function must coincide with the above-computed function $\tilde F^h(a_r^e,e_t^e)$, Eq. \eqref{Delaunaynonloc},
which came from the original  ($r_{12}^h$-defined) nonlocal Hamiltonian Eq. \eqref{delta_H_nonloc}.
 Comparison order by order (both in $1/a_r^e$ and in $e_t^e$)
 fixes all the unknown coefficients at 5PN, besides checking all the (already known \cite{Damour:2015isa}) 4PN ones. 
 The results are of the type
\beq
a_5^{\rm nl,c}=\left(\frac{128}{5}\gamma+\frac{256}{5}\ln(2)\right)\nu\,,
\eeq
where we decomposed $a_5^{\rm nonloc}= a_5^{\rm nl,c}+ a_5^{\rm nl,\ln{}} \ln u$.  
They are listed in Table \ref{table_results_nl}. Note that the nonlocal 4PN coefficients listed here slightly differ from the 
corresponding 4PN coefficients, $a_5^{\rm II, c}, \ldots$, listed in Ref. \cite{Damour:2015isa} because the latter reference had used
a fixed partie-finie scale $s/c$, and had thereby incorporated the effects linked to averaging
$${\cal F}(t,t) \ln r_{12}^h(t)$$
in the ``local'' parts of the (real and EOB) Hamiltonians.


\begin{table}
\caption{\label{table_results_nl} Coefficients of the nonlocal ($r_{12}^h$-scaled) 4+5PN part of the EOB potentials.
We suppress the label $h$ for brevity.}
\begin{ruledtabular}
\begin{tabular}{ll }
Coefficient & Expression \\
\hline
$a_5^{\rm nl,c}$ & $\left(\frac{128}{5}\gamma+\frac{256}{5}\ln(2)\right)\nu$\\
\hline
$a_5^{\rm nl,\ln{}}$ & $\frac{64}{5}\nu$\\
\hline
$d_4^{\rm nl,c}$ & $\left(-\frac{992}{5}+\frac{1184}{15} \gamma -\frac{6496}{15}\ln(2)+\frac{2916}{5}\ln(3) \right)\nu$\\
\hline
$d_4^{\rm nl,\ln{}}$ & $\frac{592}{15}\nu$\\
\hline
$ q_{43}^{\rm nl,c}$ & $\left(-\frac{5608}{15} +\frac{496256}{45}\ln(2)-\frac{33048}{5}\ln(3) \right)\nu$\\
\hline
$q_{43}^{\rm nl,\ln{}}$ & 0\\
\hline
$q_{62}^{\rm nl,c}$ & $\left(-\frac{4108}{15} -\frac{2358912}{25}\ln(2) +\frac{1399437}{50}\ln(3)\right.$\\
&$\left. +\frac{390625}{18}\ln(5)\right)\nu$\\
\hline
$q_{62}^{\rm nl,\ln{}}$ & $0$ \\
\hline
$a_6^{\rm nl,c}$ & $\left(-\frac{128}{5} -\frac{14008}{105} \gamma-\frac{31736}{105}\ln(2) +\frac{243}{7}\ln(3)\right)\nu $\\
& 
$+\left( \frac{64}{5}-\frac{288}{5} \gamma +\frac{928}{35}\ln(2)-\frac{972}{7}\ln(3)\right)\nu^2$\\
\hline
$a_6^{\rm nl,\ln{}}$ & $-\frac{7004}{105}\nu-\frac{144}{5}\nu^2$\\
\hline
$d_5^{\rm nl,c}$ & $\left(-\frac{7318}{35} -\frac{2840}{7} \gamma+\frac{120648}{35}\ln(2) -\frac{19683}{7}\ln(3)\right)\nu $\\
& $
+\left(\frac{67736}{105} -\frac{6784}{15}  \gamma-\frac{326656}{21}\ln(2)+\frac{58320}{7}\ln(3) \right)\nu^2$\\
\hline
$ d_5^{\rm nl,\ln{}}$ & $-\frac{1420}{7}\nu-\frac{3392}{15}\nu^2$\\
\hline
$q_{44}^{\rm nl,c}$ & $\left(\frac{1007633}{315} +\frac{10856}{105} \gamma -\frac{40979464}{315}\ln(2)\right.$\\
&$\left. +\frac{14203593}{280}\ln(3)+\frac{9765625}{504}\ln(5)\right)\nu$\\
& $+\left(\frac{74436}{35} -\frac{1184}{5}\gamma  +\frac{33693536}{105}\ln(2)  -\frac{6396489}{70}\ln(3)\right. $\\
& $\left.-\frac{9765625}{126}\ln(5)\right)\nu^2$\\
\hline
$q_{44}^{\rm nl,\ln{}}$ & $\frac{5428}{105}\nu-\frac{592}{5}\nu^2$\\
\hline
$q_{63}^{\rm nl,c}$ & $\left(\frac{1300084}{525} +\frac{6875745536}{4725}\ln(2) -\frac{23132628}{175}\ln(3)\right. $\\
& $\left.  -\frac{101687500}{189}\ln(5)\right)\nu$\\
& $ +\left(\frac{160124}{75}-\frac{4998308864}{1575}\ln(2) -\frac{45409167}{350}\ln(3)\right. $\\
&$\left. +\frac{26171875}{18} \ln(5)\right)\nu^2$\\
\hline
$q_{63}^{\rm nl,\ln{}}$ & $0$ 
\end{tabular}
\end{ruledtabular}
\end{table}

\section{Using self-force theory to compute the local-plus-nonlocal EOB Hamiltonian in standard gauge, at first order in $\nu$} \label{SF}

In this section, we shall use SF theory to compute the full, local-plus-nonlocal (rescaled effective) EOB Hamiltonian, at first order in $\nu$. [The
{\it rescaled} effective EOB Hamiltonian $\hat H_{\rm eff} \equiv  H_{\rm eff}/\mu$ being divided by $\mu = \nu M$, the $O(\nu)$
contributions to $\hat H_{\rm eff}$ correspond to  $O(\nu^2)$ contributions to $ H_{\rm real}= Mc^2 + \ldots$.]
It is convenient to parametrize the full, local-plus-nonlocal dynamics in terms of the various potentials entering the general
form of $\hat H_{\rm eff}^2$ in standard $p_r$-gauge, namely
\beq
\hat H_{\rm eff}^2=A(u)(1+p_\phi^2 u^2+A(u) \bar D(u) p_r^2 +Q(u,p_r))\,.
\eeq
The full knowledge of $\hat H_{\rm eff}$ means the knowledge of the various potentials: $A(u)$, $\bar D(u)$ and 
$Q(u, p_r)=p_r^4 q_4(u)+p_r^6 q_6(u)+p_r^8 q_8(u)+p_r^{10}q_{10}(u)+\ldots$.
These potentials have all, at any given PN level, a polynomial structure in $\nu$ and they can be written in the form
\begin{eqnarray}
A(u)&=&1-2u +\nu a^{\nu^1}(u)+\nu^2 a^{\nu^2}(u)+\nu^3 a^{\nu^3}(u)+\ldots\nonumber\\ 
\bar D(u)&=&1+\nu \bar d^{\nu^1}(u)+\nu^2 \bar d^{\nu^2}(u)+\nu^3 \bar d^{\nu^2}(u)+\ldots\nonumber\\
q_4(u)&=& \nu q_{4}^{\nu^1}(u)+\nu^2 q_{4}^{\nu^2}(u)+\nu^3 q_{4}^{\nu^3}(u)+\ldots\nonumber\\
q_6(u)&=& \nu q_{6}^{\nu^1}(u)+\nu^2 q_{6}^{\nu^2}(u)+\nu^3 q_{6}^{\nu^3}(u)+\ldots\nonumber\\
q_8(u)&=& \nu q_{8}^{\nu^1}(u)+\nu^2 q_{8}^{\nu^2}(u)+\nu^3 q_{8}^{\nu^3}(u)+\ldots\,,
\end{eqnarray}
etc. SF theory is an efficient tool for analytically computing (in principle, at any given PN order) the linear-in-$\nu$ pieces
of the above EOB potentials, {\it i.e.} $a^{\nu^1}(u)= 2 u^3 + a_4 u^4 + \ldots$,  $\bar d^{\nu^1}(u)$, $q_{4}^{\nu^1}(u)$, etc.
Indeed, the self-force computation of the redshift invariant $\langle\delta z_1 \rangle$ \cite{Detweiler:2008ft,Barack:2011ed} 
of a particle moving along  eccentric equatorial orbits in a perturbed Schwarzschild background, combined with the first 
law~\cite{LeTiec:2011ab,Barausse:2011dq,Tiec:2015cxa},
has already allowed one to compute the linear-in-$\nu$ pieces of most of the EOB potentials.
 More precisely, $a^{\nu^1}(u)$ is known from the redshift of a particle moving along a circular orbit, whereas $\bar d^{\nu^1}(u)$, $q_4^{\nu^1}(u)$,  $q_6^{\nu^1}(u)$ etc., are known from the averaged redshift invariant $\langle\delta z_1 \rangle$ of a particle moving along a (bound) eccentric orbit at successive orders in an expansion in powers of the eccentricity
\beq
\langle\delta z_1\rangle =\delta z_1^{e^0}+e^2\delta z_1^{e^2}+e^4\delta z_1^{e^4}+e^6\delta z_1^{e^6}+O(e^8)\,.
\eeq
$\bar d^{\nu^1}(u)$ follows from  $\delta z_1^{e^2}$, $q_{4}^{\nu^1}(u)$ follows from  $\delta z_1^{e^4}$,  $q_{6}^{\nu^1}(u)$ follows from  $\delta z_1^{e^6}$, etc.
The current self-force analytical knowledge of $\langle\delta z_1 \rangle$ is limited at order $O(e^4)$. For the purpose of the present work
 it was necessary to extend this knowledge to $O(e^6)$. We have used SF theory to compute high-PN expressions for $\delta z_1^{e^6}$, and correspondingly $q_{6}^{\nu^1}(u)$. We present in  Appendix \ref{app_self_force} our newly derived complete expression for 
 $\delta z_1^{e^6}$ up to 9.5PN as well as its transcription into the EOB potential $q_6^{\nu^1}(u)$. 
The known SF expressions for the other potentials can be found in the literature (see Refs. \cite{Bini:2013zaa,Bini:2013rfa,Bini:2015bfb,Bini:2016qtx}). We list in Table \ref{table_results_sf} the 4+5PN contributions to  $a^{\nu^1}(u)$, $\bar d^{\nu^1}(u)$, $q_4^{\nu^1}(u)$ and $q_6^{\nu^1}(u)$.
 

\begin{table}
\caption{\label{table_results_sf} Coefficients of the 4+5PN terms in the linear-in-$\nu$ (i.e., 1SF) parts of the EOB potentials. Here, 
$\gamma$ denotes Euler's constant.}
\begin{ruledtabular}
\begin{tabular}{ll }
Coefficient & Expression \\
\hline
$\nu a_{\rm 4PN+5PN}^{\nu^1} $ & $\nu \left[\left(-\frac{4237}{60}+\frac{2275}{512}\pi^2+\frac{256}{5}\ln(2)+\frac{128}{5}\gamma\right.\right. $\\
& $\left.+\frac{64}{5}\ln(u)\right) u^5+\left(-\frac{1066621}{1575}+\frac{246367}{3072}\pi^2\right.$\\
& $ -\frac{31736}{105}\ln(2)-\frac{14008}{105}\gamma-\frac{7004}{105}\ln(u)$\\
& $\left.\left.+\frac{243}{7}\ln(3)\right)u^6\right]$\\
$\nu \bar d_{\rm 4PN+5PN}^{\nu^1}$ & $\nu \left[\left(\frac{1184}{15}\gamma+\frac{2916}{5}\ln(3)-\frac{6496}{15}\ln(2)-\frac{23761}{1536}\pi^2\right. \right. $\\
& $\left.-\frac{533}{45}+\frac{592}{15}\ln(u)\right) u^4+\left(-\frac{2840}{7}\gamma-\frac{19683}{7}\ln(3)\right.$ \\
&$  +\frac{120648}{35}\ln(2)-\frac{63707}{512}\pi^2+\frac{294464}{175}$\\
& $\left.\left.-\frac{1420}{7}\ln(u)\right) u^5\right]$\\ 
$\nu q_{4\, \rm 4PN+5PN}^{\nu^1}$ & $ \nu \left[\left(\frac{496256}{45}\ln(2)-\frac{33048}{5}\ln(3)-\frac{5308}{15}\right) u^3\right.$\\
& $+\left(\frac{10856}{105}\gamma-\frac{40979464}{315}\ln(2)+\frac{14203593}{280}\ln(3) \right. $\\
& $  +\frac{9765625}{504}\ln(5)-\frac{93031}{1536}\pi^2+\frac{1295219}{350}$\\
& $\left.\left.+\frac{5428}{105}\ln(u)\right) u^4\right]$\\
$\nu q_{6\, \rm 4PN+5PN}^{\nu^1}$ & $\nu \left[\left(-\frac{2358912}{25}\ln(2)+\frac{1399437}{50}\ln(3)+\frac{390625}{18}\ln(5)\right.\right.$\\
& $\left.-\frac{827}{3}\right) u^2+\left(-\frac{101687500}{189}\ln(5)+\frac{6875745536}{4725}\ln(2)\right.$\\
& $\left.\left.
+\frac{2613083}{1050}-\frac{23132628}{175}\ln(3)\right) u^3\right]$\\
\hline
\end{tabular}
\end{ruledtabular}
\end{table}

\section{Obtaining the 5PN-accurate local EOB Hamiltonian at linear order in $\nu$} \label{5PNnu1}

In the previous Section we have used SF theory to derive the linear-in-$\nu$ local-plus-nonlocal EOB potentials.
Subtracting from  the latter local-plus-nonlocal potentials, the nonlocal part of the EOB potentials obtained in Sec. II above, allows us to  write down the {\it local} part of the EOB potentials at the first order in $\nu$ (only). At this
stage, the $O(\nu\ge 2)$ contributions to the local EOB potentials are  known only at 4PN, but not beyond. To clarify our knowledge
so far, let us parametrize the $\nu$ dependence of a generic quantity $X(\nu)$  by  the following notation
\beq
X(\nu)\equiv X^{\nu^1}\nu+X^{(\nu)}\,,
\eeq
where 
\beq
X^{(\nu)}=X^{\nu^2}\nu^2+X^{\nu^3}\nu^3+X^{\nu^4}\nu^4+\ldots\,.
\eeq
For example, the local part of the $a$ potential at 5PN ({\it i.e.} $\propto u^6$) will be written in the form
\beq
a_{\rm 5PN, loc, f}=
 \left(-\frac{1026301}{1575}\nu+\frac{246367}{3072}\nu\pi^2+a_{6, f}^{(\nu)}\right) u^6\,.
\eeq
As indicated here, as we shall always use a flexibility parameter $f(t)-1$ which is at least of order $\nu^1$, and as the
corresponding contribution to the (unrescaled) Hamiltonian \eqref{DHfh} involves an extra factor ${\mathcal F}^{\rm split}(t,t) =O(\nu^2)$,
the effect of the flexibility factor $f(t)$ on (both) the local and nonlocal EOB potentials will start at order $\nu^2$ (corresponding
to $O(\nu^3)$ in the unrescaled Hamiltonian). Our SF computation thereby uniquely determines all the linear-in-$\nu$ contributions
to the EOB potentials.

Summarizing, the local   EOB potentials at 4+5PN, obtained from our (MPM + SF) results so far, have the following form:
\begin{widetext}
\begin{eqnarray}
\label{local_pot_param}
a_{\rm 4PN+5PN,loc, f}&=&
\left[\left(\frac{2275}{512}\pi^2-\frac{4237}{60}\right)\nu +\left(\frac{41}{32}\pi^2-\frac{221}{6}\right)\nu^2\right] u^5
+ \left(-\frac{1026301}{1575}\nu+\frac{246367}{3072}\nu\pi^2+a_{6, f}^{(\nu)}\right) u^6\,,\nonumber\\
\bar d_{\rm 4PN+5PN,loc, f}&=&
\left[\left(\frac{1679}{9}-\frac{23761}{1536} \pi^2\right)\nu +\left(-260+\frac{123}{16}\pi^2\right) \nu^2\right]u^4
+\left(\frac{331054}{175}\nu-\frac{63707}{512}\nu\pi^2 +\bar d_{5, f}^{(\nu)}\right) u^5\,,\nonumber\\
q_{4\,\rm 4PN+5PN,loc, f}&=&
\left(20\nu+q_{43}^{(\nu)} \right)u^3+\left(-\frac{93031}{1536}\nu\pi^2+\frac{1580641}{3150}\nu+q_{44, f}^{(\nu)}\right) u^4\,,\nonumber\\
q_{6\, \rm 4PN+5PN,loc, f}&=&\left(-\frac{9}{5}\nu +q_{62}^{(\nu)}\right)u^2 + \left(\frac{123}{10}\nu+q_{63, f}^{(\nu)}\right) u^3\,.
\end{eqnarray}
\end{widetext}
Here, we completed our previous $O(\nu)$ SF-based results by including the 4PN-level $O(\nu^2)$ terms previously
derived in Ref. \cite{Damour:2015isa}.
 
Note the remarkable fact that the 5PN-accurate local EOB Hamiltonian is {\it logarithm free}. Not only all the $\ln u$ terms have
disappeared (as expected because they have been known for a long time to be linked to the time nonlocality), but even the
various numerical logarithms $\ln 2, \ln 3, \ldots$, as well as Euler's constant $\gamma$ have all disappeared. Only rational numbers
and $\pi^2 \sim \zeta(2)$ enter this local Hamiltonian.
The expressions above include still undetermined nonlinear-in-$\nu$ terms in the parametrized form indicated above, $a_{6, f}^{(\nu)}$, $\bar d_{5, f}^{(\nu)}$, $q_{44, f}^{(\nu)}$,  and  $q_{63, f}^{(\nu)}$. [The 4PN-level terms  $q_{43}^{(\nu)}$
and $q_{62}^{(\nu)}$ are already known and will be written below.]

Note that at the 5PN level there is  the further term in the $Q$  potential
\beq
q_{8\, \rm 5PN,loc, f}=\left(q_{82}^{\nu^1}\nu +q_{82}^{(\nu)} \right)u^2\equiv q_{82}(\nu)u^2
\,,
\eeq
which is still undetermined from our SF computation (because of its limitation to the order $O(e^6)$).
 On the other hand, there is no contribution in the local Hamiltonian of the type
$q_{10\, \rm 5PN,loc} = q_{10,1}(\nu)u$, because $Q^{\rm loc}$ starts at order $G^2$.
 Such a term can only enter the nonlocal part of the Hamiltonian, 
where it comes from the need to expand the nonlocal
Hamiltonian as a formally infinite series of powers of $p_r^2$ \cite{Damour:2015isa}.

We are going  to determine below most of the so far unknown nonlinear-in-$\nu$ coefficients by using information concerning the scattering angle for hyperbolic encounters. However, to do so it will be convenient to change the standard EOB $p_r$ gauge  used above,
into the so-called energy-gauge \cite{Damour:2017zjx}.

\section{The 5PN local EOB Hamiltonian: from the standard $p_r$-gauge to the energy-gauge}
\label{convert_eg_EOB}

In the previous section we have determined, at the linear-in-$\nu$ order, the 5PN local EOB Hamiltonian in the standard $p_r$-gauge.
We then incorporated the non-linear-in-$\nu$ contributions to the various EOB potentials $A$, $\bar D$ and $Q$ in a parametrized form 
(still in the standard $p_r$-gauge). Let us start from such a (local, $p_r$-gauge) Hamiltonian, i.e.,
\beq
\label{Heffquadloc}
\hat H_{\rm eff,loc}^2=A^{\rm loc}(1+j^2u^2+A^{\rm loc}\bar D^{\rm loc}p_r^2+Q^{\rm loc})\,,
\eeq
with 
\beq
A^{\rm loc}= 1-2u+\delta A^{\rm  loc}\,,\qquad 
\bar D^{\rm loc}=1+\delta \bar D^{\rm loc}\,,
\eeq
and
\begin{eqnarray}
\delta A^{\rm  loc}&=& 2\nu u^3 +\nu\left(\frac{94}{3}-\frac{41}{32}\pi^2\right) u^4
+a_{\rm 4PN+5PN,loc}\,,\nonumber\\
\delta \bar D^{\rm loc}&=& 6\nu u^2+(52\nu-6\nu^2) u^3+\bar d_{\rm 4PN+5PN,loc}\,,\nonumber\\
Q^{\rm loc}&=& p_r^4 \left[2 (4-3\nu)\nu u^2+q_{4 \, \rm 4PN+5PN,loc}\right]
\nonumber\\
&&+p_r^6 q_{6\, \rm 4PN+5PN,loc}
+p_r^8
(q_{82}^{\nu^1}\nu+q_{82}^{(\nu)}) u^2
\,,
\end{eqnarray}
depending on the various unknown coefficients $a_6^{(\nu)}$, $\bar d_5^{(\nu)}$, $q_{44}^{(\nu)}$, $q_{62}^{(\nu)}$, 
$q_{63}^{(\nu)}$, and $q_{82}^{(\nu)}$. Here, we did not put any explicit label $f$ or $h$ because the discussion of the present
section applies to both cases.

Let us now show how to transform the above $p_r$-gauge (local) EOB Hamiltonian, Eq. \eqref{Heffquadloc}, into its energy-gauge form, 
 i.e., into the following post-Schwarzschild  (squared) effective Hamiltonian
\beq
\label{EG-eff-sq-Ham}
\hat H^2_{\rm eff, EG}= H_S^2+(1-2 u) Q_{\rm EG}(u,H_S)\,,
\eeq
where $H_S$ denotes the (rescaled) Schwarzschild Hamiltonian, i.e. the square root of
\beq
H_S^2=(1-2 u) [1+(1-2 u) p_r^2+j^2 u^2]\,,
\eeq
and where the energy-gauge $Q$ term reads
\begin{eqnarray}
Q_{\rm EG}(u,H_S)&=&u^2 q_{2\rm EG}(H_S)+u^3 q_{3\rm EG}(H_S)\nonumber\\
&+& 
u^4 q_{4\rm EG}(H_S)+u^5q_{5\rm EG}(H_S)\nonumber\\
&+& 
u^6 q_{6\rm EG}(H_S) +\ldots\,.
\end{eqnarray}
Here, a term $q_{n\rm EG}(\gamma) u^n$, being proportional to $G^n$, describes the $n$-PM approximation. 
When working within some PN-approximation scheme, one can only determine a limited number of terms in the
PN-expansion (corresponding to an expansion in powers of $p_{\infty}^2 \equiv \gamma^2-1$)
of each separate energy-gauge coefficient $q_{n\rm EG}(\gamma)$. We will discuss below which terms in the $p_{\infty}^2$
expansion of the various $q_{n\rm EG}(\gamma)$'s correspond to the 5PN level.

The PN expansions of all the energy-gauge coefficients  $q_{n\rm EG}(\gamma)$ are determined from the corresponding 
$p_r$-gauge coefficients entering the Hamiltonian
 (notably the 5PN-level ones $a_6^{(\nu)}$, $\bar d_5^{(\nu)}$, $q_{44}^{(\nu)}$, $q_{62}^{(\nu)}$ and  $q_{63}^{(\nu)}$)
 by computing the canonical transformation connecting the two gauges. The structure of this canonical transformation is
\begin{eqnarray} \label{gauge}
g(r, p_r)&=& (r \, p_r) \frac{1}{r^2} \left[ \frac32 \eta^4\nu  +\eta^6  \left(\frac{g_1}{r}+\frac{g_2 j^2}{r^2}+g_3p_r^2\right)\right.\nonumber\\
&& +\eta^8 \left(\frac{h_1}{r^2}+\frac{h_2 j^4}{r^4}+h_3 p_r^4+\frac{h_4 j^2}{r^3}+h_5 \frac{p_r^2}{r}\right. \nonumber\\
&& \left.+\frac{h_6 j^2 p_r^2}{r^2}\right)+\eta^{10} \left(\frac{n_1}{r^3}+\frac{n_2j^2}{ r^4}\right.\nonumber\\
&& +\frac{n_3 p_r^2}{r^2} +\frac{n_4 j^4}{r^5}+\frac{n_5 j^2 p_r^2}{r^3}+\frac{n_6 p_r^4}{r}  \nonumber\\
&& \left.
+\frac{n_7 j^6}{r^6}+\frac{n_8 j^4 p_r^2}{r^4}+\frac{n_9 j^2 p_r^4}{ r^2}+n_{10} p_r^6\right)\,.
\end{eqnarray}
Here we have factored out the term $ (r \, p_r)$ that corresponds to the identity transformation, and we have ordered $g$ by means
of the PN-counting paramer $\eta = \frac1c$. In addition, we are using here
rescaled coordinates, namely $r=r^{\rm phys}/(G M)$, $p_r=p_r^{\rm phys}/\mu$, and $j \equiv p_\phi = p_\phi^{\rm phys}/(G M \mu)$.
As a consequence, each factor $\frac{p^m}{r^n}$ beyond the first factor $(r \, p_r)$ in the canonical transformation \eqref{gauge}
has to be seen as containing a factor $G^n$, and therefore to correspond to the $n$-PM approximation. [When doing this counting, one must
count each  factor $j = | {\bf r} \times {\bf p}|$ as being $\sim r p \propto G^{-1}$, {\it i.e.} use the equivalence $\frac{j}{r} \sim p \sim p_r$.]
In particular, we see that, in view of the overall prefactor $\frac{1}{r^2}= \left(\frac{GM}{r^{\rm phys}}\right)^2$, the whole canonical transformation $g$, Eq. \eqref{gauge}, starts at the 2PM ($G^2$) level. 

Previous work has determined the canonical transformation $g$, Eq. \eqref{gauge}, up to the 4PN level , i.e. $O(\eta^8)$.
The 2PN ($\frac32 \eta^4\nu$) and 3PN ($\eta^6 g_i$) gauge parameters  were derived in Ref. \cite{Damour:2017zjx}, while the 4PN 
ones ($\eta^8 h_i$)  were derived in Appendix A of Ref. \cite{Antonelli:2019ytb}. We have extended this determination
to the 5PN level, by imposing that 
 the two (effective, squared) Hamiltonians \eqref{Heffquadloc} and \eqref{EG-eff-sq-Ham} be equivalent (at 5PN accuracy), 
 through this canonical transformation.
The explicit expressions of the 5PN coefficients $n_i$ will be displayed later, in their final form,  in Table \ref{table_gauge_params},
after we determine, using our strategy, all possible unknowns. However,  as we discuss in the next section, the 
linear-in-$\nu$ results of the previous section suffice, at this stage, to uniquely determine the 3PM ($G^3$) energy-gauge
coefficient $q_{3\rm EG}(\gamma;\nu)$, and thereby to test the all-PN-orders 3PM result of \cite{Bern:2019nnu,Bern:2019crd}.

\section{Determination of the 3PM dynamics up to the 5PN (and 6PN) levels}

Let us show here how the linear-in-$\nu$ results of section \ref{5PNnu1} suffice to determine the 
the 3PM ($G^3$) energy-gauge coefficient $q_{3\rm EG}(\gamma;\nu)$, and thereby the 3PM scattering angle.
This fact follows from three other facts. First, as explicitly shown in \cite{Bini:2017wfr}, the nonlocal part
of the Hamiltonian starts contributing to the scattering angle only at the 4PM ($G^4$) level.
Second, as emphasized in Ref. \cite{Damour:2019lcq}, thanks to the
special $\nu$-dependence of the scattering angle, the knowledge of the $O(\nu^1)$ contribution to
the scattering angle suffices to know its exact-in-$\nu$ value. Third, the local PN-expanded canonical transformation
$g$, Eq. \eqref{gauge},  being, at each PN order, a polynomial in $G$ 
cannot decrease the PM order of any contribution to the Hamiltonian. 
Putting these facts together we conclude that it suffices to determine the linear-in-$\nu$ and $\leq G^3$ value of
$g$ to compute the exact-in-$\nu$ value of the 3PM-level energy-gauge  coefficient $q_{3\rm EG}(\gamma;\nu)$,
at the same PN accuracy at which we know the local linear-in-$\nu$, $p_r$-gauge Hamiltonian.

Before discussing the determination of the 3PM coefficient $q_{3\rm EG}(\gamma;\nu)$ let us recall that
the value of the 2PM coefficient $q_{2\rm EG}(\gamma)$ has been determined to all PN orders, i.e. as an exact function of $\gamma=H_S$,
 in Ref. \cite{Damour:2017zjx}. [It was then checked by other calculations \cite{Cheung:2018wkq,Bern:2019nnu}.]
It reads
\beq \label{q2EG}
q_{2\rm EG}(\gamma,\nu)=\frac{3}{2} (5\gamma^2-1) \left(1-\frac{1}{h(\gamma,\nu)} \right)\,,
\eeq
where
\beq
h(\gamma;\nu) \equiv [1+2\nu(\gamma-1)]^{1/2}\,.
\eeq
We show in Appendix \ref{Abel_tr} how one can compute in {\it closed-form}  the result of transforming the energy-gauge
2PM Hamiltonian contribution $Q_{\rm EG}^{\rm 2PM}(u,H_S)=u^2 q_{2\rm EG}(H_S)$ into its standard $p_r$-gauge version
$Q^{\rm 2PM}(u,p_r)=u^2 q_{2}(p_r^2)$. This is achieved by using an Abel transform.
Note that the knowledge of the exact $Q^{\rm 2PM}$ will be crucial for the computation of the 5PN-level term $q_{82} p_r^8 u^2$.

Let us now come to the value of the 3PM coefficient $q_{3\rm EG}(\gamma, \nu)$. 
The structure of the $\nu$-dependence of  $q_{3\rm EG}(\gamma)$  has been shown to depend on the knowledge of two
functions of $\gamma$, $A_{q3}(\gamma)$ and $B_{q3}(\gamma)$, namely
\cite{Bern:2019nnu,Bern:2019crd,Damour:2019lcq}
\beq \label{q3EG}
q_{3\rm EG}(\gamma;\nu)= A_{q3}(\gamma)+\frac{B_{q3}(\gamma)}{h(\gamma;\nu)}-\frac{A_{q3}(\gamma)+B_{q3}(\gamma)}{h^2(\gamma;\nu)} \,.
\eeq
Among the two functions of $\gamma$ entering $q_{3\rm EG}(\gamma;\nu)$, the $B_{q3}$ function is exactly known to be
\beq
\label{Bq3def}
B_{q3}(\gamma) = \frac32 \frac{(2\gamma^2-1) (5\gamma^2-1)}{(\gamma^2-1)}\,.
\eeq
Concerning the value of the other 3PM-level function $A_{q3}(\gamma)$  PN-based work previous to Ref. \cite{Bini:2019nra}
had determined its 4PN-accurate value, namely
\begin{eqnarray} \label{A3PN}
A^{\rm PN}_{q3}(\gamma)
&=&  \frac{1}{(\gamma ^2-1)} \left[2+\frac{37}{2}(\gamma^2-1)+\frac{117}{10} (\gamma^2-1)^2\right. \nonumber\\
&& \left.+  A_{\rm 5PN} (\gamma^2-1)^3 + A_{\rm 6PN} (\gamma^2-1)^4 +\ldots \right], \nonumber\\
\end{eqnarray}
where the coefficient $+\frac{117}{10}$ corresponds to the 4PN level, and where the further coefficients $ A_{\rm 5PN}$, 
$ A_{\rm 6PN}$, respectively parametrize the 5PN and 6PN contributions that we shall discuss next.

Only one line of work has so far been able to compute the exact value of the function $A_{q3}(\gamma)$,
and thereby of the 3PM coefficient $q_{3\rm EG}(\gamma;\nu)$. Namely, the  quantum-amplitude-based
computation of Bern et al. Refs.~\cite{Bern:2019nnu,Bern:2019crd} led to the following
(partly conjectural) exact expression for the function $A_{q3}(\gamma)$,
\beq \label{A3B}
A_{q3}^{\rm Bern}(\gamma) = -B_{q3}(\gamma)+\frac{{\bar C}^{B}(\gamma)}{\gamma-1}\,,
\eeq
with $B_{q3}$ given in Eq. \eqref{Bq3def} and ${\bar C}^{B}$ given by (see, e.g., Eq. (3.69) of Ref. \cite{Damour:2019lcq})
\begin{eqnarray} \label{bCB}
{\bar C}^{B}(\gamma)&=&\frac23\gamma(14\gamma^2+25)\nonumber\\
&+&
4\frac{4\gamma^4-12\gamma^2-3}{\sqrt{\gamma^2-1}}{\rm arcsinh}\left(\sqrt{\frac{\gamma-1}{2}}\right)\,.
\end{eqnarray}
The expansion in powers of $\gamma ^2-1$ of $A_{q3}^{\rm Bern}$ (describing its PN expansion)  reads
\begin{eqnarray} \label{A3Bxp}
A_{q3}^{\rm Bern}(\gamma)
&=&  \frac{1}{(\gamma ^2-1)} \left[2+\frac{37}{2}(\gamma^2-1)+\frac{117}{10} (\gamma^2-1)^2\right. \nonumber\\
&& +  \frac{219}{140} (\gamma^2-1)^3 - \frac{7079}{10080} (\gamma^2-1)^4 \nonumber\\
&&\left. +  \frac{989}{2240} (\gamma^2-1)^5+ \ldots \right]
\,.\nonumber\\
\end{eqnarray}
Let us explain how the results presented in the previous sections allows us to compute the value of the 5PN coefficient $ A_{\rm 5PN}$,
and how the recent extension of our method \cite{BDG6PN} has also allowed us to compute the 6PN coefficient $ A_{\rm 6PN}$.

Comparing the effect of the canonical transformation $g$, Eq. \eqref{gauge}, 
on the linear-in-$\nu$ local ($p_r$-gauge) Hamiltonian given in section \ref{5PNnu1} to the corresponding $O(G^3)$-truncated
energy-gauge Hamiltonian \eqref{EG-eff-sq-Ham}, with 
\beq
Q_{\rm EG}^{\leq \rm 3PM}(u,H_S)=u^2 q_{2\rm EG}(H_S, \nu)+u^3 q_{3\rm EG}(H_S, \nu) + O(G^4)\,,
\eeq
yields enough equations to determine the linear-in-$\nu$ values of the 5PN-level coefficients parametrizing the $\leq G^3$ terms
in Eq. \eqref{gauge}. In view of the overall factor $\frac1{r^2} \propto G^2$ in $g$, the only $\leq G^3$, 5PN coefficients are
the $n_i$'s with $i=4,5,6,7,8,9,10$.  For instance, the value of $n_4$ is determined to be 
\beq
n_4 =\frac{603}{1120}\nu + O(\nu^2)\,.
\eeq
See Table \ref{table_gauge_params} for the linear-in-$\nu$ values of the remaining 3PM-level (and 5PN level)
gauge parameters $n_i;  i=4,5,6,7,8,9,10$. In addition, this $\leq G^3$ determination of the gauge transformation $g$
also determines the  linear-in-$\nu$, 5PN contribution to the 3PM energy-gauge coefficient $q_{3\rm EG}(H_S, \nu)$.
In turn, as explained above (see Eq. \eqref{q3EG}) the linear-in-$\nu$ value of the 3PM coefficient $q_{3\rm EG}(H_S, \nu)$
uniquely determines a corresponding knowledge of the ($\nu$-independent) function $A_{q3}(\gamma)$. We thereby
deduced (as already announced in \cite{Bini:2019nra}) from the results of section \ref{5PNnu1} the following value
of the 5PN-level coefficient in $A_{q3}(\gamma)$:
\beq \label{A5pn}
A_{\rm 5PN}= \frac{219}{140}\,,
\eeq
in agreement with the result of Bern et al., Eq. \eqref{A3Bxp}.

Recently, we have been able to extend our computation to the 6PN level by: (i) pushing the computation of the nonlocal
action to the 6PN order; and (ii)  pushing the SF redshift computation explained in section \ref{SF} to the eighth order in eccentricity.
This has allowed us to extend the knowledge of the local Hamiltonian to the 6PN order (see below for the 5.5PN, purely
nonlocal contribution). We will report our complete 6PN results somewhere else \cite{BDG6PN}. Let us here only cite
the crucial new term allowing us to compute the 3PM-level, 6PN-accurate coefficient $ A_{\rm 6PN}$. It is the following
$O(p_r^8 u^3)$ contribution to the EOB $Q$ potential (in $p_r$-gauge):
\beq \label{q83}
Q^{\rm 6PN}_{p_r^8 u^3} = \left(- \frac{7447}{560} \nu + O(\nu^2) \right) p_r^8 u^3\,.
\eeq
Transforming the $p_r$-gauge (3PM-6PN) knowledge of Eq. \eqref{q83} into its energy-gauge correspondant (by
extending the canonical transformation \eqref{gauge} to the 6PN level) then determines the 6PN-level, linear-in-$\nu$ 
contribution to the 3PM energy-gauge coefficient $q_{3\rm EG}(H_S, \nu)$. Expressing the latter result in terms
of the parametrization Eq. \eqref{q3EG} finally leads to the value
\beq \label{A6pn}
A_{\rm 6PN}=  - \frac{7079}{10080} \,.
\eeq
for the 6PN term in the function $A_{q3}(\gamma)$. This value agrees with the result of Bern et al., Eq. \eqref{A3Bxp}.

While we were preparing our work for publication, an effective-field-theory computation of Bl\"umlein et al. \cite{Blumlein:2020znm}
reported a different, independent (purely PN-based\footnote{By contrast with our method which combines several
different approximation schemes.}) derivation of the two coefficients \eqref{A5pn}, \eqref{A6pn}. Ref. \cite{Damour:2019lcq}
had tried to reconcile an apparent tension between the high-energy limit of the result of Refs.~\cite{Bern:2019nnu,Bern:2019crd} 
and the high-energy behavior of an older SF computation \cite{Akcay:2012ea} by conjecturing another value of the function
$A_{q3}(\gamma)$, which has a softer high-energy behavior, and which starts to differ from Eq. \eqref{A3Bxp} at the 6PN level. 
The latter conjecture is now disproved by the
result \eqref{A6pn}. [See, however, Ref. \cite{Damour:2019lcq}, for other conjectural possibilities for reconciling the results of
Refs.~\cite{Bern:2019nnu,Bern:2019crd} and Ref. \cite{Akcay:2012ea}.]


\section{Nonlocal part of the scattering angle}

Ref.~\cite{Damour:2019lcq} has recently pointed out the existence of a restricted functional dependence of the
 scattering angle $\chi(\gamma, j; \nu)$ on the symmetric mass ratio $\nu$ \footnote{
 Note that it is important here to use the effective EOB energy
 $\gamma = \e$  as energy argument, besides $j = J/(GM\mu)$ and $\nu= \mu/M$.}.
This generic constraint applies to the total (local-plus-nonlocal) scattering angle.
In the present section we compute the nonlocal contribution, $\chi_{\rm nonloc, f}(\gamma,j;\nu)$, to the scattering angle
with sufficient accuracy to be able to fully exploit the structure pointed out in Ref.~\cite{Damour:2019lcq}.
More precisely, when considering the large-$j$ expansion of $\chi_{\rm nonloc, f}(\gamma,j;\nu)$, namely (see  \cite{Bini:2017wfr}),
\bea
 \chi_{\rm nonloc, f}(\gamma,j;\nu) &=& \frac{\nu \, p_\infty^4}{j^4} \left[A_0(p_\infty;\nu) + \frac{A_1(p_\infty;\nu) }{p_\infty j}\right. \nonumber\\
 &+&\left. \frac{A_2(p_\infty;\nu) }{(p_\infty j)^2}+O\left( \frac{1 }{(p_\infty j)^3}\right)   \right]\,,
\eea
(where we recall that $p_\infty^2 \equiv \gamma^2-1$)
we shall see below that it is enough, for our present 5PN accuracy, to compute the coefficient $A_0(p_\infty;\nu)$ entering
the leading order in  $1/j$.
The difficulty, however, is to compute it at the 1PN fractional accuracy: $A_0(p_\infty;\nu)= A_{00} + \eta^2A_{02}\, p_\infty^2+ O(\eta^4 p_\infty^4)$. As will become clear the small expansion parameter $\frac1{p_\infty j}$ (which happens to be of
{\it Newtonian} order $\sim \eta^0$) is equivalent to the inverse of the Newtonian eccentricity.

In order to compute the nonlocal contribution, $\chi_{\rm nonloc, f}$, to the scattering angle we extend the strategy used at the 
leading PN order  in \cite{Bini:2017wfr}. It was shown there that
\beq
\frac12 \chi_{\rm nonloc, f}(\gamma,j;\nu) = \frac{1}{2\nu}\frac{\partial}{\partial j}W_{\rm nonloc, f}(\gamma,j;\nu)\,,
\eeq
where
\beq
W_{\rm nonloc, f}(\gamma,j;\nu) = \int dt \delta H_{\rm nonloc, f}\,.
\eeq
Inserting the expression, Eq. \eqref{deltaHnlf}, of $\delta H_{\rm nonloc, f}$ then leads to expressing $W_{\rm nonloc, f}$
as a sum of three terms, say
\beq
W_{\rm nonloc, f}= W^{\rm flux \, split} + W^{\rm flux} + W^{\rm f-h}\,,
\eeq
where
\beq \label{Wfluxsplit}
 W^{\rm flux \, split} = -\frac{G^2 H_{}}{c^{5}}\int d t {\rm Pf}_{2s/c}\int \frac{d t'}{|t'-t|}{\mathcal F}^{\rm split}_{\rm 1PN}(t,t'),
 \eeq
 \beq \label{Wflux}
 W^{\rm flux}=+\frac{2G^2H }{c^{5}}\int dt {\mathcal F}^{\rm 1PN}(t,t)\ln \left(\frac{r^h(t)}{s}\right)\,,
\eeq
and
\beq \label{Wf-h}
W^{\rm f-h}= + 2\frac{G^2 H_{}}{c^{5}}\int dt {\mathcal F}^{\rm split}_{\rm 1PN}(t,t)  \ln \left( f(t)\right)\,.
\eeq

To this end we need to evaluate the flux-split, as given in Eq. \eqref{flux1PNdef}, along hyperbolic orbits at the
fractional 1PN order. We then need (as will be made clear below) to compute the first few terms of the  expansion
of  $W_{\rm nonloc, f}(\gamma,j;\nu)$ in the large-$j$ limit, corresponding to a large-eccentricity limit for the 
considered hyperbolic orbit. 

In order to evaluate such a large-eccentricity limit, we start from the 1PN-accurate harmonic-coordinate quasi-Keplerian 
parametrization \cite{DD1} of the hyperbolic motion, namely
\begin{eqnarray} \label{QK}
r&=& \bar a_r (e_r \cosh v-1)\,,\nonumber\\
\ell &=& \bar n (t-t_P)=e_t \sinh v-v \,,\nonumber\\
\bar \phi &=&\frac{\phi-\phi_P}{K}=V=2 {\rm arctan}\left[\sqrt{\frac{e_\phi+1}{e_\phi-1}}\tanh \frac{v}{2}  \right] \,,
\end{eqnarray}
where the orbital parameters $\bar n$, $\bar a_r$, $e_r$, $e_t$, $e_\phi$ are the functions of  
$\bar E=(E_{\rm tot}-Mc^2)/\mu$ and $j$ listed in Table \ref{table_relations_hyp}.
As shown in Ref. \cite{DD1}, the hyperbolic representation \eqref{QK} is an  analytic continuation of the 
corresponding ellipticlike quasi-Keplerian parametrization.


\begin{table}
\caption{\label{table_relations_hyp} Quasi-Keplerian representation of the hyperbolic 1PN motion, in terms of  $\bar E=(E_{\rm tot}-Mc^2)/\mu$ and $j$. At this level the periastron advance, as stated above, is simply $K=1+\frac{3}{j^2}\eta^2$.}
\begin{ruledtabular}
\begin{tabular}{|l|l|}
$\bar n$&$(2\bar E)^{3/2}\left[1+\eta^2\frac{\bar E}{4}(15-\nu)\right]$\\
\hline
$\bar a_r$&$\frac{1}{2\bar E}\left[1+\eta^2\frac{\bar E }{2}(7-\nu)\right]$\\
\hline
$e_t^2$& $1+2\bar Ej^2 + \bar E\left[-\bar Ej^2(-17+7\nu) +4(1-\nu) \right] \eta^2   $\\
\hline
$e_r^2$& $1+2\bar Ej^2+ \bar E \left[-5\bar E j^2(3-\nu) +2(-6+\nu)\right] \eta^2$\\
\hline
$e_\phi^2$&$ 1+2\bar Ej^2+ \bar E \left[ -\bar Ej^2(15-\nu)-12\right]\eta^2$\\
\end{tabular}
\end{ruledtabular}
\end{table}

It is then useful to change the integration variables $t, t'$ entering the definition of $W_{\rm nonloc, f}$
into the variables $T=\tanh v/2$ and $T'=\tanh v'/2$, where $v$ and $v'$ are the variables
entering the quasi-Keplerian representation \eqref{QK}.
This operation maps the original integration domain  $(t,t') \in R\times R$ onto the compact domain  $(T,T') \in [-1,1]\times [-1,1]$.
It also transforms the singular line $t=t'$ into $T=T'$, together with  a transformation of the constant  cutoff $|t'-t| = 2 s/c$ implied
by the Pf operation into a corresponding $T$-dependent cutoff (see below).

We succeeded in computing the large eccentricity limit of $W^{\rm flux \, split}$, at the leading order in eccentricity but including the 
fractional 1PN contribution. Both integrals in $T'$ (with Pf) and in $T$ can be performed exactly, within this limit.
Note that during the various computational steps we take $e_r$ as fundamental eccentricity, and expand in powers of 
$\frac1{a_r} \sim p_\infty^2$ .
Some details follow. 

The formal structure of ${\mathcal F}^{\rm 1PN}(T,T')$ is
\begin{eqnarray}
{\mathcal F}^{\rm 1PN}(T,T')&=&\frac{\nu^2}{\bar a_r^5e_r^4}\left[ {\mathcal F}_{00}+\eta^2 ({\mathcal F}_{20}+\nu {\mathcal F}_{21})\right]\,,
\end{eqnarray}
where, for example,
\begin{eqnarray}
{\mathcal F}_{00}(T,T')&=&
3(1-T'{}^2)(1-T^2)\times \nonumber\\
&& \times (T'{}^4-4T'{}^2+1)(T^4-4T^2+1)\nonumber\\
&&
+TT'\{37(T'{}^4+1)(T^4+1)\nonumber\\
&&-52[T^2(T'{}^4+1)+T'{}^2(T^4+1)]\nonumber\\
&& +76T'{}^2T^2\}\,.
\end{eqnarray}

Similarly, the expression for  the integration measure $d{\mathcal M}_{(t,t')}=dtdt'/|t-t'|$, at 1PN, transformed in the variables $T,T'$ is
\begin{eqnarray}
d{\mathcal M}_{(T,T')}&&= 2e_r\bar a_r^{3/2}\left[1-\frac{1+2\nu}{2\bar a_r}\eta^2\right]\times \nonumber\\
&& \times \frac{(1+T'^2) (1+T^2)dT dT'}{(1-T'^2)(1-T^2)(1+TT')|T-T'|}\,,
\end{eqnarray}
at the leading order in a large-eccentricity expansion.

The (PN-expanded) transformed integrand $d{\mathcal M}_{(T,T')} \times {\mathcal F}(T,T')$ is then written as
\beq
d{\mathcal M}_{(T,T')} {\mathcal F}(T,T')={\mathcal G}(T,T') \frac{dT dT'}{|T-T'|}\,.
\eeq

The original integral was singular at $t=t'$, i.e., along the bisecting line of the $t-t'$ plane.
This singularity line   becomes the bisecting line in the plane $T-T'$, but endowed with a $T-$dependent slit (equivalent to a Pf scale
$2 f(T) s/c$, where $f(T)$ is identified from the relation $dT =f(T) dt$). In the large eccentricity limit, one finds 
\beq
f(T)=\frac{\bar n} {2e_t }\frac{(1-T^2)^2} {1+T^2}\,.
\eeq
In other words, the integration domain of the flux-split integral is divided  into the following parts
\beq
[-1,T-\epsilon f(T)] \quad \cup \quad [T+\epsilon f(T), 1]
\eeq
that is
\begin{eqnarray}
{\mathcal I}&=& {\rm Pf}_{\epsilon}\int_{-1}^1 dT' \frac{{\mathcal G}(T,T')}{|-T'+T|}\nonumber\\
&=& \int_{-1}^1 dT' \frac{{\mathcal G}(T,T')-{\mathcal G}(T,T)}{|-T'+T|}\nonumber\\
&&+  {\mathcal G}(T,T)\left[\int_{-1}^{T-\epsilon f(T)} \frac{dT'}{T-T'}\right.\nonumber\\
&& \left. + \int_{T+\epsilon f(T)}^1 \frac{dT'}{T'-T}\right]\nonumber\\
&=& \int_{-1}^1 dT' \frac{q(T,T')}{|-T'+T|}\nonumber\\
&&+  {\mathcal G}(T,T) \left[ -\ln\left(\frac{\epsilon f(T)}{1+T} \right) + \ln\left( \frac{1-T}{\epsilon f(T)} \right) \right]\nonumber\\
&=& \int_{-1}^1 dT' \frac{q(T,T')}{|-T'+T|}+  {\mathcal G}(T,T)  \ln\left( \frac{1-T^2}{\epsilon^2 f^2(T)} \right) \,,
\end{eqnarray}
where $\epsilon=2s/c$ and
\beq
q(T,T')={\mathcal G}(T,T')-{\mathcal G}(T,T)\,.
\eeq
[One formally considers $\epsilon=2s/c$ as being infinitesimal, before replacing it by a finite value at the end.]
Further integration in $T$ then gives 
\begin{eqnarray}
{\mathcal J}&\equiv & \int_{-1}^1 {\mathcal I} dT \nonumber\\
&=& \int_{-1}^1 dT \int_{-1}^1 dT' \frac{q(T,T')}{|-T'+T|}\nonumber\\
&& -2 \ln \epsilon  \int_{-1}^1 dT  {\mathcal G}(T,T)\nonumber\\
&& +\int_{-1}^1 dT   {\mathcal G}(T,T)  \ln\left( \frac{1-T^2}{f^2(T)}\right)\,,
\end{eqnarray}
so that the first term in $W^{\rm nonloc, f}$ is given by
\beq
W^{\rm flux split }=-H_{\rm real}{\mathcal J}\,.
\eeq
We find
\begin{widetext}
\begin{eqnarray}
W^{\rm flux\, split}&=&\frac{2}{15} \frac{\pi \nu^2}{e_r^3\bar a_r^{7/2}}H_{\rm real}\left\{ 100 + 37 \ln \left(\frac{s}{4e_r\bar a_r^{3/2}}\right) 
+\left[\frac{685}{4}-\frac{1017}{14}\nu+\left(\frac{3429}{56}-\frac{37}{2}\nu\right)\ln \left(\frac{s}{4e_r\bar a_r^{3/2}}\right)\right] \frac{\eta^2}{\bar a_r}
\right\}\,.
\end{eqnarray}
\end{widetext}
Note the presence of logarithmic terms.

The second contribution to  $W^{\rm nonloc, f}$, namely  Eq.~\eqref{Wflux},
 can be similarly computed, leading to 
\begin{widetext}
\begin{eqnarray}
W^{\rm flux}
&=& \frac{2}{15} \frac{\pi \nu^2}{e_r^3\bar a_r^{7/2}}H_{\rm real}\left\{
-\frac{85}{4}-37\ln \left(\frac{s}{2e_r\bar a_r}\right)
+\left[
-\frac{9679}{224}+\frac{981}{56}\nu+\left(-\frac{3429}{56}+\frac{37}{2}\nu\right)\ln \left(\frac{s}{2e_r\bar a_r}\right)
\right] \frac{\eta^2}{\bar a_r}
\right\}\,.
\end{eqnarray}
\end{widetext}

Summarizing, and re-expressing $e_r$ and $\bar a_r$ in terms of $p_\infty$ and $j$, we find for the h-contribution to $W^{\rm nonloc, f}$:
\begin{widetext}
\begin{eqnarray}
W^{\rm nonloc, h}&\equiv&W^{\rm flux\, split}+W^{\rm flux}\nonumber\\
&=& \frac{2}{15}\frac{\nu^2 p_\infty^4\pi}{ j^3}\left[
\frac{315}{4}+37\ln \left(\frac{p_\infty}{2} \right)
+\left[\frac{2753}{224}-\frac{1071}{8}\nu+\left(\frac{1357}{56}-\frac{111}{2}\nu\right)\ln \left(\frac{p_\infty}{2} \right)\right]p_\infty^2\eta^2
\right]\,.
\end{eqnarray}
\end{widetext}
This result is accurate modulo two types of corrections: $O(\eta^4)$ and $O(\frac1{p_\infty j})$.
This suffices to compute the $1/j^4$ contribution to $\chi^{\rm nonloc, h}$ at the fractional 1PN accuracy, namely
\begin{eqnarray}
\frac12 \chi^{\rm nonloc, h} &=& \frac{1}{2\nu}\frac{\partial}{\partial j}W^{\rm nonloc, h}
=\frac{\chi_4^{\rm nonloc}}{j^4} + O\left(\frac1{j^5}\right)\,, \nonumber\\
\end{eqnarray}
where (indicating for clarity the $\geq$4PN nature of $\chi_4^{\rm nonloc, h}$)
\begin{eqnarray}
\label{final_chi_4}
\chi_4^{\rm nonloc, h}&=& \nu \,\eta^8 p_\infty^4 \pi\left[a_{\rm 4PN}+\eta^2 a_{\rm 5PN}\right. \nonumber\\
&+ &\left. \ln \left(\frac{p_\infty}{2} \right)  \left(b_{\rm 4PN}+\eta^2 b_{\rm 5PN}\right)\right]\,,
\end{eqnarray}
with
\begin{eqnarray}
a_{\rm 4PN}&=&-\frac{63}{4}\,,\nonumber\\ 
a_{\rm 5PN}&=&-\frac{2753}{1120}+\frac{1071}{40}\nu\,,\nonumber\\ 
b_{\rm 4PN}&=&-\frac{37}{5}\,, \nonumber\\
b_{\rm 5PN}&=&-\frac{1357}{280}+\frac{111}{10}\nu\,.
\end{eqnarray}
The meaning of this result will be further discussed in the next section.

\section{Use and determination of  the flexibility factor $f(t)$} \label{f1}

The general rule uncovered in Ref.~\cite{Damour:2019lcq} restricts, at each PM order $G^n$, the $\nu$-dependence of the
rescaled scattering angle $ \widetilde \chi_n \equiv h^{n-1} \chi_n$, where we recall the definition,
\beq
h(\gamma, \nu) \equiv \sqrt{1+2\nu(\gamma-1)}\,,\qquad \gamma =\sqrt{1+p_\infty^2 \eta^2}\,.
\eeq
In the present case we are interested in the $O(G^4)$ contribution $\chi_4$ to the scattering angle.
Let us then reexpress the result \eqref{final_chi_4} in terms of the quantity $ \widetilde \chi_4 =h^3 \chi_4$.
We find
\begin{widetext}
\begin{eqnarray}
\label{h3chi4_nonloch}
h^3 \chi_4^{\rm nonloc, h}(\gamma,\nu) &=& \nu \,\eta^8 p_\infty^4  \pi \left\{ -\frac{63}{4}-\frac{37}{5} \ln \left(\frac{p_\infty}{2} \right) +\eta^2 p_\infty^2\left[-\frac{2753}{1120}-\frac{1357}{280} \ln \left(\frac{p_\infty}{2} \right)+\frac{63}{20}\nu \right] 
 \right\}\,.
\end{eqnarray}
\end{widetext}
The general rule of Ref.~\cite{Damour:2019lcq} states, in this case, that the product $(h^3 \chi_4)(\gamma,\nu)$ should be
(at most) {\it linear} in $\nu$. Taking into account the overall factor $\nu$ in  $\chi_4^{\rm nonloc, h}$, we see that this is true, 
at the fractional 1PN accuracy ({\it i.e.}, at 5PN) for the logarithmic contributions to $\chi_4^{\rm nonloc, h}$. However, the last term,
$ \propto \frac{63}{20}\nu$, in the expression of $h^3\chi_4^{\rm nonloc, h}$ corresponds to a 5PN-level contribution equal to
\beq \label{chi4nu2}
\delta^{\nu^2} \left[\frac{h^3\chi_4^{\rm nonloc, h}}{j^4}\right]= \nu^2 \,\pi \frac{63}{20} \eta^{10} \frac{ p_\infty^6}{j^4}\,.
\eeq
The latter contribution is quadratic in $\nu$. However, the general rule of Ref.~\cite{Damour:2019lcq} applies to the {\it total}
scattering angle, and therefore says that
\bea
\label{eq:6.4_n}
h^3 \chi_4^{\rm tot}&=& h^3 \chi_4^{\rm loc, f} + h^3 \chi_4^{\rm nonloc, f} \nonumber\\
&=& h^3\chi_4^{\rm loc, f} + h^3 \chi_4^{\rm nonloc, h} + h^3\chi_4^{\rm f-h}
\eea
should be linear in $\nu$. The presence of the $O(\nu^2)$ contribution \eqref{chi4nu2} in $h^3 \chi_4^{\rm nonloc, h}$ 
is then telling us that there should be compensating $O(\nu^2)$ contributions in the other terms 
$ h^3\chi_4^{\rm loc, f} + h^3\chi_4^{\rm f-h}$. This can be arranged in many ways. On the one hand, if we were to insist on 
defining the nonlocal action by the h-route, {\it i.e.}, by systematically using $2 r_{12}^h/c$ as Pf scale, we just need
to adapt the $O(\nu^2)$ structure of the local (5PN-level) EOB potentials (so far only determined at the $O(\nu^1)$ level)
so as to absorb the term \eqref{chi4nu2}. On the other hand, it seems advantageous to use the natural flexibility in the
definition of the Pf scale ({\it i.e.}, in using a flexed Pf scale $2 r_{12}^f/c$) to make 
$ h^3\chi_4^{\rm nonloc, f} = h^3 \chi_4^{\rm nonloc, h} + h^3\chi_4^{\rm f-h}$ separately linear in $\nu$. This allows
to better separate the determinations of the local and nonlocal parts of the dynamics. We shall see below that this
second route has several nice properties.

If we choose the f-route (as we shall do here), we need to determine the coefficients $c_1, c_2, c_3$ entering the
5PN-relevant flexibility factor $f(t)= 1+ \eta^2 f_1(t) + O(\eta^4)$, namely,
\beq  \label{f_1_param}
f_1=c_1 p_r^2+ c_2 p^2 + c_3\frac{1}{r}\,,
\eeq
 so that
\beq \label{chi4f-h}
 h^3\chi_4^{\rm f-h} =\chi_4^{\rm f-h} + O(\eta^{12}) = - \nu^2 \,\pi \frac{63}{20} \eta^{10}  p_\infty^6\,.
\eeq
Here, we used the facts that $h=1+ O(\eta^2)$, and that  we require $\chi_4^{\rm f-h} =  O(\eta^{10})$.
It is not difficult to write the constraint on the coefficients $c_1, c_2, c_3$ implied by the equation \eqref{chi4f-h}.
Indeed, we can write
\beq \label{chif-h}
\frac12 \chi^{\rm f-h} = \frac{1}{2\nu}\frac{\partial}{\partial j}W^{\rm f-h}
\eeq
where (see Eq. \eqref{Wf-h})
\bea \label{Wf1}
W^{\rm f-h}&=& + 2\frac{G^2 H_{}}{c^{5}}\int dt {\mathcal F}^{\rm split}_{\rm 1PN}(t,t)  \ln \left(1+ f_1(t)\right)\nonumber\\
&=& + 2\frac{G^2 H_{}}{c^{5}}\int dt {\mathcal F}^{\rm GW}_{\rm 1PN}(t)  f_1(t)\,.
\eea

Recalling that the leading-order GW flux reads (in terms of scaled variables, and henceforth using $G=c=1$)
\beq
{\mathcal F}^{\rm GW}_{\rm 1PN}(t) = \nu^2 \frac{8}{15} \frac{1}{r^4}\left(12 p^2 - 11 p_r^2  \right)\,,
\eeq
we have
\beq \label{intWf1}
\Delta W^{\rm f-h}= \nu^2 \frac{16}{15} \int \frac{dt }{r^4}\left(12 p^2 - 11 p_r^2  \right) \left(c_1 p_r^2+ c_2 p^2 + \frac{c_3}{r}\right).
\eeq
This integral  should be evaluated at Newtonian order. Though, for our present purpose of compensating the term \eqref{chi4f-h},
we only need to compute the integral \eqref{intWf1} to leading order in inverse eccentricity, let us give its exact value, as computed
along a Newtonian orbit of  squared effective energy $\gamma^2=1+ p_{\infty}^2$ and angular momentum $j$,
with associated eccentricity $e^2 \equiv 1+  p_{\infty}^2 j^2$ and associated Newtonianlike energy 
$\bar E \equiv \frac12 p_{\infty}^2$:  
\begin{eqnarray}\label{intWhyperbolic}
\Delta W^{\rm f-h}&=& \frac{16 \nu^2 \bar E^{9/2}}{15}\left[\frac{A_1}{(e^2-1)^{9/2}}{\rm arctan} \sqrt{\frac{e+1}{e-1}} \right. \nonumber\\
&&\left.+\frac{A_2}{(e^2-1)^4} \right]\,,
\end{eqnarray}
where
\begin{eqnarray}
A_1 &=&4 \sqrt{2}[(13 c_1+74 c_2)e^6+(150 c_1+1242 c_2+366 c_3)e^4\nonumber\\
&&+(96 c_1+1544 c_2+968 c_3)e^2+192(c_2+c_3)]\,, \nonumber\\
A_2 &=& \frac{2\sqrt{2}}{15}[(1437 c_1+9866 c_2+1568 c_3)e^4\nonumber\\
&&+(2356 c_1+28758 c_2 +14734 c_3)e^2\nonumber\\
&&+92 c_1+7156 c_2 +6588 c_3]\,.
\end{eqnarray}
The beginning of the more immediately relevant large-$j$ expansion of $\Delta W^{\rm f-h}$ reads 
\begin{eqnarray} \label{intWf2}
\Delta W^{\rm f-h}&=& \frac{\nu^2}{15} \Big(\pi  (13 c_1 + 74 c_2)\frac{p_{\infty}^6}{j^3} \nonumber\\
&&+ \frac{64}{15} (51c_1+343 c_2+49 c_3)  \frac{p_{\infty}^5}{j^4}\nonumber\\
&&+ 3\pi (63c_1+488c_2+122c_3)  \frac{p_{\infty}^4}{j^5}+ \ldots \Big)\,.\nonumber\\
\end{eqnarray}

Let us now compare the result \eqref{intWf2} to the additional contribution to $W = \int dt H$, namely
\beq \label{Wfhneeded}
W^{\rm f-h, needed}= \nu^3 \, \pi \frac{21}{10} \eta^{10}  \frac{p_{\rm inf}^6}{j^3}\,,
\eeq
 which, according to  Eqs. \eqref{chi4f-h}, \eqref{chif-h}, is needed to compensate for the undesired $\nu^2/j^4$ contribution to $\chi^{\rm nonloc, h}$.
 
 By comparing Eq. \eqref{intWf2} to Eq. \eqref{Wfhneeded}, we see that it is enough that the coefficients $c_1, c_2, c_3$
 parametrizing $f_1$ satisfy the single constraint
 \beq \label{eq_c1_c2}
13 c_1+ 74 c_2 = \frac{63}{2} \nu \,. 
\eeq
 The third flexibility parameter $c_3$ does not enter this constraint because it starts contributing to $W$
 at the $\propto  \frac{\nu^2 c_3}{j^4}$ level, corresponding to a term
 $\propto  \frac{\nu^2 c_3}{j^5}$ in $\chi$.  Such a flexed contribution is not needed at 5PN.
 The choice of the value of $c_3$ is free. We could simply take $c_3 =0$. See below for the effect of choosing
 a non zero value of $c_3$.

Eq. \eqref{eq_c1_c2} yields only one constraint on the two flexibility parameters $c_1, c_2$.
The numerically simplest solution of Eq. \eqref{eq_c1_c2}
(having the  smallest denominators) would be $c_1=\frac{39}{2}\nu $, $c_2=-3 \nu$. On the other hand,
similarly to the choice of $p_r$-gauge, or energy-gauge, for
the EOB Hamiltonian, we could choose here, respectively, a flexibility factor $f_1$ containing either only $p_r^2$, namely
\beq
f_1 =  \frac{63}{26} \nu p_r^2 \,,
\eeq
or, only $p^2$, namely
\beq
f_1 =  \frac{63}{148} \nu p^2 \,.
\eeq
By straightforward computations, we showed that the additional  contribution $\Delta^{\rm f-h} H$ to the 
(nonlocal) Hamiltonian is equivalent, modulo
a canonical transformation, to the following $p_r$-gauge-type Hamiltonian
\begin{eqnarray}
\label{dHf-h}
\Delta^{\rm f-h} H&=& \frac{16}{15}\nu^2 \left[(13 c_1+74 c_2)\frac{p_r^4}{r^4}\right. \nonumber\\
&&  +(12 c_1+121 c_2+49 c_3) \frac{p_r^2}{r^5}\nonumber\\
&&\left.+ 12(c_2+c_3)\frac1{r^6}\right]\,.
\end{eqnarray}
Let us note in passing, that an efficient way of showing that $\Delta^{\rm f-h} H$ is canonically equivalent to Eq. \eqref{dHf-h}
is to compute its integral along an ellipticlike orbit (instead of an hyperboliclike one, as  in Eq. \eqref{intWhyperbolic}). 
[The time integral of the change of an Hamiltonian under an infinitesimal canonical transformation
vanishes both along hyperbolic orbits and along elliptic motions.]
The latter integral
is much simpler than Eq. \eqref{intWhyperbolic} and reads
\begin{eqnarray}
\Big[\oint dt\Delta H^{\rm f-h}\Big]_{\rm elliptic}&=&
\frac{\nu^2}{15 j^9}[(74c_2+13c_1)e_r^6\nonumber\\
&&+(1242 c_2+366 c_3+150 c_1) e_r^4\nonumber\\
&&+(96 c_1+968 c_3+1544 c_2) e_r^2\nonumber\\
&&+192(c_2+c_3)] \,.
\end{eqnarray}
In view of Eq. \eqref{dHf-h}, it is easy to see that the Hamiltonian variation $\Delta^{f-h} H$ associated with
a general $f_1$ (with arbitrary parameters $c_1, c_2, c_3$) is  equivalent to varying the potentials $A, {\bar D}, Q$
parametrizing a $p_r$-gauge EOB Hamiltonian by the amounts
\bea
\Delta^{\rm f} A   &=& \Delta^{\rm f} a_6 \, u^6 ,\nonumber\\
\Delta^{\rm f}{\bar D}  &=& \Delta^{\rm f} \bar d_{5} \, u^5 ,\nonumber\\
\Delta^{\rm f} Q &=& \Delta^{\rm f}  q_{44} \, p_r^4 \, u^4,
\eea
where
\bea \label{delf-hHc1c2c3}
\Delta^{\rm f} a_6 &=&\frac{128}{5}\nu (c_2+c_3) ,\nonumber\\
\Delta^{\rm f} \bar d_{5} &=& \frac{32}{15}\nu (12 c_1+121 c_2+49 c_3),\nonumber\\
\Delta^{\rm f}  q_{44} &=&\frac{32}{15}\nu (13 c_1 + 74 c_2)  \,.
\eea
The latter changes parametrize the contribution $\Delta^{f-h} H$ which is  a part of the {\it nonlocal} Hamiltonian, $H^{\rm nonloc, f}$,
see Eq. \eqref{deltaHnlf}. They are absent in the $h$-part of the nonlocal Hamiltonian $H^{\rm nonloc, h}$. More generally,
both parts of the $h$-type Hamiltonian, the local one, $H^{\rm loc, h}$, and the nonlocal one $H^{\rm nonloc, h}$ are
totally independent of the choice of the flexibility factor $f$. Therefore, when one decides (as  is our preferred choice)
to  use the f-route\footnote{We mean by  f-route the use of a tuned flexibility factor $f$ such that $\chi^{\rm nonloc, f}$
separately satisfies the rule of Ref.~\cite{Damour:2019lcq}.} for computing the local Hamiltonian, one ends up with $f$-type EOB potentials (say in $p_r$-gauge)
parametrizing the complementary {\it local} Hamiltonian that are related to the corresponding $h$-type ones in the following way
\bea \label{eobfvseobh}
a_6^{\rm loc, f} &=& a_6^{\rm loc, h} - \Delta^{\rm f} a_6 \,,\nonumber\\
\bar d_5^{\rm loc, f} &=& \bar d_5^{\rm loc, h} - \Delta^{\rm f} \bar d_5 \,,\nonumber\\
 q_{44}^{\rm loc, f} &=&  q_{44}^{\rm loc, h} - \Delta^{\rm f}  q_{44} \,.
\eea
The minus signs on the right-hand sides are needed because $\Delta^{\rm f} a_6$, etc. parametrize the additional
contribution $+ \Delta^{f-h} H \in H^{\rm nonloc, f}$, see Eq. \eqref{deltaHnlf}.

The changes \eqref{eobfvseobh} have been written for a general flexibility factor of the form \eqref{f_1_param}.
Let us now apply these general results to the relevant case where the parameters $c_1, c_2, c_3$ satisfy the constraint
\eqref{eq_c1_c2}. We are going to see below that, when using the f-route, the 5PN-level value of the EOB coefficient
$ q_{44}^{\rm loc, f}$ is fully determined, and takes the value indicated in Table  \ref{table_eob_loc}.
We therefore conclude from the last Eq. \eqref{eobfvseobh} that the 5PN value of the  EOB coefficient
$ q_{44}^{\rm loc, h}$ that would be derived by using the $h$-route is also fully determined, and differs from the
$f$-one by
\beq  \label{q44hvsq44f}
 q_{44}^{\rm loc, h} =  q_{44}^{\rm loc, f} +  \frac{336}{5} \nu^2\,.
\eeq
This has the effect of changing the rational $O(\nu^2)$ contribution $- \frac{2075}{3} \nu^2$ in  $q_{44}^{\rm loc, f}$
into $-\frac{9367}{15}\nu^2$.

The corresponding changes in the values of $a_6^{\rm loc} $ and $\bar d_5^{\rm loc}$ are (currently) irrelevant
because they only shift the two $O(\nu^2)$ parameters $a_6^{\nu^2}$ and  $\bar d_5^{\nu^2}$ that
are left undetermined by our method. If wished, the three flexibility parameters $c_1, c_2, c_3$ can be chosen
so as to satisfy, besides the  constraint Eq. \eqref{eq_c1_c2},  the two other equations $\Delta^{\rm f} a_6=0$,
$\Delta^{\rm f} \bar d_5=0$, ensuring that the two undetermined $O(\nu^2)$ parameters of the f-route
coincide with their corresponding h-route values. This yields the following specific values
\bea
c_1 &=& \frac{189}{4} \nu \, , \nonumber\\
c_2&=&  -\frac{63}{8} \nu \, ,\nonumber\\
c_3&=&\frac{63}{8} \nu\,.
\eea
These values define a sort of {\it minimal} choice for the flexibility factor, ensuring that the corresponding nonlocal
scattering angle $h^3 \chi_4^{\rm nonloc, f} $ be linear in $\nu$, while leaving fixed the two $O(\nu^2)$ parameters 
$a_6^{\nu^2}$ and  $\bar d_5^{\nu^2}$ entering the local dynamics.

Let us mention at this point that the formulation used in the published version of Ref. \cite{Bini:2019nra}
contains an inconsistency related to the present discussion. Indeed, the value of the local Hamiltonian defined
(in $p_r$-gauge) by Eqs. (17) there, is the f-route value, while Eq. (5)  states that one was using the h-route.
The simplest way to correct this inconsistency is to multiply the Pf scale $r_{12}^h$  entering Eq. (5) by
a factor $f=1+f_1$, solution of Eq. \eqref{eq_c1_c2}.
Alternatively,  if one insists on using the h-route ({\it i.e.},  $r_{12}^h$ as Pf length scale in the nonlocal action),  
one should replace the value  of $q_{44}^{\rm loc}= q_{44}^{\rm loc, f}$ given in Eqs. (17) there, by  $q_{44}^{\rm loc, h}$,
as given in Eq. \eqref{q44hvsq44f} above. [Correlatively,  the f-route value of
$\chi^{\rm 5PN}_{4 \rm loc}=\chi^{\rm 5PN}_{4 \rm loc, f} $
given in Eq. (19) there should then be changed into
its h-route value 
$\chi^{\rm 5PN}_{4 \rm loc,h} = \chi^{\rm 5PN}_{4 \rm loc, f} + \chi^{\rm f-h}$,
where $\chi^{\rm f-h} =-\frac{63}{20}\pi  p_{\infty}^6 \nu^2$ (see Eqs. \eqref{eq:6.4_n}, \eqref{chi4f-h})].

\section{Using the mass-ratio dependence of the scattering angle to determine most of the $\nu^{n\geq 2}$ structure
of the $f$-route local Hamiltonian}

In the following, we assume that we define the nonlocal Hamiltonian by using  a flexed Pf scale $r_{12}^f= f(t) r_{12}^h$,
with a flexibility factor $f(t)= 1 + \eta^2 f_1(t)$ satisfying the constraint discussed in the previous section. 
This allows us to separately apply the constraints
found in Ref. \cite{Damour:2019lcq} to the scattering angle deriving from the corresponding {\it local} Hamiltonian, $H^{\rm loc, f}$.
We are going to see that these constraints determine most of the nonlinear-in-$\nu$ contributions to $H^{\rm loc, f}$.

Let us start by recalling that, given any (local) Hamiltonian, the scattering angle of hyperboliclike motions is given by the integral 
 ($ u = 1/r$) \cite{Damour:2016gwp}
 \beq \label{chiintegral}
\frac12 (\chi(E,j)+\pi)=-\int_0^{u_{\rm max}} \frac{\partial}{\partial j}p_r(u; E,j) \frac{du}{u^2}\,,
\eeq
where $u_{\rm max}=u_{\rm max}(E,j)=1/r_{\rm min}$ corresponds to the distance of closest approach of the two bodies, and where
 the radial momentum $p_r=p_r(u; E,j)$ is obtained from writing the energy conservation at a given angular momentum.
As $j=J/(G M\mu)$, the PM expansion of the scattering angle is an expansion in powers of $1/j \propto G$:
\begin{eqnarray}
\frac{\chi^{\rm loc}(\e,j,\nu)}{2}&=& \chi_1^{\rm loc}(p_\infty,\nu) \frac{1}{j} +\chi_2^{\rm loc}(p_\infty,\nu) \frac{1}{j^2}\nonumber\\
&&+\chi_3^{\rm loc}(p_\infty,\nu) \frac{1}{j^3}
+\ldots \nonumber\\
\end{eqnarray}
where we replaced $\gamma=\e$ by  $p_\infty\equiv \sqrt{\e^2-1}$.

The test-mass (Schwarzschild) limit $\chi_n^{\rm Sch}$ corresponds to setting $\nu=0$.
 
With this notation, let us consider the function (which vanishes for $\nu=0$)
\beq \label{Tn}
T_n(p_\infty, \nu)=h^{n-1}(p_\infty, \nu) \chi_n^{\rm loc}-\chi_n^{\rm Sch}(p_\infty)\,,\qquad n\ge 2\,,
\eeq
where $h \equiv \sqrt{1+ 2 \nu (\e-1)}$. Ref. \cite{Damour:2019lcq} has shown that $T_n$ must be a polynomial
in $\nu$ of order (at most) $d_n =[\frac{n-1}{2}]$, where $[x]$ denotes the integer part of $x$: $T_n \sim \nu + \nu^2 + \ldots+ \nu^{d_n}$.
 Therefore we have the following conditions $C_n$:
\begin{itemize}
\item[{$C_2$}:] $T_2=0$ does not depend on $\nu$;
\item[{$C_3$}:] $T_3 \sim \nu$;
\item[{$C_4$}:] $T_4 \sim \nu$;
\item[{$C_5$}:] $T_5 \sim \nu + \nu^2$;
\item[{$C_6$}:] $T_6 \sim \nu + \nu^2$.
\end{itemize}

Here, we shall apply these results at the 5PN level, using the 5PN expansion of $h$, namely,
\begin{eqnarray}
h&=&\{1+2\nu [(1+p_\infty^2)^{1/2}-1]\}^{1/2}\nonumber\\
&=& 1+\frac12 \nu p_\infty^2 +\left(-\frac18 \nu-\frac18\nu^2\right) p_\infty^4 \nonumber\\
&+&\left(\frac{1}{16}\nu^2+\frac{1}{16}\nu+\frac{1}{16}\nu^3\right) p_\infty^6 \nonumber\\
&+&\left(-\frac{5}{128}\nu^2-\frac{5}{128}\nu-\frac{3}{64}\nu^3-\frac{5}{128}\nu^4\right) p_\infty^8 \nonumber\\
&+&\left(\frac{7}{256}\nu^2+\frac{7}{256}\nu+\frac{9}{256}\nu^3+\frac{5}{128}\nu^4+\frac{7}{256}\nu^5\right) p_\infty^{10}\nonumber\\
&+& O(p_\infty^{12})\,. 
\end{eqnarray}
and, correspondingly, of $T_n$:
\beq
T_n^{5\rm PN }(p_\infty, \nu)=[h^{n-1}(p_\infty, \nu) \chi_n^{\rm loc}-\chi_n^{\rm Sch}]^{5\rm PN }\,,\qquad n\ge 2\,.
\eeq
Our SF-based computation above has heretofore  determined only the coefficients of the $O(\nu^1)$ terms in the 
($\mu$-rescaled) local Hamiltonian. The determination of most of the $O(\nu^{\geq 2})$ coefficients in the local EOB Hamiltonian 
will now be obtained by first computing the 
PN expansion of the (local part of the conservative) scattering angle, $\chi^{\rm loc}$ (using in Eq. \eqref{chiintegral}
a  PN-expanded expression for $p_r$), and then computing the various $T_n$'s at the 5PN accuracy.

From the condition $C_2$ we find
\begin{eqnarray}
\label{cond_C2}
q_{62}^{\rm loc} &=& -\frac95 \nu-\frac{27}{5}\nu^2+6\nu^3\nonumber\\ 
q_{82}^{\rm loc} &=& \frac{18}{7}\nu^2+\frac{6}{7}\nu+\frac{24}{7}\nu^3-6\nu^4\,.
\end{eqnarray}

From the condition $C_3$ we find
\begin{eqnarray}
\label{cond_C3}
q_{43}^{\rm loc} &=& 20\nu-83\nu^2+10\nu^3\nonumber\\ 
q_{63}^{\rm loc} &=& \frac{123}{10}\nu-\frac{69}{5}\nu^2+116\nu^3-14\nu^4\,.
\end{eqnarray}

From the condition $C_4$ we find
\begin{eqnarray}
\label{cond_C4}
q_{44}^{\rm loc}&=&\left( \frac{1580641}{3150}-\frac{93031}{1536}\pi^2 \right)\nu\nonumber\\
&&+\left(-\frac{2075}{3}+\frac{31633}{512}\pi^2\right)\nu^2\nonumber\\
&&+\left(640-\frac{615}{32}\pi^2\right)\nu^3\,.
\end{eqnarray}

From the condition $C_5$ we fix $\bar d_5^{\nu^3}$ (and  $\bar d_5^{\nu^4}=0$)
so that
\begin{eqnarray}
\label{cond_C5}
\bar d_5^{\rm loc} &=& \left( \frac{331054}{175}-\frac{63707}{512}\pi^2 \right)\nu\nonumber\\
&& + \bar d_5^{\nu^2}\nu^2\nonumber\\
&& +\left(\frac{1069}{3}-\frac{205}{16}\pi^2\right)\nu^3\,,
\end{eqnarray}
where the $O(\nu^2)$ coefficient $\bar d_5^{\nu^2}$ remains undetermined.

Finally, from the condition $C_6$ we fix $a_6^{\nu^3}$ (and  $a_6^{\nu^4}=0$)
so that
\beq
\label{cond_C6} 
a_6^{\rm loc} =\left( -\frac{1026301}{1575}-\frac{246367}{3072}\pi^2 \right)\nu+ a_6^{\nu^2}\nu^2+4\nu^3\,,
\eeq
where the $O(\nu^2)$ coefficient $a_6^{\nu^2}$ remains undetermined.

The additional condition $C_7$ (meaning that $T_7 \sim \nu +\nu^2 +\nu^3$) does not carry any new information.

Summarizing: The conditions $C_n$ \cite{Damour:2019lcq} has allowed us to determine all the terms in the 
5PN-accurate (gauge-fixed) $f$-flexed local effective EOB Hamiltonian apart from the two $O(\nu^2)$ terms parametrized by $a_6^{\nu^2}$
and $\bar d_5^{\nu^2}$.

\section{Values of the  5PN-accurate $f$-{\it route local} scattering angle at PM orders $G^3$, $G^4$, $G^5$ and $G^6$} \label{scatteringangle}

Having determined most of the coefficients parametrizing the local Hamiltonian we can write down the following (PN-expanded) values
for  the $f$-flexed  local parts of the successive $n$-PM contributions, $\chi_n$, to the scattering angle (subtracted by their Schwarzschild values): 
\begin{widetext}
\begin{eqnarray} \label{chi2-chi6}
\pi^{-1}\left(\chi_2^{\rm loc}-\chi_2^{\rm Sch}\right)&=& -\frac34 p_\infty^2 \nu +\left(\frac{9}{16}\nu^2-\frac34\nu\right) p_\infty^4 \nonumber\\
&+&
\left(\frac{9}{64}\nu+\frac{27}{64}\nu^2-\frac{15}{32}\nu^3\right) p_\infty^6 
+\left(-\frac{45}{256}\nu^2-\frac{15}{256}\nu-\frac{15}{64}\nu^3+\frac{105}{256}\nu^4\right) p_\infty^8 
\,,\nonumber\\
\chi_3^{\rm loc}-\chi_3^{\rm Sch}&=& -8 p_\infty\nu +(8\nu^2-36\nu) p_\infty^3 +
\left(-\frac{91}{5}\nu+34\nu^2-8\nu^3\right) p_\infty^5 
+\left(\frac{69}{70}\nu + \frac{51}{5}\nu^2 -32\nu^3+8\nu^4\right) p_\infty^7 
\,,\nonumber\\
\pi^{-1}\left(\chi_4^{\rm loc}-\chi_4^{\rm Sch}\right)&=& -\frac{15}{4}\nu 
+\left(\frac{45}{8}\nu^2-\frac{109}{2}\nu+\frac{123}{256}\pi^2\nu\right) p_\infty^2 \nonumber\\
&+&\left(-\frac{225}{32}\nu^3+\frac{33601}{16384}\pi^2\nu-\frac{19597}{192}\nu+\frac{4827}{64}\nu^2-\frac{369}{512}\pi^2\nu^2\right) p_\infty^4 \nonumber\\
&+&\left(-\frac{94899}{32768}\pi^2\nu^2+\frac{93031}{32768}\pi^2\nu-\frac{1945583}{33600}\nu+\frac{1937}{16}\nu^2-\frac{2895}{32}\nu^3+\frac{525}{64}\nu^4+\frac{1845}{2048}\pi^2\nu^3\right) p_\infty^6 
\,,\nonumber\\
\chi_5^{\rm loc}-\chi_5^{\rm Sch}&=& -\frac{8\nu}{p_\infty} +\left(\frac{41}{8}\pi^2\nu-\frac{1168}{3}\nu+24\nu^2\right) p_\infty\nonumber\\ 
&+&\left(-\frac{227059}{135}\nu+\frac{5069}{144}\pi^2\nu-\frac{287}{24}\nu^2\pi^2+\frac{7342}{9}\nu^2-40\nu^3\right) p_\infty^3 \nonumber\\
&+&\left(-\frac{11108}{9}\nu^3-\frac{1460479}{525}\nu+\frac{41026}{15}\nu^2+56\nu^4-\frac{4}{15}\bar d_5^{\nu^2}\nu^2+\frac{451}{24}\nu^3\pi^2-\frac{40817}{640}\nu^2\pi^2\right. \nonumber\\
&+&\left.\frac{111049}{960}\pi^2\nu \right) p_\infty^5 
\,,\nonumber\\
\pi^{-1}\left(\chi_6^{\rm loc}-\chi_6^{\rm Sch}\right)&=& -\frac{625}{4}\nu+\frac{615}{256}\pi^2\nu+\frac{105}{16}\nu^2  
+\left(\frac{35065}{64}\nu^2+\frac{257195}{8192}\pi^2\nu-\frac{293413}{192}\nu-\frac{615}{64}\pi^2\nu^2-\frac{525}{32}\nu^3\right) p_\infty^2\nonumber\\
&+& \left(\frac{3675}{128}\nu^4-\frac{15}{32} \bar d_5^{\nu^2}\nu^2+\frac{2321185}{16384}\pi^2\nu-\frac{15}{32} a_6^{\nu^2} \nu^2+\frac{39975}{2048}\pi^2\nu^3-\frac{63277573}{13440}\nu-\frac{34325}{32}\nu^3\right. \nonumber\\
&+& \left.\frac{444955}{128}\nu^2-\frac{2584605}{32768}\pi^2\nu^2\right) p_\infty^4 \,.
\end{eqnarray}
\end{widetext}

The corresponding (5PN-accurate) Schwarzschild terms ($\nu=0$) are given by
\begin{eqnarray}
\chi_1^{\rm Schw}(p_\infty)&=& \frac{1}{p_\infty}+2 p_\infty 
\,,\nonumber\\
\chi_2^{\rm Schw}(p_\infty)&=& \left(\frac32 + \frac{15}{8}  p_\infty^2 \right)\pi
\,,\nonumber\\
\chi_3^{\rm Schw}(p_\infty)&=& -\frac{1}{ 3 p_\infty^3}+\frac{4}{p_\infty} +24 p_\infty + \frac{64}{3}  p_\infty^3
\,,\nonumber\\ 
\chi_4^{\rm Schw}(p_\infty)&=& \left(\frac{105}{8}+  \frac{315}{8} p_\infty^2+\frac{3465}{128} p_\infty^4\right)\pi
\,,\nonumber\\
\chi_5^{\rm Schw}(p_\infty)&=&\frac{1}{5 p_\infty^5}-\frac{2}{p_\infty^3}+\frac{ 32 }{p_\infty} +320p_\infty\nonumber\\
&& + 640 p_\infty^3 
+\frac{1792}{5} p_\infty^5
\,,\nonumber\\
\chi_6^{\rm Schw}(p_\infty)&=& \left(
\frac{1155}{8}+\frac{45045}{64} p_\infty^2  + \frac{135135}{128} p_\infty^4\right)\pi
\,,\nonumber\\
\chi_7^{\rm Schw}(p_\infty)&=&
-\frac{1}{7 p_\infty^7}+\frac{8}{5 p_\infty^5}  -  \frac{16}{p_\infty^3}  + \frac{320}{p_\infty} +4480 p_\infty\nonumber\\
&& +14336 p_\infty^3 
\,,\nonumber\\
\chi_8^{\rm Schw}(p_\infty)&=& \left(\frac{225225}{128} +\frac{765765}{64} p_\infty^2 \right)\pi
\,,\nonumber\\
\chi_9^{\rm Schw}(p_\infty)&=& \frac{1}{9 p_\infty^9}-\frac{10}{7p_\infty^7} +  \frac{96}{7}\frac{1}{p_\infty^5} -\frac{448}{3}\frac{1}{p_\infty^3}\nonumber\\
&&+\frac{3584}{p_\infty}+64512 p_\infty 
\,,\nonumber\\
\chi_{10}^{\rm Schw}(p_\infty)&=&\frac{2909907}{128}\pi\,.
\end{eqnarray}

These results for the  scattering angle provide a lot of new information that offers gauge-invariant checks for future 
independent computations of the dynamics of binary systems. 

In particular, using the fact (explicitly proven in Ref.~\cite{Bini:2017wfr}) that the nonlocal dynamics starts contributing to the
scattering angle only at $O(G^4)$, so that $\chi_3= \chi_3^{\rm loc} +\chi_3^{\rm nonloc}=\chi_3^{\rm loc}$,
our result above for $\chi_3^{\rm loc}$ actually describes the total 3PM-level scattering angle. Its explicit expression
(when combining the test-mass and $\nu^{\geq 1}$ piece, and adding our recent 6PN extension, embodied in Eq. \eqref{A6pn}) reads
\begin{eqnarray} \label{chi36pn}
\chi_3
&=&
-\frac{1}{3p_\infty^3}+\frac{4}{p_\infty} +(24-8\nu)p_\infty \nonumber\\
&&+\left(\frac{64}{3}-36\nu+8\nu^2\right)p_\infty^3\nonumber\\
&&+\left(-\frac{91}{5}\nu+34\nu^2-8\nu^3\right)p_\infty^5\nonumber\\
&&+\left(\frac{69}{70}\nu+\frac{51}{5}\nu^2-32\nu^3+8\nu^4\right)p_\infty^7\nonumber\\
&& +\left(\frac{1447}{5040}\nu-\frac{93}{56}\nu^2-\frac{27}{10}\nu^3+30\nu^4-8\nu^5\right)p_\infty^9\nonumber\\
&&+O(p_\infty^{11})\,,
\end{eqnarray}
In this expression the last term $\propto p_\infty^9$ is the 6PN contribution to $\chi_3$. As already mentioned,
 this  result is in agreement with the corresponding 6PN-level term in the PN expansion
of the 3PM-level  recent result of ~\cite{Bern:2019nnu,Bern:2019crd}. It has also been recently obtained in 
Ref. \cite{Blumlein:2020znm}.
Let us note in passing that all the rather complicated $\nu$ structure of $\chi_3$ is actually described by the simple rule $C_3$ mentioned above (i.e. the linearity of $T_3$, Eq. \eqref{Tn}, in $\nu$). Indeed, we have
\beq
h^2\chi_3 = \left(1 + 2\nu(\gamma-1) \right) \chi_3^{\rm Schw}(p_\infty)- 2 \nu p_\infty \overline C(p_\infty)
\eeq
where 
\beq
\chi_3^{\rm Schw}(p_\infty)=\frac{1}{ 3 p_\infty^3}\Big(  -1+12 p_\infty^2 +72 p_\infty^4 + 64 p_\infty^6 \Big)
\eeq
and
\beq
{\overline C}(\gamma) \equiv  (\gamma-1) \left(A_{q3}(\gamma) + B_{q3}(\gamma) \right)
\eeq
whose 6PN-accurate expansion reads
\bea\label{CB6PN}
{\overline C}^{ 6 \rm PN}(p_\infty)&=& 4 + 18 p_\infty^2 + \frac{91}{10} p_\infty^4  - \frac{69}{140} p_\infty^6  \nonumber\\
&-& \frac{1447}{10080} p_\infty^8 + O(p_\infty^{10})\,.
\eea
In addition, our results also provide a complete,  5PN-accurate value for the 4PM-level scattering angle 
$\chi_4= \chi_4^{\rm loc, f} + \chi_4^{\rm nonloc, f}$. It is convenient to reexpress the result for $\chi_4$
in terms of its rescaled version, $\widetilde \chi_4=h^3 \chi_4$. We have evaluated in section \ref{f1}
the nonlocal contribution to $h^3 \chi_4$, namely (when using a flexibility factor $f_1$ of the type discussed there)
\begin{widetext}
\begin{eqnarray}
\label{h3chi4_nonolocf}
h^3 \chi_4^{\rm nonloc, f}(\gamma,\nu) &=& \nu \,\eta^8 p_\infty^4  \pi \left\{ -\frac{63}{4}-\frac{37}{5} \ln \left(\frac{p_\infty}{2} \right) +\eta^2 p_\infty^2\left[-\frac{2753}{1120}-\frac{1357}{280} \ln \left(\frac{p_\infty}{2} \right) \right] 
 \right\}\,.
\end{eqnarray}
\end{widetext}
Concerning the corresponding complementary f-type local contribution $\chi_4^{\rm loc, f}$
we have already given its explicit value in Eqs. \eqref{chi2-chi6} above.  Let us also cite
the much simpler expression of its rescaled version, which is linear in $\nu$.
Similarly to the  rescaled version of $\chi_3$ written above, it  can be written as
\beq
h^3 \chi_4^{\rm loc, f}(\gamma,\nu) =\left(1 + 2\nu(\gamma-1) \right) \chi_4^{\rm Schw}(p_\infty) + \nu \widehat \chi_4^{\rm loc, f}(p_\infty)\,,
\eeq 
where 
\begin{eqnarray}
&&\widehat \chi_4^{\rm loc, f}(p_\infty)=
\pi\left[-\frac{15}{4}+\left(\frac{123}{256}\pi^2-\frac{767}{16}\right) p_\infty^2\right. \nonumber\\
&& +\left(-\frac{4033}{48}+\frac{33601}{16384}\pi^2\right) p_\infty^4\nonumber\\
&&\left. +\left(-\frac{6514457}{134400}+\frac{93031}{32768}\pi^2\right) p_\infty^6\right]\nonumber\\
&&+ O(p_{\infty}^{8})\,.
\end{eqnarray}

Finally, concerning the 5PM and 6PM local scattering angles, $\chi_5^{\rm loc, f}$, $\chi_6^{\rm loc, f}$,
most of the information displayed in Eqs. \eqref{chi2-chi6} above comes from the  $h^{n-1}$-rescaling rule.
[It is, however, important to confirm this rule by explicit computations.] Apart from the latter rule, the
new 5PM and 6PM information derived here concerns the linear-in-$\nu$ contributions. Indeed, we
can write the  expressions
\begin{eqnarray}
&&\nu^{-1}h^4 [\chi_5^{\rm loc, f}(\gamma,\nu) -\chi_5^{\rm Schw}(p_\infty)]=\nonumber\\
&&\qquad -\frac{8}{p_\infty}+\left(8\nu+\frac{41}{8}\pi^2-\frac{1168}{3}\right) p_\infty \nonumber\\
&& \qquad+\left(\frac{370}{9}\nu+\frac{5069}{144}\pi^2-\frac{41}{24}\nu\pi^2-\frac{227059}{135}\right)p_\infty^3\nonumber\\
&& \qquad +\left(-\frac{4}{15}\nu\bar d_5^{\nu^2}+\frac{23407}{5760}\nu\pi^2\right. \nonumber\\
&& \left.\qquad+\frac{111049}{960}\pi^2-\frac{1460479}{525}-\frac{58874}{135} \nu\right)p_\infty^5\nonumber\\
&&\qquad + O(p_\infty^7)\,,\nonumber\\
&&\nu^{-1}\pi^{-1}h^5 [\chi_6^{\rm loc, f}(\gamma,\nu) -\chi_6^{\rm Schw}(p_\infty)]=\nonumber\\
&&\qquad \frac{615}{256}\pi^2+\frac{105}{16}\nu-\frac{625}{4} \nonumber\\
&&+\left(\frac{10065}{64}\nu-\frac{293413}{192} +\frac{257195}{8192}\pi^2-\frac{1845}{512}\nu\pi^2\right)p_\infty^2\nonumber\\
&&+\left(-\frac{15}{32}\nu (\bar d_5^{\nu^2}+ a_6^{\nu^2}) 
-\frac{23675}{96}\nu -\frac{61855}{32768}\nu\pi^2\right. \nonumber\\
&&\left.+\frac{2321185}{16384}\pi^2-\frac{63277573}{13440}
 \right)p_\infty^4\nonumber\\
&&+O(p_\infty^6)\,,
\end{eqnarray}
which exhibit the simple $\sim 1+ \nu + \nu^2$ structure, and emphasize that the coefficients of the $\nu^2$
are currently not fully determined, since they involve the $O(\nu^2)$ terms $\bar d_5^{\nu^2}$ and $a_6^{\nu^2}$.

\section{Final results for the 5PN-accurate $f$-route local EOB Hamiltonian}

Let us gather the 5PN-accurate results (for the f-type local dynamics) obtained so far in the previous sections.
They concern various forms of the  local  Hamiltonian:  (i) the energy-gauge
version of the local effective EOB Hamiltonian; (ii) the $p_r$-gauge version of the local effective EOB Hamiltonian;
 (iii) the local {\it real} Hamiltonian; and  (iv) the canonical transformation connecting the $p_r$-gauge to the
 energy-gauge. Before listing our results, let us recall again the link between the usual ``real'' Hamiltonian, $H$,
 and the dimensionless ($\mu$-rescaled) effective EOB Hamiltonian $\widehat H_{\rm eff}$:
\beq
\label{Hrealdef}
H^{\rm loc}= H_{\rm eob}^{\rm loc}=M \sqrt{1+ 2 \nu (\widehat H_{\rm eff}^{\rm loc} -1)}\,.
\eeq
Note that we sometimes (as indicated here) add a subscript eob to the real, local Hamiltonian $H^{\rm loc}$
 when we wish to emphasize that it is expressed in terms of  EOB canonical coordinates. But, numerically,
 $ H_{\rm eob}^{\rm loc}$ is equal to the usual (local) Hamiltonian, whose conserved value is equal to the
 total, c.m. conserved energy of the binary system (minus the nonlocal 4+5PN contribution linked to Eq. \eqref{Snonloc}).

\subsection{5PN-accurate $f$-flexed local effective EOB Hamiltonian in energy-gauge}

We recall that the energy-gauge, squared effective EOB Hamiltonian is written as
\beq
\hat H^2_{\rm eff, EG}= H_S^2+(1-2 u) Q_{\rm EG}(u,H_S)\,,
\eeq
where the rescaled Schwarzschild Hamiltonian  $H_S = \sqrt{(1-2 u) [1+(1-2 u) p_r^2+j^2 u^2]}$ and where 
the energy-gauge $Q$ potential is written as
\begin{eqnarray} \label{QEG1}
Q_{\rm EG}(u,H_S)&=&u^2 q_{2\rm EG}(H_S; \nu)+u^3 q_{3\rm EG}(H_S; \nu)\nonumber\\
&+& 
u^4 q_{4\rm EG}(H_S; \nu)+u^5q_{5\rm EG}(H_S; \nu)\nonumber\\
&+& 
u^6 q_{6\rm EG}(H_S; \nu) +\ldots\,.
\end{eqnarray}
The exact value of the 2PM coefficient $q_{2\rm EG}(\gamma;\nu)$ is given by Eq. \eqref{q2EG} (where we recall that 
$h(\gamma;\nu) \equiv [1+2\nu(\gamma-1)]^{1/2}$).
The exact $\nu$-dependence of the 3PM coefficient $q_{3\rm EG}(\gamma;\nu)$ is described by 
\bea \label{q3EG1}
q_{3\rm EG}(\gamma;\nu)&=& A_{q3}(\gamma) \left(1- \frac1{h^2(\gamma;\nu)} \right) \nonumber\\
&&+\frac{B_{q3}(\gamma)}{h(\gamma;\nu)}\left(1- \frac1{h(\gamma;\nu)} \right)\,,
\eea
where the exact value of $B_{q3}(\gamma)$ is given in Eq. \eqref{Bq3def}, and where our new method
has allowed us to compute the 6PN-accurate value of the function $A_{q3}(\gamma)$, as given
by Eqs. \eqref{A3PN},\eqref{A5pn}, \eqref{A6pn}. According to Refs.~\cite{Bern:2019nnu,Bern:2019crd},
the exact value of  $A_{q3}(\gamma)$, is given by Eqs. \eqref{A3B}, \eqref{bCB}.

Less PN information is known about the higher PM coefficients $q_{n\rm EG}(\gamma;\nu)$, though
the analog of the exact $\nu$-structure displayed for $n=3$ in Eq. \eqref{q3EG1} has been given in Ref. \cite{Damour:2019lcq}.
Here, we shall parametrize their PN expansions as follows
\begin{eqnarray}
\label{q_energy_gauge_fun}
q_{4\rm EG}(\gamma;\nu) &=& \nu \left(\frac{175}{3}-\frac{41}{32}\pi^2\right) -\frac{7}{2}\nu^2\nonumber\\
 &&  +q_{4\rm EG}^1(\nu) (\gamma^2-1)+q_{4\rm EG}^2(\nu) (\gamma^2-1)^2 , \nonumber\\ 
q_{5\rm EG}(\gamma;\nu)&=&q_{5\rm EG}^0(\nu)+q_{5\rm EG}^1(\nu)(\gamma^2-1),\nonumber\\ 
q_{6\rm EG}(\gamma;\nu)&=&q_{6\rm EG}^0(\nu)  \,.
\end{eqnarray}
Here, the first term in $q_{4\rm EG}$ is at the 3PN level, the first term in  $q_{5\rm EG}$ is at the 4PN level, and the first, and only, 
term in $q_{6\rm EG}$ is at the 5PN level.

The final form of the $\nu$-dependent PN-expansion parameters $q_{n\rm EG}^p(\nu)$ entering the $f$-flexed energy-gauge (squared) effective  Hamiltonian, 
Eqs. \eqref{QEG1}, \eqref{q_energy_gauge_fun}, is the following
\begin{eqnarray}
\label{en_gauge_pot}
q_{4\rm EG}^1(\nu) &=& \left(\frac{5632}{45}-\frac{33601}{6144}\pi^2\right)\nu \nonumber\\
&&+\left(-\frac{405}{4}+\frac{123}{64}\pi^2\right)\nu^2+\frac{13}{2}\nu^3\,, \nonumber\\
q_{4\rm EG}^2(\nu)  
&=& \left(\frac{699761}{7200}-\frac{93031}{12288}\pi^2\right)\nu\nonumber\\
&& +\left(-\frac{77443}{480}+\frac{31633}{4096}\pi^2\right)\nu^2\nonumber\\
&& +\left(130-\frac{615}{256}\pi^2\right)\nu^3-\frac{293}{32}\nu^4\,,\nonumber\\
q_{5\rm EG}^0 (\nu)
&=& \left(\frac{44357}{360}-\frac{29917}{6144}\pi^2\right)\nu\nonumber\\
&& +\left(\frac{205}{64}\pi^2-\frac{2387}{24}\right)\nu^2\nonumber\\
&& +\frac94 \nu^3\,,\nonumber\\
q_{5\rm EG}^1(\nu)   
&=& \left(\frac{15540691}{25200}-\frac{2590847}{61440}\pi^2\right)\nu\nonumber\\
&&+\left(-\frac{15581}{80}+\frac15\bar d_5^{\nu^2}+\frac{347673}{20480}\pi^2\right)\nu^2\nonumber\\
&&+\left(\frac{5131}{24}-\frac{1763}{256}\pi^2\right)\nu^3\nonumber\\
&&-\frac{93}{16}\nu^4\,,\nonumber\\
q_{6\rm EG}^0 (\nu)  
&=& \left(-\frac{69733}{350}+\frac{541363}{10240} \pi^2\right)\nu\nonumber\\
&& +\left(\frac{11717}{60}+\frac15\bar d_5^{\nu^2}+\frac{17857}{5120}\pi^2
+a_6^{\nu^2}\right)\nu^2\nonumber\\
&& +\left(\frac{326}{3}-\frac{287}{64}\pi^2\right)\nu^3\nonumber\\
&&-\frac{11}{8}\nu^4\,.\nonumber\\
\end{eqnarray}
 
\subsection{5PN-accurate $f$-type local effective EOB Hamiltonian in $p_r$-gauge}

In the standard $p_r$-gauge, the final form of the 5PN-accurate building blocks $A(u; \nu)$,  ${\bar D}(u; \nu)$ and $Q(u,p_r;\nu)$
of the $f$-type local effective EOB Hamiltonian $\widehat H_{\rm eff}^{\rm loc}$ are:
\begin{eqnarray}
A_{\rm loc} &=& 1-2 u+2\nu u^3+\nu \left(\frac{94}{3}-\frac{41}{32}\pi^2\right)u^4\nonumber\\
&&+a_5^{\rm loc} u^5+a_6^{\rm loc} u^6\,,\nonumber\\ 
\bar D_{\rm loc} &=& 1+6\nu u^2+(52\nu-6\nu^2) u^3+\bar d_4^{\rm loc} u^4+\bar d_5^{\rm loc} u^5\,,\nonumber\\
Q_{\rm loc}&=& p_r^4 [2  (4-3\nu)\nu u^2+q_{43}^{\rm loc}u^3 +q_{44}^{\rm loc}u^4 ]\nonumber\\
&&+p_r^6 (q_{62}^{\rm loc}u^2 +q_{63}^{\rm loc}u^3 )+q_{82}^{\rm loc}p_r^8  u^2\,.
\end{eqnarray}
The values of the coefficients 
$a_5^{\rm loc}$, $a_6^{\rm loc}$, $\bar d_4^{\rm loc}$, $\bar d_5^{\rm loc}$, $q_{43}^{\rm loc}$, $q_{44}^{\rm loc}$, $q_{62}^{\rm loc}$, $q_{63}^{\rm loc}$, $q_{82}^{\rm loc}$ parametrizing the 4+5PN structure of $\widehat H_{\rm eff}^{\rm loc}$ 
are summarized in Table \ref{table_eob_loc}.


\begin{table}
\caption{\label{table_eob_loc} Coefficients of the f-route local 4+5PN part of the EOB potentials. 
}
\begin{ruledtabular}
\begin{tabular}{ll}
\hline
Coefficient &  Expression \\ 
$a_5^{\rm loc}$ & $\left(\frac{2275}{512}\pi^2-\frac{4237}{60}\right)\nu +\left(\frac{41}{32}\pi^2-\frac{221}{6}\right)\nu^2$\\
$a_6^{\rm loc}$ &$\left( -\frac{1026301}{1575}+\frac{246367}{3072}\pi^2 \right)\nu+ a_6^{\nu^2}\nu^2+4\nu^3$\\
$\bar d_4^{\rm loc}$ & $\left(\frac{1679}{9}-\frac{23761}{1536} \pi^2\right)\nu +\left(-260+\frac{123}{16}\pi^2\right) \nu^2$\\
$\bar d_5^{\rm loc}$ & $\left( \frac{331054}{175}-\frac{63707}{512}\pi^2 \right)\nu+ \bar d_5^{\nu^2}\nu^2$\\
& $+\left(\frac{1069}{3}-\frac{205}{16}\pi^2\right)\nu^3$\\
$q_{43}^{\rm loc}$ & $20\nu-83\nu^2+10\nu^3$\\ 
$q_{44}^{\rm loc}$ & $\left( \frac{1580641}{3150}-\frac{93031}{1536}\pi^2 \right)\nu+\left(-\frac{2075}{3}+\frac{31633}{512}\pi^2\right)\nu^2$\\
&$+\left(640-\frac{615}{32}\pi^2\right)\nu^3$\\
$q_{62}^{\rm loc}$ & $-\frac95 \nu-\frac{27}{5}\nu^2+6\nu^3$\\ 
$q_{63}^{\rm loc}$ & $\frac{123}{10}\nu-\frac{69}{5}\nu^2+116\nu^3-14\nu^4$\\
$q_{82}^{\rm loc}$ & $\frac{18}{7}\nu^2+\frac{6}{7}\nu+\frac{24}{7}\nu^3-6\nu^4$\\
\hline
\end{tabular}
\end{ruledtabular}
\end{table}

\subsection{The standard ($f$-type local) EOB Hamiltonian at 5PN}

For completeness, let also display the  5PN ($f$-type local) real  Hamiltonian as function of $u,p_r$, and $p^2$, where 
$$p^2\equiv p_r^2+j^2u^2 \,.$$
Inserting the results of the previous subsection in the EOB energy map \eqref{Hrealdef},
one gets the following explicit (real) EOB Hamiltonian
\beq \label{Heobfin}
H_{\rm eob}^{\rm loc,5PN}=\sum_{0\leq k \leq 6, 0\leq l\leq 4, k+l\leq4}  C^{(2 l)}_{2k}(\nu) p^{2k} p_r^{2l}u^{6-k-l}\,.
\eeq
The  $\nu$-dependent coefficients  $C^{(2 l)}_{2k}(\nu)= \sum_n C^{(2 l)}_{2k, n} \nu^n$ are listed in Table \ref{table_Hreal}.


\begin{table}[h]
\caption{\label{table_Hreal} Coefficients entering the real EOB Hamiltonian.
All coefficients $C^{(2l)}_{2k}(\nu)$ start linearly in $\nu$, except $C^{(6)}_{2}(\nu)$ and $C^{(8)}_{0}(\nu)$, which begin instead at $O(\nu^2)$.
Furthermore, the highest power of $\nu$ in the coefficients $C^{(0)}_{2k}(\nu)$ is $\nu^6$, whereas the remaining coefficients stop at $\nu^5$.
The number of non-zero coefficients is then 95.}
\begin{ruledtabular}
\begin{tabular}{lll}
Coefficient & Powers & Value \\
\hline
$C^{(0)}_{12}(\nu)$ & $p_r^0 p^{12} u^0$ & $  -\frac{21}{1024}\nu-\frac{21}{1024}\nu^2-\frac{7}{256}\nu^3-\frac{35}{1024}\nu^4$\\
& &$-\frac{35}{1024}\nu^5-\frac{21}{1024}\nu^6$\\
$C^{(0)}_{10}(\nu)$ & $p_r^0 p^{10} u^1$ & $  -\frac{7}{256}\nu-\frac{7}{256}\nu^2-\frac{3}{128}\nu^3+\frac{5}{256}\nu^4$\\
&&$+\frac{35}{256}\nu^5+\frac{63}{256}\nu^6 $\\ 
$C^{(0)}_{8}(\nu)$ & $p_r^0 p^{8} u^2$ & $ \frac{5}{256}\nu+\frac{5}{256}\nu^2+\frac{3}{64}\nu^3+\frac{35}{256}\nu^4$\\
&&$+\frac{35}{256}\nu^5-\frac{315}{256}\nu^6  $\\
$C^{(0)}_{6}(\nu)$ & $p_r^0 p^{6} u^3$ & $ -\frac{1}{32}\nu+\frac{1}{32}\nu^2-\frac{5}{32}\nu^4-\frac{45}{32}\nu^5 +\frac{105}{32}\nu^6 $\\
$ C^{(0)}_{4}(\nu)$ &$p_r^0 p^{4} u^4$ & $\frac{5}{64}\nu+\left(\frac{41}{512}\pi^2-\frac{385}{192}\right)\nu^2$\\
&&$+\left(-\frac{91}{48}+\frac{41}{512}\pi^2\right)\nu^3  $ \\
& &$ +\left(-\frac{123}{512}\pi^2+\frac{363}{64}\right)\nu^4+\frac{155}{64}\nu^5-\frac{315}{64}\nu^6$\\
$C^{(0)}_{2}(\nu)$ &$p_r^0 p^{2} u^5$ &$-\frac{7}{16}\nu+\left(\frac{1619}{2048}\pi^2-\frac{1141}{120}\right)\nu^2$\\
&&$ +\left(-\frac{2275}{2048}\pi^2+\frac{3997}{240}\right)\nu^3 $ \\
& &$+\left(-\frac{671}{48}+\frac{41}{64}\pi^2\right)\nu^4-\frac{25}{16}\nu^5 +\frac{63}{16}\nu^6$\\
$C^{(0)}_{0}(\nu)$ &$p_r^0 p^{0} u^6$ & $-\frac{21}{16}\nu+\left(\frac{254113}{6144}\pi^2-\frac{8478053}{25200}\right)\nu^2  $\\
& &$+\left(\frac12 a_6^{\nu^2}-\frac{1199}{40}+\frac{1947}{1024}\pi^2\right)\nu^3$\\
& &$ +\left(\frac{355}{48}-\frac{41}{128}\pi^2\right)\nu^4+\frac{5}{16}\nu^5-\frac{21}{16}\nu^6$\\
\hline
$C^{(2)}_{8}(\nu)$ & $p_r^2 p^{8} u^1$ &$ -\frac{35}{128}\nu-\frac{35}{128}\nu^2-\frac{45}{128}\nu^3-\frac{25}{64}\nu^4-\frac{35}{128}\nu^5  $\\
$C^{(2)}_{6}(\nu)$  &$p_r^2 p^{6} u^2$ &$  -\frac{5}{16}\nu-\frac{5}{4}\nu^2-\frac{9}{8}\nu^3-\frac12 \nu^4+\frac{5}{4}\nu^5 $\\
$C^{(2)}_{4}(\nu)$ & $p_r^2 p^{4} u^3$ &$ +\frac{3}{16}\nu +\frac{111}{16}\nu^2+\frac{99}{16}\nu^3+\frac{69}{8}\nu^4-\frac{33}{16}\nu^5$\\
$C^{(2)}_{2}(\nu)$ & $p_r^2 p^{2} u^4$ &$-\frac14 \nu+\left(-\frac{611}{36}+\frac{25729}{6144}\pi^2\right)\nu^2 $\\
& &$ +\left(\frac{1549}{36}+\frac{13921}{6144}\pi^2\right)\nu^3$\\
&&$+\left(\frac{59}{2}-\frac{123}{64}\pi^2\right)\nu^4+2\nu^5$\\
$C^{(2)}_{0}(\nu)$ & $p_r^2 p^{0} u^5$ &$\frac{5}{8}\nu+\left(\frac{447313}{700}-\frac{36359}{1024}\pi^2\right)\nu^2 $\\
&&$+\left(-\frac{61153}{3072}\pi^2+\frac{31397}{72}+\frac12  \bar d_5^{\nu^2}\right)\nu^3$\\
&&$+\left(-\frac{41}{16}\pi^2+\frac{1031}{12}\right)\nu^4 -\frac{11}{8}\nu^5$\\
\hline 
$C^{(4)}_{4}(\nu)$  & $p_r^4 p^{4} u^2$ &$-\frac{15}{16}\nu+ \frac{9}{16}\nu^2-\frac{3}{4}\nu^3 -\frac{9}{16}\nu^4-\frac{9}{8}\nu^5$\\
$C^{(4)}_{2}(\nu)$ & $p_r^4 p^{2} u^3$ &$-\frac{3}{4}\nu-\frac{33}{4}\nu^2+\frac{47}{4}\nu^3 +10\nu^4+2\nu^5$\\
$C^{(4)}_{0}(\nu)$ &$p_r^4 p^{0} u^4$ &$+\frac{1}{4}\nu+\left(-\frac{93031}{3072}\pi^2+\frac{405004}{1575}\right)\nu^2 $ \\
&&$+\left(\frac{31633}{1024}\pi^2-\frac{1697}{6}\right)\nu^3$\\
&&$+\left(-\frac{615}{64}\pi^2+\frac{1115}{4}\right)\nu^4+\frac12 \nu^5$\\
\hline
$C^{(6)}_{2}(\nu)$&$p_r^6 p^{2} u^2$ &$ \frac{9}{20}\nu^2+\frac{9}{5}\nu^3 -\frac{3}{20}\nu^4-\frac{3}{2}\nu^5 $\\
$C^{(6)}_{0}(\nu)$&$p_r^6 p^{0} u^3$ &$  -\frac{1}{2}\nu+\frac{211}{20}\nu^2-\frac{23}{5}\nu^3+\frac{493}{10}\nu^4-4\nu^5$\\
\hline
$C^{(8)}_{0}(\nu)$ &$p_r^8 p^{0} u^2$ &$  \frac{3}{7}\nu^2+\frac{9}{7}\nu^3+\frac{12}{7}\nu^4-3\nu^5 $\\
\end{tabular}
\end{ruledtabular}
\end{table}

\begin{figure*}
\includegraphics[scale=0.75]{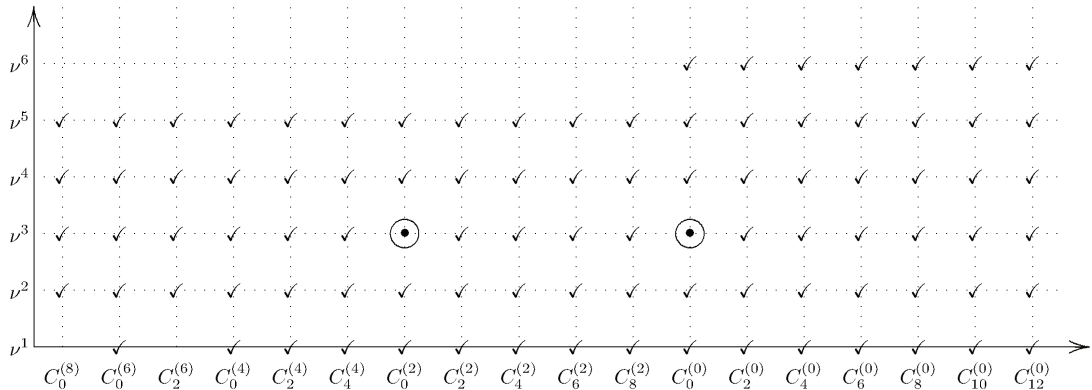}
\caption{\label{fig:1}
Matrix of the 97 non-zero numerical coefficients $C^{(2 l)}_{2k, n}$ encoding the various powers of $\nu$ in
the Hamiltonian \eqref{Heobfin}.}
\end{figure*}

Let us recall once more that, modulo the only two undetermined coefficients  $a_6^{\nu^2}$ and $\bar d_5^{\nu^2}$,
the full 5PN-accurate dynamics has been determined here. It is given by adding to the local action defined by $H_{\rm loc}^{\leq 5 \rm PN}$ the $f$-flexed 4+5PN nonlocal one written down in Eqs. \eqref{delta_H_nonloc}, \eqref{DHfh}. We find remarkable that though the real local Hamiltonian finally involves
97 different numerical coefficients keying the various powers of $u$, $p^2$, $p_r^2$ and $\nu$ (as listed in Table \ref{table_Hreal}),
our combination of tools has allowed us to determine all these coefficients, except for two of them. 
To help vizualizing the structure of the 5PN Hamiltonian (encoded in the $\nu$-dependent coefficients $C^{(2 l)}_{2k}(\nu)$)
we present the matrix of the non-zero numerical coefficients $C^{(2 l)}_{2k, n}$ entering  
$C^{(2 l)}_{2k}(\nu)= \sum_n C^{(2 l)}_{2k, n} \nu^n$ in Fig. \ref{fig:1}


We summarize in 
Fig. 2  the source of information having allowed us to determine each one of these 97 coefficients. Fig. 2 is a schematic version
of Fig. \ref{fig:1} in which we do not distinguish
 $p^2$ from $p_r^2$, so that there seems to appear only 36 coefficients: 
the test-particle limit determines the  $\nu^1$ row; the 1SF computations determine the  $\nu^2$ row; the first two columns
are respectively determined by the 1PM and 2PM exact EOB Hamiltonians; the  $\nu^{\geq 3}$ dependence of the next third and fourth
columns (respectively corresponding to 3PM and 4PM) is completely determined by the EOB-PM  result concerning
the $\nu$-polynomial structure of $T_n$, Eq. \eqref{Tn}.
 The latter result also determines
the coefficients in the last two columns (5PM and 6PM) except for the two coefficients
$h_{2 \,5}^{\nu^3}$ and $h_{0 \,6}^{\nu^3}$. 


\begin{figure}
\includegraphics[scale=0.75]{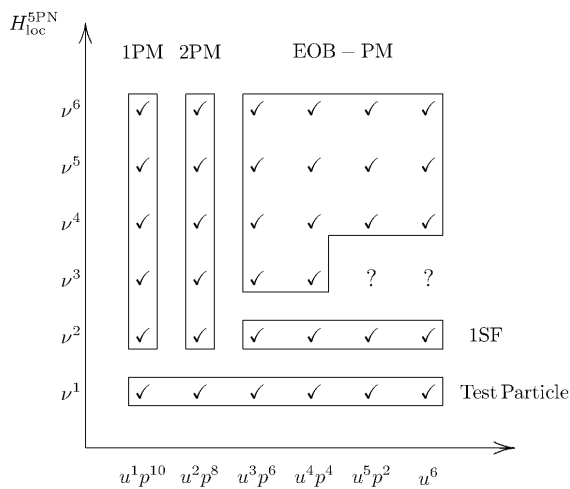}
\caption{Schematic representation of the theoretical tools used to obtain the various contributions to the 5PN-accurate local Hamiltonian, adapted from Ref. \cite{Bini:2019nra}. 
These contributions are keyed,  on the horizontal axis, by powers of $u=GM/r$ and squared momentum  
$p^2 \sim p_r^2 $, and, on the 
the vertical axis, by powers of $\nu \equiv m_1 m_2/(m_1+m_2)^2$. The checks indicate the coefficients determined  in the present work. The question marks denote the only two missing coefficients.
Note that even if certain coefficients in Table \ref{table_Hreal} only include terms up to $O(\nu^5)$, the identification $p^2 \sim p_r^2 $ done in this schematic figure  lumps terms together so that $O(\nu^6)$ terms arise in each column.}
\end{figure}

\subsection{Canonical transformation between the $p_r$-gauge and the energy-gauge}

Let us finally give the values of the parameters $g_i$, $h_i$, $n_i$ entering the generating function $g(q,p)$, Eq. \eqref{gauge},
of the canonical transformation connecting the $p_r$-gauge and the energy-gauge $f$-flexed local Hamiltonians. If we denote the $p_r$-gauge phase-space
variables as  $(r,p_r)$ (with Hamiltonian $H(r,p_r)$) and the energy-gauge ones as $(r',p_r')$ (with Hamiltonian $H'(r',p_r')$), we have
$H(r,p_r)= H'(r',p_r')$ with the following link between the phase-space variables (besides $p_\phi=j=p_{\phi}'$)
\beq
r'=r+\partial_{p_r'}g(r,p_r') \; ; \; p_r=p_r'+\partial_r g(r,p_r')\,.
\eeq
In Table \ref{table_gauge_params} we list the final form of the gauge parameters, necessary to pass from the standard EOB gauge to the energy gauge.

\begin{widetext}


\begin{table}
\caption{\label{table_gauge_params} Gauge parameters entering the canonical transformation \eqref{gauge}}
\begin{ruledtabular}
\begin{tabular}{lll }
PN order & parameter & value \\
\hline
&$g_1$ &  $\frac{13}{2}\nu+ \frac12 \nu^2$\\ 
3PN & $g_2$ & $\frac{3}{2}\nu -\frac{9}{8}\nu^2$\\ 
(See Ref. \cite{Damour:2017zjx})) &$g_3$  & $\frac{5}{2}\nu -\frac{15}{8}\nu^2$\\
\hline
&$h_1$ & $\left(\frac{2419}{144}-\frac{25729}{12288}\pi^2\right)\nu+\left(-\frac{353}{16}+\frac{123}{128}\pi^2\right)\nu^2+\frac{5}{8}\nu^3 $\\ 
&$h_2$ & $-\frac{9}{32}\nu -\frac{27}{32}\nu^2 +\frac{15}{16}\nu^3$\\
&$h_3$ & $-\frac{99}{160}\nu -\frac{297}{160}\nu^2+\frac{33}{16}\nu^3 $\\
4PN & $h_4$ & $\frac{263}{80}\nu-\frac{101}{16}\nu^2-\frac{13}{8}\nu^3 $\\
(See Ref. \cite{Antonelli:2019ytb}) &$h_5$ & $ \frac{281}{80}\nu -\frac{199}{16}\nu^2 -\frac{11}{8}\nu^3$\\
&$h_6$ & $-\frac{3}{4}\nu-\frac{9}{4}\nu^2+\frac{5}{2}\nu^3$\\ 
\hline
&$n_1$ & $\left(\frac{3082519}{25200}-\frac{224053}{30720}\pi^2\right)\nu+\left(\frac{15499}{240}+\frac{1}{10}\bar d_5^{\nu^2}-\frac{96211}{20480}\pi^2\right)\nu^2+\left(-\frac{7}{24}+\frac{41}{256}\pi^2\right)\nu^3+\frac78 \nu^4$\\
&$n_2$ & $\left(\frac{798353}{25200} -\frac{93031}{24576}\pi^2\right)\nu+\left(\frac{31633}{8192}\pi^2-\frac{11227}{240}\right)\nu^2+\left(\frac{357}{16}-\frac{615}{512}\pi^2\right)\nu^3-\frac{21}{8}\nu^4$\\
&$n_3 $ & $\left(\frac{758123}{10800}-\frac{651217}{73728}\pi^2\right)\nu+\left(-\frac{71857}{720}+\frac{221431}{24576}\pi^2\right)\nu^2+\left(\frac{3331}{48}-\frac{1435}{512}\pi^2\right)\nu^3-\frac73 \nu^4$\\
&$n_4$ & $\frac{603}{1120}\nu-\frac{321}{160}\nu^2+\frac{91}{16}\nu^3+\frac{41}{16}\nu^4 $\\
&$n_5$ & $\frac{97}{35}\nu -\frac{51}{20}\nu^2+\frac{31}{2}\nu^3+\frac{11}{2}\nu^4 $\\
5PN &$n_6$ &$\frac{3657}{1120}\nu-\frac{171}{160}\nu^2 +\frac{273}{16}\nu^3+\frac{39}{16}\nu^4 $\\
&$n_7$ &$ \frac{15}{128}\nu+\frac{45}{128}\nu^2+\frac{15}{32}\nu^3-\frac{105}{128}\nu^4 $\\
&$n_8$ &$ \frac{55}{128}\nu+\frac{165}{128}\nu^2+\frac{55}{32}\nu^3-\frac{385}{128}\nu^4 $\\
&$n_9$ &$ \frac{73}{128}\nu+\frac{219}{128}\nu^2+\frac{73}{32}\nu^3-\frac{511}{128}\nu^4 $\\
&$n_{10}$ &$ \frac{279}{896}\nu+\frac{837}{896}\nu^2+\frac{279}{224}\nu^3-\frac{279}{128}\nu^4$\\
\hline
\end{tabular}
\end{ruledtabular}
\end{table}

\end{widetext}

\section{5.5PN-level action and its transcription into the EOB standard gauge Hamiltonian} 

A somewhat surprising result of SF computations was the discovery \cite{Shah:2013uya} of half-integer-power PN contributions
(starting at the 5.5PN level) to the near-zone metric and to the Hamiltonian. This was quickly understood 
\cite{Shah:2013uya,Bini:2013rfa,Blanchet:2013txa} as coming from second-order tail (or tail-of-tail, or simply tail$^2$) effects.
The conservative action term associated with such tail$^2$ effects was obtained in 
 Ref. \cite{Damour:2015isa} (see Section IXB there, Eq. (9.19)). It reads
 \beq \label{S5.5pn}
S_{\rm 5.5PN}=-\int dt H_{\rm 5.5 PN}\,,
\eeq
where the 5.5 PN Hamiltonian is given by the following nonlocal tail$^2$ expression
\begin{eqnarray} \label{Htail2}
H_{\rm 5.5 PN} = H_{{\rm tail}^2}&=&\frac{B}{2}\left(\frac{G {\cal M}}{c^3} \right)^2 \int_{-\infty}^\infty \frac{d\tau}{\tau}[{\mathcal G}^{\rm split}(t,t+\tau)\nonumber\\
&&
-{\mathcal G}^{\rm split}(t,t-\tau)]\,,
\end{eqnarray}
with $B=-\frac{107}{105}$. Similarly to the tail$^1$ effect discussed above, this action
 involves a time-split bilinear  function of the multipole moments that is closely linked to the gravitational-wave flux,
 namely
 \beq
{\mathcal G}^{\rm split}(t,t')=\frac{G}{5c^5} I_{ij}^{(3)}(t) I_{ij}^{(4)}(t') + \ldots\,.
\eeq
 At the present 5.5PN accuracy, it is enough to use the leading-order version of the time-split function ${\mathcal G}^{\rm split}(t,t')$, 
 obtained by keeping only  the lowest-order quadrupolar contribution (neglecting higher multipole terms)
with $I_{ij} \approx \mu r_{\langle ij\rangle}$ evaluated at the Newtonian level. In addition, we can also neglect
the difference between ${\cal M}$ and $M$.

An important conceptual point is that though Eqs. \eqref{S5.5pn}, \eqref{Htail2}, seem to define only the nonlocal part of the 5.5PN action,
actually they give the {\it complete} 5.5PN action. Indeed,  the usual PN-expanded way of computing 
the local part of the action ({e.g.} by integrating the near-zone Hamiltonian density, as in \cite{Jaranowski:2015lha}) cannot generate 
any half-integral PN contribution. In addition, the nonlocal action,  Eqs. \eqref{S5.5pn}, \eqref{Htail2}, has (contrary to the
4+5 PN one) no ultraviolet divergence at small $\tau=t'-t$. This indicates the completeness of the 5.5PN action written above.
Actually, the correctness of this action has been directly checked by satisfactorily comparing its predictions with SF computations 
(that automatically include all local and nonlocal effects), see \cite{Damour:2015isa}.

As before, we can use the Delaunay averaging technique to relate the 5.5PN Hamiltonian \eqref{Htail2} to
its EOB counterpart. The time-average of $H_{\rm 5.5 PN}$ was already considered in  Ref. \cite{Damour:2015isa} and
shown there to be expressible as
\beq
\langle H_{\rm 5.5 PN}\rangle =- \pi \frac{2B}{5}\frac{G}{c^5} \left(\frac{G {\cal M}}{c^3} \right)^2 n_{\rm phys}^7 S_7^{\rm quad}\,,
\eeq
where $n_{\rm phys}=2\pi/P_{\rm phys}$ is the (physical) orbital frequency, and where
\beq
S_7^{\rm quad}= \sum_{p=1}^{\infty} p^7 |I_{ij}(p)|^2\,,
\eeq
with $I_{ij}(p)$ denoting the Fourier coefficients of the quadrupole moment $I_{ij}(t)$.

Extending the results of Ref. \cite{Damour:2015isa}, we have computed (starting directly from the integral expression  \eqref{Htail2}) 
the orbital average of $H_{\rm 5.5 PN}$ to the 16th order in eccentricity, with the result:
\beq
\langle H_{\rm 5.5 PN}\rangle = -\frac{\mu^2}{M} c^2  \frac{6848}{525} \frac{\pi}{a_r^{13/2}} \varphi(e)\,,
\eeq
where $a_r$ is dimensionless and
\begin{eqnarray}
\varphi(e)&=&  1+\frac{2335}{192}e^2+\frac{42955}{768}e^4+\frac{6204647}{36864}e^6\nonumber\\
&& +\frac{352891481}{884736}e^8+\frac{286907786543}{353894400}e^{10}\nonumber\\
&&+\frac{6287456255443}{4246732800}e^{12}+
\frac{5545903772613817}{2219625676800}e^{14}\nonumber\\
&&+
\frac{422825073954708079}{106542032486400}e^{16}+
O(e^{18})\,.
\end{eqnarray}
The last two terms will not be used below. [The first two terms in $\varphi(e)$ were previously computed in Refs. \cite{Arun:2007rg,Arun:2009mc}.]
Note that the rescaled function $\tilde \varphi(e)=\varphi(e)(1-e^2)^{13/2}$, once re-expanded in $e$, becomes 
\begin{eqnarray}
\tilde \varphi(e)&=& 1+\frac{1087}{192} e^2-\frac{4027}{768} e^4-\frac{172009}{36864} e^6\nonumber\\
&&+\frac{1758725}{884736} e^8+\frac{211269943}{353894400} e^{10}+\frac{976098889}{4246732800} e^{12}\nonumber\\
&& +\frac{796425035243}{6658877030400} e^{14}+
\frac{2583007392829}{35514010828800}e^{16}\nonumber\\
&&+
O(e^{18})\,,
\end{eqnarray}
with coefficients which remain of order 1.

Let us now transcribe the 5.5PN-level tail$^2$ time-averaged nonlocal original $H_{\rm 5.5 PN}$ into its corresponding EOB version, parametrized 
(in standard $p_r$-gauge) by an  effective EOB Hamiltonian expanded as a series in powers of $p_r^2$:
\begin{eqnarray}
\delta {\hat H^2}_{{ \rm eff}\, {\rm 5.5 PN}}&=& A_{65}u^{13/2}+\bar D_{55} u^{11/2} p_r^2+q_{4,4.5}p_r^4 u^{9/2}\nonumber\\
&+&q_{6,3.5}p_r^6 u^{7/2}+q_{8,2.5}p_r^8 u^{5/2} + O(p_r^{10})\,.
\end{eqnarray}
We can compute the orbital average of $\delta \hat H_{\rm eff}^2$ (henceforth omitting the additional 5.5 PN subscript),
by writing
\beq
\langle \delta \hat H_{\rm eff}^2 \rangle =\frac{n}{2\pi} \int  \frac{\delta \hat H_{\rm eff}^2}{\dot \phi}  d\phi\,,
\eeq
where, at this leading order, we can use the Newtonian relations for $r=r(\phi)$ and $p_r=\dot r$
\beq
r = \frac{a_r(1-e^2)}{1+e\cos(\phi)}\,,\quad
p_r=\frac{e}{\sqrt{a_r (1-e^2)}}\sin\phi\,,
\eeq
with the (rescaled) orbital frequency of the radial motion given by $GM n_{\rm phys}= n=a_r^{-3/2}$.
The result reads
\begin{eqnarray}
\langle \delta \hat H_{\rm eff}^2 \rangle &=&\frac{1}{a_r^{13/2}} 
\left[A_{6.5} 
+\left(\frac{143}{16} A_{6.5} +\frac12 \bar D_{5.5} \right) e^2\right. \nonumber\\
&+&
\left(\frac{36465}{1024} A_{6.5} +\frac{3}{8}  q_{4,4.5}+\frac{195}{64}\bar D_{5.5} \right) e^4\nonumber\\
&+&\left(\frac{20995}{2048}\bar D_{5.5} +\frac{1616615}{16384} A_{6.5}\right. \nonumber\\
&& \left. +\frac{5}{16}  q_{6,3.5}+\frac{255}{128}  q_{4,4.5}\right) e^6\nonumber\\
&+&\left(\frac{929553625}{4194304} A_{6.5} +\frac{3380195}{131072}\bar D_{5.5}\right. \nonumber\\ 
&+&\left.\left.\frac{101745}{16384} q_{4,4.5}+\frac{1615}{1024}  q_{6,3.5}+\frac{35}{128}  q_{8,2.5}\right) e^8\right]\nonumber\\
&+&
O(e^{10})\,.
\end{eqnarray}
Comparison (at the Newtonian level) among these two gauge-invariant quantities
\beq
\langle \delta \hat H_{\rm eff}^2 \rangle =\frac{2}{\mu c^2} \langle H_{\rm 5.5 PN}\rangle\,,
\eeq
allows us to determine all tail$^2$ coefficients
\begin{eqnarray}
A_{6.5} &=& \nu\frac{13696}{525}\pi\nonumber\\ 
\bar D_{5.5} &=& \nu\frac{264932}{1575}\pi\nonumber\\ 
q_{4,4.5} &=& \nu\frac{88703}{1890}\pi\nonumber\\  
q_{6,3.5}&=& -\nu\frac{2723471}{756000}\pi\nonumber\\ 
q_{8,2.5} &=& \nu\frac{5994461}{12700800}\pi\,. 
\end{eqnarray}
The coefficients $A_{6.5}$, $\bar D_{5.5}$, $q_{4,4.5}$ and $q_{6,3.5}$ agree with previous results (both from
Ref. \cite{Damour:2015isa} and from self-force computations). The last coefficient, $q_{8,2.5}$,  is instead new and 
constitutes a prediction for future self-force computations of the averaged redshift invariant at order $O(e^8)$.
Note that the entire 5.5 PN action is linear in $\nu$ (and proportional to $\nu$). Therefore, self-force computations
at the 5.5 PN level allow one to compute  exact, $\nu$-dependent 5.5 PN observables.

In the present section, we have considered 5.5PN-level gauge-invariant quantities linked to ellipticlike  motions.
We shall leave to future work the 5.5PN contribution to the scattering angle implied by the action \eqref{S5.5pn}.

\section{Action variables and Delaunay Hamiltonian for the ($f$-route) local effective 5PN dynamics} \label{delaunay}

We have derived above the 5PN-accurate local Hamiltonian (in its f-version), notably by making use of the special
$\nu$-dependent structure of the scattering angle \cite{Damour:2019lcq}. The so-obtained local 5PN dynamics
has been so far expressed within the EOB formalism, using two special gauges ($p_r$-gauge and energy-gauge).
As these gauges are uniquely fixed by their definitions, all our results above can be considered as being
gauge-invariant. Our discussion above of the gauge-invariant  scattering angle has, in particular, confirmed the fully gauge-fixed
nature of the  $p_r$ gauge. The same holds for the energy gauge (as shown in Refs. \cite{Damour:2017zjx,Damour:2019lcq}).
It is, however, interesting to complete our study of the 5PN local dynamics by discussing another gauge-invariant description
of the dynamics, applicable to bound-state motions (rather than scattering motions),
whose usefulness for relativistic gravity was first emphasized in Ref. \cite{Damour:1988mr}, namely
the Delaunay Hamiltonian,
\beq \label{Hdelaunay}
H^{\rm loc, f}= H(I_r, I_\phi)\,,
\eeq
{\it i.e.}, the Hamiltonian expressed in terms of the action variables
\begin{eqnarray} \label{Irj}
I_r &=& \frac1{2\pi} \oint p_r dr ,\nonumber\\
 I_\phi&=&  \frac1{2\pi} \oint p_\phi d\phi= p_\phi= j\,.
\end{eqnarray}
Note that we work here with dimensionless scaled variables $I_r = I_r ^{\rm phys}/(GM\mu)$, $I_\phi=j=J/GM\mu)$.

Equivalently (modulo solving Eq. \eqref{Hdelaunay} with respect to $I_r$), one can consider the
gauge-invariant functional link between the radial action $I_r$ and the energy and the angular momentum, say
\beq \label{Ir}
I_r= I_r(\gamma, j)\,.
\eeq
As indicated here, we are going to see that great simplifications are reached if we use as energy variable
the (scaled) effective EOB energy 
\beq
\gamma\equiv \e = \widehat H_{\rm eff},
\eeq
which is related to the total local c.m. energy by the usual EOB energy map
\beq
H^{\rm loc, f}=  M \sqrt{1+ 2 \nu (\gamma -1)}\,.
\eeq
We use the same notation $\gamma$ as in our previous discussion of scattering states, but one must
note that we are now going to consider bound states for which $\gamma <1$. This implies that the
above defined squared asymptotic EOB momentum $p_{\infty}^2$, is now a negative quantity:
\beq
 \gamma^2-1 \equiv p_{\infty}^2  \equiv - |p|^2\,.
\eeq
Before studying the precise structure of the gauge-invariant function $I_r= I_r(\gamma, j)$, let us recall 
how this function acts as a potential for deriving both the periastron advance ($\Phi$) and the radial  period ($P$):
\bea
\frac{\Phi}{2\pi}&=&K=-\frac{\partial I_r(\gamma ,j)}{\partial j}\,, \nonumber\\
\frac{P}{2\pi GM}&=& + h(\gamma, \nu) \frac{\partial I_r(\gamma,j)}{\partial \gamma}\,. 
\eea
The factor $h(\gamma;\nu) \equiv [1+2\nu(\gamma-1)]^{1/2}$ in the last equation comes from the ``redshift''
factor $dH/dH_{\rm eff}$ connecting the real-time period to the effective-time period \cite{Buonanno:1998gg}.

We have computed the function $I_r(\gamma,j)$ associated with the $f$-route local 5PN-accurate Hamiltonian
by using the technique explained in Ref. \cite{Damour:1988mr} (and used there at the 2PN level).
We start from the  local  effective EOB ($p_r$-gauge) Hamiltonian at 5PN
\begin{eqnarray}
\widehat H_{\rm eff}^2&=&A \left[1+ u^2 j^2+p_r^2 A  \bar D + q_4  p_r^4\right. \nonumber\\
&& \left. + q_6  p_r^6+ q_8  p_r^8\right]
\end{eqnarray}
where
\begin{eqnarray}
A(u,\nu) &=& 1-2 u+2\nu u^3+a_4(\nu) u^4+ a_5(\nu) u^5+a_6(\nu) u^6\nonumber\\
\bar D(u,\nu) &=& 1+6\nu u^2+\bar d_3(\nu) u^3+\bar d_4(\nu) u^4+\bar d_5(\nu)  u^5\nonumber\\
q_4(u,\nu)&=& q_{42}(\nu) u^2+q_{43}(\nu)u^3+q_{44}(\nu)u^4\nonumber\\
q_6(u,\nu)&=& q_{62}(\nu) u^2+ q_{63}(\nu) u^3\nonumber\\
q_8(u,\nu)&=& q_{82}(\nu)u^2\,.
\end{eqnarray}
We then use the energy conservation law
\beq
\gamma^2= \widehat H_{\rm eff}^2(p_r^2, j^2,u)
\eeq
to iteratively solve for the radial momentum $p_r$ as a function of $\gamma$, $j$ and $u=1/r$.
This is done in a PN-expanded way, after restoring a place holder $\eta=1/c$ for PN orders,
with the following PN orders:
\beq \label{etascaling}
p_r \mapsto \eta p_r\, ,\, j \mapsto \frac{j}{\eta}\, ,\, u\mapsto \eta^2 u\, , p_{\infty}^2 \mapsto \eta^2 p_{\infty}^2\,.
\eeq
Under this scaling the quantity
\beq
e^2 \equiv 1+ p_{\infty}^2 j^2\,,
\eeq
is fixed as $\eta \to 0$, and describes the  eccentricity of the limiting Newtonianlike dynamics. 
[We are considering the case where $0< - p_{\infty}^2j^2 <1$, so that $0<e^2<1$.]
Indeed, the PN-expanded
value of the radial momentum has the structure
\beq
p_r(u; p_{\infty}^2 ,j)=\sum_{k=0}^5 p_r^{(2k)}(u; p_{\infty}^2 ,j)\eta^{2k}+O(\eta^{12})\
\eeq
with leading-order contribution
\beq
p_r^{(0)}(u; p_{\infty}^2 ,j)= (p_{\infty}^2+2 u-u^2 j^2)^{1/2}\,,
\eeq
and 1PN correction given by
\beq
p_r^{(2)}(u; p_{\infty}^2 ,j)= \frac{u^3 j^2}{p_r^{(0)}}+2up_r^{(0)}\,.
\eeq
The roots of the second-order polynomial $p_{\infty}^2+2 u-u^2 j^2$,
\beq
u_{\pm }=\frac{ 1\pm e }{j^2}\,,
\eeq
are  the Newtonianlike values associated with the periastron and apoastron passages.

Following \cite{Damour:1988mr}, one can compute the PN-expansion of the radial integral
\beq
I_r =\frac1{2\pi} \oint dr \left(\sum_{k=0}^5 p_r^{(2k)}(u; p_{\infty}^2 ,j)\eta^{2k}\right)
\eeq
by taking the Hadamard partie finie of the resulting integrals. This leads to an explicit PN-expanded expression
for the radial integral:
\beq
I_r(p_{\infty}^2 ,j;\nu)=\sum_{k=0}^{5} \eta^{2k} I_r^{(2k)}(p_{\infty}^2,j;\nu)+O(\eta^{12})\,,
\eeq
starting with the Newtonianlike value ($k=0$):
\beq
 I_r^{(0)}(p_{\infty}^2,j;\nu)= -j+\frac{1}{\sqrt{- p_{\infty}^2}}=  -j+\frac{1}{\sqrt{1-\gamma^2}}\,.
\eeq
We recall that we are here considering ellipticlike motions with $\gamma^2<1$.

The  function $I_r(\gamma ,j;\nu)$  exhibits a remarkably simple structure, which is the reflection of the 
simple $\nu$-dependence of the PM-expanded scattering angle \cite{Damour:2019lcq}.[The latter structure separately applies to the presently considered f-route local dynamics.] We can write the 5PN-accurate local 
$I_r(\gamma ,j;\nu)$ in the form
\begin{eqnarray} \label{jxp}
&&I_r^{\rm 5PN, loc}(\gamma,j,\nu)= -j+I_0(\gamma) \nonumber\\
&& + \frac{I_1^S(\gamma)}{hj}  \nonumber\\
&&+ \frac{ I_3^S(\gamma)+ \nu I_3^{\nu^1}(\gamma) }{h^3j^3}\nonumber\\
&& +  
\frac{ I_5^S(\gamma)+ \nu I_5^{\rm \nu^1}(\gamma)+\nu^2 I_5^{\nu^2}(\gamma)}{h^5j^5} \nonumber\\
&&+\frac{ I_7^S(\gamma)+ \nu I_7^{\rm \nu^1}(\gamma)+\nu^2 I_7^{\nu^2}(\gamma)+ \nu^3 I_7^{\nu^3}(\gamma)}{h^7j^7}\nonumber\\
&&+\frac{ I_9^S(\gamma)+ \nu I_9^{\rm \nu^1}(\gamma)+\nu^2 I_9^{\nu^2}(\gamma)+ \nu^3 I_9^{\nu^3}(\gamma) + \nu^4 I_9^{\nu^4}(\gamma)}{h^9j^9}.\nonumber\\
\end{eqnarray}
Here, each line does not correspond to a well-defined  PN order, though the successive lines start at some minimum PN order which increases
linearly with the power of $j$ present in the denominators. 
On the first line the term $-j$ can be considered to be of Newtonian order, while the second term is a function of $\gamma$
(given below) which, when it is expanded in powers of $\gamma-1=O(\eta^2)$, starts at the Newtonian order but then contains higher
PN corrections of arbitrarily high PN orders. Similarly, the next line (proportional to $1/j$) starts at 1PN order, but includes higher
PN orders when expanded in powers of $\gamma-1=O(\eta^2)$. Each 
 extra power of $1/j^2$ represents an extra PN order. The last term, $\propto 1/j^9$ is $1/j^{10}$ smaller than the first,
Newtonian term $-j$, which corresponds (in view of the scaling  $ j \mapsto \frac{j}{\eta}$) to a relative factor $\eta^{10}$,
corresponding indeed to a 5PN accuracy. 

There are several remarkable features in the structure \eqref{jxp}. First,
the only $j$-independent term in this PN expansion (on the first line) starts at the Newtonian order and can be proven to be exactly given by
the simple formula
\beq
I_0(\gamma) = \frac{ 2\gamma^2-1 }{\sqrt{1-\gamma^2}}.
\eeq
Note that this is the analytic continuation (from $\gamma >1$ to $\gamma<1$) of the 1PM scattering coefficient $\chi_1$.
Second, all the powers of $j$ in the denominators are accompanied by the same power of $h(\gamma;\nu) \equiv [1+2\nu(\gamma-1)]^{1/2}$.
This factoring absorbs most of the complicated $\nu$-dependence of the PN-expanded $I_r$ to leave only the simple
polynomial $\nu$-dependence exhibited by the numerators. Indeed, these exhibit the simple rule that the numerator $I_{2n+1}$ corresponding
to the denominator $(h j)^{2n+1}$ is a polynomial in $\nu$ of order $n$. [The latter rule follows from the rule about $h^{n-1}\chi_n$ via
the analytic continuation in $\gamma$ allowing one to identify $\Phi(\gamma,j)$ with a suitably defined analytic
continuation of $\chi(\gamma,j)+ \chi(\gamma,-j)$ \cite{Kalin:2019rwq}.]
The last remarkably simple feature of the expansion \eqref{jxp}
is that the $\nu \to 0$ limits of each numerator, {\it i.e.}, the coefficients $ I_{2n+1}^S(\gamma)$ are very simple polynomial functions of $\gamma$,
which are given by the following expressions
\bea
I_1^S(\gamma) &=& -\frac{3}{4}+\frac{15}{4}\gamma^2, \nonumber\\
I_3^S(\gamma) &=& \frac{35}{64}-\frac{315}{32}\gamma^2+\frac{1155}{64}\gamma^4, \nonumber\\
I_5^S(\gamma) &=&  -\frac{231}{256}+\frac{9009}{256}\gamma^2-\frac{45045}{256}\gamma^4+\frac{51051}{256}\gamma^6 \nonumber\\
I_7^S(\gamma) &=&  \frac{32175}{16384}-\frac{546975}{4096}\gamma^2+\frac{10392525}{8192}\gamma^4\nonumber\\
&&-\frac{14549535}{4096}\gamma^6+\frac{47805615}{16384}\gamma^8\nonumber\\
I_9^S(\gamma) &=& -\frac{323323}{65536}+\frac{33948915}{65536}\gamma^2-\frac{260275015}{32768}\gamma^4\nonumber\\
&&+\frac{1301375075}{32768}\gamma^6-\frac{5019589575}{65536}\gamma^8\nonumber\\
&&+\frac{3234846615}{65536}\gamma^{10}\,.
\eea
Note the $\gamma \to 1$ values of the latter test-mass polynomials
\begin{eqnarray}
&&I_1^S(1)=3\,,\qquad I_3^S(1)=\frac{35}{4}\,,\qquad 
I_5^S(1)=\frac{231}{4}\,,\nonumber\\
&&I_7^S(1)=\frac{32175}{64}\,,\qquad
I_9^S(1)=\frac{323323}{64}\,.
\end{eqnarray}
The simple ``Schwarzschild'' polynomials $I_{2n+1}(\gamma)$
can be exactly computed by considering the  $\nu \to 0$ limit of the radial action (Schwarzschild limit).
Indeed, the test-particle limit of the radial action, 
\beq
I_{r}^{\rm Sch}(\e,j) = \frac1{2\pi} \oint dr p_r^{\rm Sch}(\e,j)\,,
\eeq
is easily written down by solving $\e^2=H_S^2=(1-2 u) [1+(1-2 u) p_r^2+j^2 u^2]$, and reads
(remembering $\e=\gamma$)
\beq
\label{IrS}
I_{r}^{\rm Sch}(\gamma,j)= \frac1{2\pi} \oint du \frac{\sqrt{\gamma^2-(1-2u)(1+j^2u^2)}}{u^2(1-2u)}\,,
\eeq
where the integral is taken around the two roots of the cubic polynomial $P_3(u)=\gamma^2-(1-2u)(1+j^2u^2)$
that are close to the Newtonian roots $u_\pm$ used in our PN-expanded computation above.
We see that $I_{r}^{\rm Sch}(\gamma,j)$ is a {\it complete} elliptic integral (i.e. a period of an elliptic curve),
so that it can be written down explicitly, e.g., in terms of a combination of usual Legendre complete elliptic integrals
(however, the third type of Legendre elliptic integral appears). The latter exact, elliptic-integral representation
is rather complex, but it is relatively easy to compute both its PN expansion ({\it i.e.}, its expansion in powers of $\eta$, 
see Eq. \eqref{etascaling}), and its expansion in inverse powers of $j$. See Appendix \ref{C}, which also
includes a discussion of 
the simpler complete elliptic integral giving the test-mass periastron advance.

Finally, the primitive information (beyond the test-mass limit) contained in the 5PN radial action $I_r(\gamma ,j;\nu)$
is fully described  by the small number of $\gamma$-dependent coefficients of the various powers of $\nu$ in the numerators of Eq. \eqref{jxp}.
These coefficients (contrary to their corresponding  $\nu \to 0$ limits $I_{2n+1}^S(\gamma)$)
are not known as exact functions of $\gamma$ but only as limited expansions in powers of $\gamma-1=O(\eta^2)$.
For instance, $I_3^{\nu^1}(\gamma)$ is known to fractional 3PN accuracy, {\it i.e.} up to the third order in $\gamma-1$.
The PN knowledge of the higher terms $I_{2n+1}^{\nu^p}(\gamma)$ linearly decreases as $n$ increases, until the last
terms $I_{9}^{\nu^p}(\gamma)$ which are only known at the lowest (Newtonian) accuracy, {\it i.e.} only for $\gamma=1$.
The known information carried by all these $I_{2n+1}^{\nu^p}(\gamma)$ is gathered in table \ref{table_Ir_large_j}.
\begin{table*}[h]
\caption{\label{table_Ir_large_j} Coefficients entering the $j$-expansion of $I_r(\gamma,j;\nu)$}
\begin{ruledtabular}
\begin{tabular}{ll}
Coefficient & Value \\
\hline
$I_3^{\nu^1}(\gamma)$ & $  -\frac52 +\left(-\frac{557}{12}+\frac{41}{64}\pi^2\right)(\gamma-1) 
+\left(-\frac{10873}{72}+\frac{35569}{6144}\pi^2\right)(\gamma-1)^2  
+ \left(-\frac{7199407}{25200}+\frac{15829}{768}\pi^2\right)(\gamma-1)^3$ \\ 
$I_5^{\nu^1}(\gamma)$ & $\left(-\frac{125}{2}+\frac{123}{128}\pi^2\right) 
+\left(-\frac{224113}{240}+\frac{51439}{2048}\pi^2\right)(\gamma-1)
+ \left(-\frac{89527351}{16800}+\frac{979913}{4096}\pi^2\right)(\gamma-1)^2 $\\
$I_5^{\nu^2}(\gamma)$ &  $\frac{21}{8}+\left(\frac{2013}{16}-\frac{369}{128}\pi^2\right)(\gamma-1) 
+\left( \frac{9739}{96}-\frac34 \bar d_5^{\nu^2}-\frac{18275}{4096}\pi^2-\frac{3}{4}a_6^{\nu^2}\right)(\gamma-1)^2$\\
$I_7^{\nu^1}(\gamma)$ & $ \left(-\frac{248057}{288}+\frac{425105}{24576}\pi^2\right) + 
\left(-\frac{99111883}{6720}+\frac{2310485}{8192}\pi^2 \right)(\gamma-1) $\\ 
$I_7^{\nu^2}(\gamma)$ & $ \left(-\frac{1025}{256}\pi^2+\frac{18925}{96}\right) +\left(-\frac{5}{4}\bar d_5^{\nu^2} -\frac{15}{4}a_6^{\nu^2}-\frac{1290275}{12288}\pi^2+\frac{1089349}{288}\right)(\gamma-1)  $\\
$I_7^{\nu^3}(\gamma)$ & $-\frac{45}{16}+\left(\frac{3075}{512}\pi^2-\frac{7595}{32}\right)(\gamma-1) $\\
$I_9^{\nu^1}(\gamma)$ & $  -\frac{6817563}{640}+\frac{121807}{1024}\pi^2$\\
$I_9^{\nu^2}(\gamma)$ &  $-\frac{7}{16}\bar d_5^{\nu^2}+\frac{572999}{128}-\frac{1755159}{16384}\pi^2-\frac{35}{16} a_6^{\nu^2} $\\
$I_9^{\nu^3}(\gamma)$ &  $ -\frac{42665}{96}+\frac{10045}{1024}\pi^2$\\  
$I_9^{\nu^4}(\gamma)$ & $\frac{385}{128}$\\  
\end{tabular}
\end{ruledtabular}
\end{table*}

By inverting the functional relation $ I_r= I_r(\e,j)$, one can finally obtain the explicit value
of the  corresponding (effective) Delaunay Hamiltonian, $\widehat H_{\rm eff}(I_r,j)$.
 This is conveniently done by defining the variables
\beq
I_3 \equiv I_r+j \,,\qquad j=I_2\,,
\eeq
in terms of which one can get the PN expansion of  $H_{\rm eff}(I_2,I_3)/\mu = \gamma c^2$ in the form
\beq \label{Hdelaunay}
\frac{ H_{\rm eff}^{\rm 5PN, loc,f}(I_2,I_3;\nu)}{\mu}=\eta^{-2}+\sum_{k=0}^5\eta^{2k} \bar E_{\rm eff}^{2k}(I_2,I_3;\nu)+O(\eta^{12})\,.
\eeq
\begin{table*}[h]
\caption{\label{table_barE} Coefficients entering the PN expansion of the Delaunay Hamiltonian $\hat H_{\rm eff}(I_2,I_3;\nu)-1$}
\begin{ruledtabular}
\begin{tabular}{ll}
Coefficient & Value \\
\hline
$\bar E_{\rm eff}^{0}$&$ -\frac{1}{2 I_3^2}$\\ 
$\bar E_{\rm eff}^{2}$&$\frac{15}{8I_3^4}-\frac{3}{I_2I_3^3}$\\
$\bar E_{\rm eff}^{4}$&$ (\frac52 \nu-\frac{35}{4})\frac{1}{ I_2^3 I_3^3}-\frac{27}{2}\frac{1}{I_2^2 I_3^4}+(\frac{105}{4}
-\frac32 \nu)\frac{1}{I_2I_3^5}-\frac{145}{16}\frac{1}{I_3^6}$\\
$\bar E_{\rm eff}^{6}$&$ [-\frac{231}{4}-\frac{21}{8}\nu^2+(\frac{125}{2}-\frac{123}{128}\pi^2)\nu]\frac{1}{I_2^5I_3^3}
+(-\frac{315}{4}+\frac{45}{2}\nu)\frac{1}{I_2^4I_3^4}
+[\frac{303}{8}+\frac{15}{4}\nu^2+(\frac{41}{128}\pi^2-\frac{661}{12})\nu]\frac{1}{I_2^3I_3^5}$\\
&$
+(-\frac{45}{2}\nu+225)\frac{1}{I_2^2I_3^6}
+(-\frac{9}{8}\nu^2-\frac{825}{4}+\frac{75}{4}\nu)\frac{1}{I_2I_3^7}+\frac{6363}{128}\frac{1}{I_3^8}$\\
$\bar E_{\rm eff}^{8}$&$ [-\frac{32175}{64}+\frac{45}{16}\nu^3+(-\frac{18925}{96}+\frac{1025}{256}\pi^2)\nu^2+(\frac{248057}{288}-\frac{425105}{24576}\pi^2)\nu]\frac{1}{I_2^7I_3^3}$\\
&$+[-\frac{20307}{32}-33\nu^2+(-\frac{1107}{128}\pi^2+\frac{5025}{8})\nu]\frac{1}{I_2^6I_3^4}
+[\frac{7749}{32}-\frac{105}{16}\nu^3+(\frac{7643}{32}-\frac{123}{32}\pi^2)\nu^2+(-\frac{453613}{480}
+\frac{80959}{4096}\pi^2)\nu]\frac{1}{I_2^5I_3^5}$\\
&$+[\frac{21435}{16}+75\nu^2+(-\frac{3755}{4}+\frac{615}{128}\pi^2)\nu]\frac{1}{I_2^4I_3^6}
+[\frac{46275}{64}+\frac{75}{16}\nu^3+(-\frac{2989}{32}+\frac{123}{256}\pi^2)\nu^2+(-\frac{124129}{24576}\pi^2+\frac{120763}{288})\nu]\frac{1}{I_2^3 I_3^7}$\\
&$+(-\frac{63}{2}\nu^2+\frac{3465}{8}\nu-\frac{85365}{32})\frac{1}{I_2^2I_3^8}
+(\frac{585}{32}\nu^2-\frac{5745}{32}\nu+\frac{50703}{32}-\frac{15}{16}\nu^3)\frac{1}{I_2I_3^9}-\frac{75303}{256}\frac{1}{I_3^{10}}$\\
$\bar E_{\rm eff}^{10}$&$ [-\frac{323323}{64}-\frac{385}{128}\nu^4+(-\frac{10045}{1024}\pi^2+\frac{42665}{96})\nu^3+(\frac{7}{16}\bar d_5^{\nu^2}-\frac{572999}{128}+\frac{35}{16}a_6^{\nu^2}
+\frac{1755159}{16384}\pi^2)\nu^2+(-\frac{121807}{1024}\pi^2$\\
&$+\frac{6817563}{640})\nu]\frac{1}{I_2^9I_3^3}$\\
&$+[-\frac{386595}{64}+45\nu^3+(-\frac{18495}{8}+\frac{5535}{128}\pi^2)\nu^2+(\frac{314417}{32}-\frac{1481955}{8192}\pi^2)\nu]\frac{1}{I_2^8 I_3^4}$\\
&$+[\frac{344637}{128}+\frac{315}{32}\nu^4+(-\frac{79655}{96}+\frac{17425}{1024}\pi^2)\nu^3+(-\frac{2344095}{16384}\pi^2-\frac{5}{8}\bar d_5^{\nu^2}+\frac{392325}{64}-\frac{15}{8}a_6^{\nu^2})\nu^2$\\
&$+(\frac{1792917}{8192}\pi^2-\frac{76818229}{6720})\nu]\frac{1}{I_2^7 I_3^5}$\\
&$+[\frac{769545}{64}-165\nu^3+(\frac{116065}{24}-\frac{8815}{128}\pi^2)\nu^2+(-\frac{1907369}{96}+\frac{1345585}{4096}\pi^2)\nu]\frac{1}{I_2^6I_3^6}$\\
&$+[\frac{125235}{16}-\frac{735}{64}\nu^4+(\frac{16145}{32}-\frac{7995}{1024}\pi^2)\nu^3+(\frac{1241145}{16384}\pi^2+\frac{3}{16}\bar d_5^{\nu^2}+\frac{3}{16}a_6^{\nu^2}-\frac{205425}{64})\nu^2$\\
&$+(-\frac{860567}{4096}\pi^2+\frac{47882269}{16800})\nu]\frac{1}{I_2^5 I_3^7}$\\
&$+[-\frac{715575}{64}+\frac{315}{2}\nu^3+(-\frac{5677}{2}+\frac{861}{64}\pi^2)\nu^2+(\frac{1418221}{96}-\frac{937783}{8192}\pi^2)\nu]\frac{1}{I_2^4 I_3^8}$\\
&$+[-\frac{2464245}{128}+\frac{175}{32}\nu^4+(\frac{615}{1024}\pi^2-\frac{4255}{32})\nu^3+(\frac{149563}{192}-\frac{153649}{16384}\pi^2)\nu^2+(\frac{1388971}{24576}\pi^2-\frac{21149141}{100800})\nu]\frac{1}{I_2^3I_3^9}$\\
&$+(-\frac{181845}{32}\nu-\frac{81}{2}\nu^3+\frac{1769931}{64}+\frac{2835}{4}\nu^2)\frac{1}{I_2^2 I_3^{10}}$\\
&$+(-\frac{105}{128}\nu^4-\frac{26595}{128}\nu^2+\frac{200445}{128}\nu+\frac{75}{4}\nu^3-\frac{1550595}{128})\frac{1}{I_2 I_3^{11}}+\frac{1874587}{1024}\frac{1}{I_3^{12}}$\\
\end{tabular}
\end{ruledtabular}
\end{table*}
The values of the coefficients $\bar E_{\rm eff}^{2k}(I_2,I_3;\nu)$ are displayed in Table \ref{table_barE}. Note, however,
that the structure of this (effective) gauge-invariant Delaunay Hamiltonian is not particularly illuminating.
The simple $\nu$-structure exhibited by the radial action function \eqref{jxp} is lost in the Delaunay Hamiltonian \eqref{Hdelaunay}.
Indeed, the hidden simplicity of the 5PN local dynamics is  more transparent when encoding it either in the EOB potentials displayed above,
or in the radial action \eqref{jxp}.
Let us emphasize again that, given a specific gauge choice (say, $p_r$ gauge, or energy gauge), the corresponding EOB
potentials are completely gauge-fixed, and can therefore be considered as being as gauge-invariantly defined
as the more traditional gauge-invariant functions  $ I_r= I_r(E,j;\nu)$ or $H(I_r,j;\nu)$.

\section{Conclusions}

We have shown how to successfully combine several different theoretical tools to develop a new methodology \cite{Bini:2019nra} for extending the analytical computation of the conservative two-body dynamics beyond the current post-Newtonian knowledge (4PN). 
Our approach has allowed us to derive an almost complete expression for the 5PN-level
 action, given by the sum of a 4PN+5PN nonlocal action, Eq. \eqref{Snonloc}, and of a local one $\int pdq- H_{\rm loc, f}^{\leq {\rm 5PN}} dt$. 
 We succeeded in determining the full functional structure of $H_{\rm loc, f}^{\leq {\rm 5PN}}$ (which contains 97  numerical coefficients), except for   two ($\nu^3$-level) unknown coefficients ($\nu^2$-level in the EOB potentials $A$ and $\bar D$). The two main derivations underlying our new results are:
 (i) the computation of the Delaunay average of the nonlocal action around eccentric orbits to the tenth order in eccentricity included;
 (ii) the self-force computation of the redshift along eccentric orbits (around a Schwarzschild black hole) to sixth order in eccentricity.
 
 We completed our results beyond the 5PN level in two different directions. On the one hand, we added the 5.5PN contribution to the action (which is purely
 nonlocal) and transcribed it into its EOB ($p_r$-gauge) form up to the eight order in $p_r$. On the other hand, we used a recent extension of our
 self-force computation to the eigth order in eccentricity to improve the determination of the third post-Minkowskian ($O(G^3)$) 
 part of the dynamics to the 6PN-level. This allowed us to compute the $O(G^3)$ contribution to the scattering angle up to the 6PN-level
 included. Our 6PN-accurate  $O(G^3)$ scattering angle agrees with the recent third post-Minkowskian ($O(G^3)$)
result of Bern et al \cite{Bern:2019nnu,Bern:2019crd}.

We computed both the nonlocal, and the local,  contributions to the 5PN-accurate, $O(G^4)$ scattering angle.
As our 5PN (and 5.5PN) results are complete at the $O(G^4)$ order, the latter result offers
checks for future  fourth post-Minkowskian calculations.
We could conveniently separate the study of the nonlocal
 versus local contributions to the scattering angle by flexing (at the 5PN level) the scale $2 r_{12}^f/c$ entering the definition of the nonlocal action.

We point out a remarkable hidden simplicity of the local 5PN dynamics. This hidden simplicity only manifests itself when
using a gauge-invariant description of the dynamics. There are several (complementary) ways of viewing the (local) 5PN
dynamics in a gauge-invariant fashion. One can use the EOB description, in one of its gauge-fixed versions ($p_r$-gauge or energy-gauge).
When comparing the EOB encoding of 5PN-level information (and $\nu$-structure) to  the (simplified) $h^{n-1}\chi_n$ scattering
encoding, one can see not only that they are one-to-one, but that the EOB encoding is as minimal as the $h^{n-1}\chi_n$ one.
[See, section \ref{scatteringangle}.] An alternative gauge-invariant approach is to focus on gauge-invariant observables. Two of them
have a particularly interesting structure: the scattering function $\chi(\e,j)$, and the radial action $I_r(\e,j)$. We have emphasized that
the ($f$-flexed local) radial action (when expressed in terms of the EOB effective energy $\e$ and of the product $h j$, where $h=E^{\rm tot}/M$) has a 
remarkably simple, see Eq. \eqref{jxp}, which parallels the simple structure of $\chi(\e,j)$. This simplicity is, essentially, already
automatically incorporated in the structure of the EOB Hamiltonian  (see Table \ref{table_eob_loc} and Eq. \eqref{en_gauge_pot}). Let us also 
note that the local 5PN dynamics is completely logarithm-free, and that all its numerical coefficients are rational at PM orders $G^{\leq 3}$,
and include $\pi^2$ at PM orders  $G^{\geq 4}$. 
We have relegated most of the technical details of our computation to various Appendices. More precisely:
\begin{enumerate}
\item Appendix \ref{app_self_force} displays our new self-force result on the time-averaged redshift $\langle z_1 \rangle$ at the
sixth order in eccentricity, $O(e^6)$, and its conversion into the corresponding EOB potential $q_6(u)$. 
\item Appendix \ref{Abel_tr} shows how to obtain a closed-form expression for the 2PM Hamiltonian in the
standard ($p_r$) EOB gauge by computing  the (inverse) Abel transform of its corresponding (closed-form) energy-gauge expression. 
\item Appendix \ref{C} discusses the radial action, and the Delaunay Hamiltonian, for the test-mass limit. 
\end{enumerate}
 
Most of the coefficients entering long expressions, like the redshift invariant at the sixth order in eccentricity, have been
given in the form of Tables. [They are available in electronic format upon request.].

Standard PN approaches to binary dynamics (in their various flavours: Hamiltonian, Lagrangian or effective-field-theory)  
have reached their limits, in view of the  complexity of the required computations, and of the subtle infrared
issues linked to time nonlocality.
Our work, which tackles nonlocality from the beginning, offers an alternative approach to standard computations,
combining information from different contexts and using it in a synergetic way.
It is therefore expected that it may lead to further progress in analytically controlling the dynamics of binary systems. 
It would be interesting to explore combining our new approach with the recently pioneered new approach to
binary dynamics based on focussing on (classical or quantum) scattering motions 
\cite{Damour:2017zjx,Cheung:2018wkq,Bern:2019nnu,Bern:2019crd}.

The techniques we have been defining here can be extended to higher PN orders. We will separately present our
complete, recent 6PN-level results \cite{BDG6PN}.

Two coefficients are still missing to have the complete 5PN Hamiltonian of a two-body system. 
Several routes for determining the two missing coefficients are conceivable, notably:
second-order self-force computations;  or partial standard PN computations of the 5PN dynamics targeted towards  
 a selected mass dependence. [The recent progress in computer-aided evaluation of the PN-expanded interaction potential 
 of binary systems \cite{Blumlein:2019zku,Blumlein:2020pog,Blumlein:2020znm} gives hope that the two missing coefficients
 might be soon derived.] Also high-accuracy numerical simulations might enter the game.

\appendix

\section{The time-averaged redshift $\langle z_1 \rangle$  at $O(e^6)$ and its EOB transcription $q_6(u)$.}
\label{app_self_force}

A redshift invariant for slightly eccentric orbits was introduced in the spacetime of a non-rotating black hole by Barack and Sago \cite{Barack:2011ed} as the orbital averaged value of the linear-in-mass-ratio correction $\delta U$ to the coordinate time component of the particle's 4-velocity.
The latter has been computed through the 9.5PN level in Ref. \cite{Bini:2016qtx} up to the fourth order in the eccentricity, improving the previous analytical knowledge at 6.5PN for $\delta U^{e^2}$ \cite{Bini:2015bfb} and at 4PN for $\delta U^{e^4}$  \cite{Damour:2015isa,Bini:2015bfb}.
Higher-order terms in the eccentricity expansion have been obtained in Refs. \cite{Hopper:2015icj,Bini:2016qtx} up to the order $O(e^{20})$, but at the 4PN level of approximation only, by  combining the 4PN results of Ref. \cite{Damour:2015isa} with the first law for eccentric orbits \cite{Tiec:2015cxa}.
The $O(e^{4})$ 9.5PN-accurate results of Ref. \cite{Bini:2016qtx} have also been transcribed there in terms of the corresponding EOB potentials $\bar d(u)$ and $q(u) \equiv q_4(u)$.

We have extended here the calculation of Ref. \cite{Bini:2016qtx} by including contributions of sixth order  in eccentricity through the same, 
9.5PN level.
Our analytical computation of the conservative SF effects along an eccentric orbit in a Schwarzschild background follows the same approach as in Ref. \cite{Bini:2016qtx}, to which we refer for a full account of intermediate steps.
We work with the redshift function $z_1=U^{-1} $ and its first-order SF perturbation $\delta z_1=-\delta U/U_0^2$
(with $U_0$ denoting the corresponding background value).
The small eccentricity expansion of the time-averaged value $\langle \delta z_1 \rangle$, expressed in terms
of the (Schwarzschild-background) inverse parameter $u_p\equiv 1/p$, and eccentricity $e$, reads
 \begin{eqnarray}
\langle\delta z_1\rangle &=&\delta z_1^{e^0}(u_p)+e^2\delta z_1^{e^2}(u_p)+e^4\delta z_1^{e^4}(u_p)\nonumber\\
&&+e^6\delta z_1^{e^6}(u_p)+O(e^8)\,.
\end{eqnarray}
New with this work is the computation of the 9.5PN accurate  $O(e^6)$ contribution, namely,
\begin{eqnarray}
\delta z_1^{e^6} &=&c_3^{\rm c} u_p^3+c_4^{\rm c} u_p^4 
+(c_5^{\rm c}+c_5^{\ln{}}\ln(u_p))u_p^5\nonumber\\
&&+(c_6^{\rm c}+c_6^{\ln{}}\ln(u_p))u_p^6+c_{6.5}^{\rm c}u_p^{13/2}\nonumber\\
&&+(c_7^{\rm c}+c_7^{\ln{}}\ln(u_p))u_p^7+c_{7.5}^{\rm c}u_p^{15/2}\nonumber\\
&&+(c_8^{\rm c}+c_8^{\ln{}}\ln(u_p)+c_8^{\ln^2{}}\ln^2(u_p))u_p^8+c_{8.5}^{\rm c}u_p^{17/2}\nonumber\\
&&+(c_9^{\rm c}+c_9^{\ln{}}\ln(u_p)+c_9^{\ln^2{}}\ln^2(u_p))u_p^9\nonumber\\
&&+(c_{9.5}^{\rm c}+c_{9.5}^{\ln{}}\ln(u_p)u_p^{19/2}
+O(u_p^{10})\,,
\end{eqnarray}
with coefficients listed in Table \ref{coeffs_z1e6}.

The improved knowledge of the redshift function can then be converted into the EOB potential $ q_6(u) \,p_r^6 \in \widehat Q(u, p_r)$ by using the following
relation (obtained in extending to the $O(e^6)$ level the $O(e^4)$-level results of Ref. \cite{Tiec:2015cxa})
\begin{eqnarray}
q_6(u_p)
&=& B(u_p)+\sum_{k=0}^3 \left( \sum_{n=0}^{6-2k}C^{e^{2k}}_n(u_p)\frac{d^n}{du_p^n}\delta z_1^{e^{2k}}(u_p) \right)
\nonumber\,,
\end{eqnarray}
where we have used the notation $\frac{d^0}{du_p^0}f=f$. The coefficients $B(u_p)$ and $C^{e^{2k}}_n(u_p)$  are listed in Table \ref{coeffs_q6vsz1}.

The PN expansion of $q_6(u)$ then reads
\begin{eqnarray}
q_6(u) &=&b_2^{\rm c} u^2+b_3^{\rm c} u^3+b_{3.5}^{\rm c}u^{7/2}
+(b_4^{\rm c}+b_4^{\ln{}}\ln(u)) u^4\nonumber\\
&& +b_{4.5}^{\rm c}u^{9/2} +(b_5^{\rm c}+b_5^{\ln{}}\ln(u))u^5
+b_{5.5}^{\rm c}u^{11/2}\nonumber\\
&&+(b_6^{\rm c}+b_6^{\ln{}}\ln(u)+b_6^{\ln^2{}}\ln^2(u))u^6\nonumber\\
&&
+b_{6.5}^{\rm c}u^{13/2}
+O(u^{7})\,,
\end{eqnarray}
with coefficients listed in Table \ref{coeffs_q6}.

\begin{table*}[h]
\caption{\label{coeffs_z1e6} Coefficients entering the PN expansion of $\delta z_1^{e^6}$}
\begin{ruledtabular}
\begin{tabular}{ll}
Coefficient & Value \\
\hline
$c_3^{\rm c}$ & $\frac14 $\\
$c_4^{\rm c}$ & $-\frac{53}{12}-\frac{41}{128}\pi^2$\\ 
$c_5^{\rm c}$ & $-\frac{178288}{5}\ln(2)+\frac{1994301}{160}\ln(3)-\frac{38471}{360}+\frac{6455}{4096}\pi^2+16\gamma+\frac{1953125}{288}\ln(5)$\\
$c_5^{\ln{}}$ & $8$\\
$c_6^{\rm c}$ & $-\frac{1694}{5}\gamma+\frac{66668054}{135}\ln(2)-\frac{29268135}{448}\ln(3)+\frac{782899}{4096}\pi^2-\frac{2027890625}{12096}\ln(5)-\frac{17344111}{5040}$\\
$c_6^{\ln{}}$ & $-\frac{847}{5}$\\
$c_{6.5}^{\rm c} $ & $-\frac{18404963}{151200}\pi$\\
$c_7^{\rm c}$ & $-\frac{10083929027}{2835}\ln(2)+\frac{11019270343}{340200}+\frac{10727453}{2835}\gamma+\frac{4609218071875}{2612736}\ln(5)$\\
&$+\frac{4663687}{524288}\pi^4-\frac{130309059379}{28311552}\pi^2-\frac{14238373347}{17920}\ln(3)+\frac{96889010407}{373248}\ln(7)$\\
$c_7^{\ln{}}$ & $\frac{10727453}{5670}$\\
$c_{7.5}^{\rm c} $ & $\frac{629926159}{470400}\pi$\\
$c_8^{\rm c}$ & $-\frac{138663992506361}{20528640}\ln(7)+\frac{1097743020107}{4026531840}\pi^4-\frac{1044921875}{3024}\ln(5)^2-\frac{159152}{15}\zeta(3)
$\\
&$ -\frac{4891192867}{70875}\gamma-\frac{160452171}{400}\ln(3)^2 $\\
&$ -\frac{133972817261}{2764800}\pi^2+\frac{9033082952}{1575}\ln(2)^2+\frac{1216376}{225}\gamma^2+\frac{105459653332171}{16372125}\ln(2)$\\
&$-\frac{160452171}{200}\gamma\ln(3)$\\
&$+\frac{84510345271221}{4928000}\ln(3)-\frac{116526439405625}{18289152}\ln(5)-\frac{1044921875}{1512}\gamma\ln(5)$\\
&$+\frac{1912624751720539}{3929310000}-\frac{160452171}{200}\ln(2)\ln(3)$\\
&$  +\frac{4574838928}{1575}\gamma\ln(2)\ln(5)-\frac{1044921875}{1512}\ln(2)\ln(5)$\\
$c_8^{\ln{}}$ & $-\frac{4891192867}{141750} +\frac{1216376}{225}\gamma -\frac{160452171}{400} \ln(3)-\frac{1044921875}{3024} +\frac{2287419464}{1575}\ln(2) $\\
$c_8^{\ln^2{}}$ & $+\frac{304094}{225} $\\
$c_{8.5}^{\rm c} $ & $\frac{17411624626943}{22632825600}\pi$\\
$c_9^{\rm c}$ & $-\frac{2325452157955686875}{36614882304}\ln(5)-\frac{189182288}{3675}\gamma^2+\frac{9958568909678}{38201625}\gamma-\frac{35883263448213399}{448448000}\ln(3)-\frac{904011369824}{19845}\gamma\ln(2)$\\
&$+\frac{17956170280520566}{343814625}\ln(2)-\frac{13247132039065189}{69526957500}+\frac{919213293396729}{85899345920}\pi^4-\frac{45448745981842837}{1109812838400}\pi^2+\frac{3089591667}{1400}\ln(3)^2$\\
&$+\frac{262937890625}{31752}\ln(5)^2-\frac{1255868873488}{14175}\ln(2)^2+\frac{2825264}{35}\zeta(3)+\frac{1975634469}{4900}\ln(2)\ln(3)$\\
&$+\frac{3089591667}{700}\gamma\ln(3)+\frac{262937890625}{15876}\ln(2)\ln(5)+\frac{16842023587039315}{213497856}\ln(7)+\frac{262937890625}{15876}\gamma\ln(5)$\\
$c_9^{\ln{}}$ & $-\frac{452005684912}{19845}\ln(2) +\frac{4964431663039}{38201625} +\frac{3089591667}{1400} \ln(3)-\frac{189182288}{3675}\gamma+\frac{262937890625}{31752}\ln(5) $\\ 
$c_9^{\ln^2{}}$ & $-\frac{47295572}{3675}$\\ 
$c_{9.5}^{\rm c}$ & $-\frac{151427301903}{98000}\pi\ln(3)-\frac{768417611}{113400}\pi^3-\frac{609707863599642191}{6590678814720}\pi+\frac{82220684377}{3969000}\pi\gamma$\\
&$+\frac{25820141287513}{3969000}\pi\ln(2)-\frac{111806640625}{63504}\pi\ln(5)$\\
$c_{9.5}^{\ln{}}$ & $+\frac{82220684377}{7938000}\pi$\\
\end{tabular}
\end{ruledtabular}
\end{table*}


\begin{table*}[h]
\caption{\label{coeffs_q6vsz1} Coefficients entering the expression for $q_6(u_p)$ in terms of the redshift function and its derivatives.}
\begin{ruledtabular}
\begin{tabular}{ll}
Coefficient & Value \\
\hline
$B(u_p)$&$-
\frac{1}{1024}(3276u_p^4-7371u_p^3+6212u_p^2-2312u_p+320)\frac{u_p(1-2u_p)^3}{(1-3u_p)^{9/2}}
$\\
\hline
$C^{e^0}_0(u_p)$&$
\frac{3}{1024}(207u_p^2-216u_p+56)\frac{u_p(1-2u_p)^3}{(1-3u_p)^{7/2}}
$\\
$C^{e^0}_1(u_p)$&$-
\frac{1}{1280}(-2048-568038697344u_p^{12}+169966688256u_p^{13}-223561417224u_p^8+73260864684u_p^7$\\
&$
+3259361067u_p^5-17888024322u_p^6+42346416u_p^3+501408678672u_p^9-802643130720u_p^{10}$\\
&$
+867902879808u_p^{11}-2782272u_p^2-438377232u_p^4+111360u_p)\frac{(1-2u_p)^3}{u_p^2(1-6u_p)^8(1-3u_p)^{5/2}}
$\\
$C^{e^0}_2(u_p)$&$
-\frac{1}{1280}(291931776u_p^{11}-1340770752u_p^{10}+2510150688u_p^9-2598166704u_p^8+1680719416u_p^7$\\
&$
-721475988u_p^6+211137750u_p^5-42247977u_p^4+5663472u_p^3-483984u_p^2$\\
&$
+23808u_p-512)\frac{(1-2u_p)^3}{u_p(1-6u_p)^7(1-3u_p)^{3/2}}
$\\
$C^{e^0}_3(u_p)$&$
-\frac{1}{160}(1918080u_p^8-4203792u_p^7+3933936u_p^6-2053224u_p^5+656676u_p^4-132441u_p^3+16376u_p^2$\\
&$
-1112u_p+32)\frac{(1-2u_p)^4}{(1-3u_p)^{1/2}(1-6u_p)^6}
$\\
$C^{e^0}_4(u_p)$&$
-\frac{1}{960}(45720u_p^5-50316u_p^4+20554u_p^3-4349u_p^2+552u_p-24)\frac{u_p(1-2u_p)^5(1-3u_p)^{1/2}}{(1-6u_p)^5}
$\\
$C^{e^0}_5(u_p)$&$
-\frac{3}{80}\frac{u_p^3(1-2u_p)^6(1-3u_p)^{3/2}}{(1-6u_p)^4}
$\\
$C^{e^0}_6(u_p)$&$
-\frac{1}{720}\frac{u_p^3(1-2u_p)^6(1-3u_p)^{5/2}}{(1-6u_p)^3}$\\
\hline 
$C^{e^2}_0(u_p)$&$
-\frac{1}{320}(779683968u_p^9-1886037696u_p^8+1975861728u_p^7-1178577360u_p^6+442967544u_p^5$\\
&$
-109412372u_p^4+17816962u_p^3-1833219u_p^2+106288u_p-2672)\frac{(1-2u_p)^3}{u_p(1-6u_p)^7(1-3u_p)^{3/2}}
$\\
$C^{e^2}_1(u_p)$&$
\frac{1}{120}(16132608u_p^8-33912864u_p^7+30156192u_p^6-14824092u_p^5+4434916u_p^4-837157u_p^3$\\
&$
+98524u_p^2-6600u_p+192)\frac{(1-2u_p)^4}{u_p^2(1-6u_p)^6(1-3u_p)^{1/2}}
$\\
$C^{e^2}_2(u_p)$&$
\frac{1}{120}(26640u_p^5-10776u_p^4-6970u_p^3+4001u_p^2-552u_p+24)\frac{(1-2u_p)^5(1-3u_p)^{1/2}}{u_p(1-6u_p)^5}
$\\
$C^{e^2}_3(u_p)$&$
-\frac{1}{30}(4-45u_p+72u_p^2)\frac{(1-2u_p)^6(1-3u_p)^{3/2}}{(1-6u_p)^4}
$\\
$C^{e^2}_4(u_p)$&$
\frac{1}{60}\frac{u_p(1-2u_p)^6(1-3u_p)^{5/2}}{(1-6u_p)^3}
$\\
\hline 
$C^{e^4}_0(u_p)$&$
-\frac{1}{15}(49032u_p^5-43812u_p^4+11586u_p^3-609u_p^2-88u_p+8)\frac{(1-2u_p)^5(1-3u_p)^{1/2}}{u_p^3(1-6u_p)^5}
$\\
$C^{e^4}_1(u_p)$&$
\frac{4}{15}(8-81u_p+144u_p^2)\frac{(1-2u_p)^6(1-3u_p)^{3/2}}{u_p^2(1-6u_p)^4}
$\\
$C^{e^4}_2(u_p)$&$
-\frac{4}{15}\frac{(1-2u_p)^6(1-3u_p)^{5/2}}{u_p(1-6u_p)^3}
$\\
\hline
$C^{e^6}_0(u_p)$&$
\frac{16}{5}\frac{(1-2u_p)^6(1-3u_p)^{5/2}}{u_p^3(1-6u_p)^3}$\\
\end{tabular}
\end{ruledtabular}
\end{table*}


\begin{table*}[h]
\caption{\label{coeffs_q6} Coefficients entering the PN expansion of $q_6(u)$}
\begin{ruledtabular}
\begin{tabular}{ll}
Coefficient & Value \\
\hline
$b_2^{\rm c}$ &$-\frac{827}{3}+\frac{1399437}{50}\ln(3)-\frac{2358912}{25}\ln(2)+\frac{390625}{18}\ln(5)$\\
$b_3^{\rm c}$ & $\frac{2613083}{1050}+\frac{6875745536}{4725}\ln(2)-\frac{23132628}{175}\ln(3)-\frac{101687500}{189}\ln(5) $\\
$b_{3.5}^{\rm c}$ & $-\frac{2723471}{756000}\pi  $\\
$b_4^{\rm c}$ & $\frac{153776136875}{23328}\ln(5)+\frac{447248}{1575}\gamma-\frac{9678652821}{5600}\ln(3)+\frac{96889010407}{116640}\ln(7)$\\
& $-\frac{41589250561}{7938000}-\frac{9733841}{327680}\pi^2-\frac{211076833264}{14175}\ln(2) $\\
$b_4^{\ln{}}$ & $+\frac{223624}{1575} $\\
$b_{4.5}^{\rm c}$ & $ +\frac{1783458013}{56448000}\pi $\\
$b_5^{\rm c}$ & $\frac{3651910996}{86625}\gamma-\frac{7733712492302375}{201180672}\ln(5)+\frac{912077147376081}{15680000}\ln(3)-\frac{211655031897463}{9331200}\ln(7)$\\
&$+\frac{5043177377399716}{81860625}\ln(2)+\frac{38342542739}{7864320}\pi^2+\frac{15438788608}{875}\ln(2)^2$\\
&$-\frac{1061386821}{875}\ln(3)^2-\frac{830563821453539}{1746360000}$\\
&$-\frac{208984375}{189}\ln(5)^2-\frac{417968750}{189}\gamma \ln(5)-\frac{2122773642}{875}\ln(2) \ln(3)-\frac{2122773642}{875}\gamma \ln(3)$\\
&$+\frac{70193205248}{7875}\gamma \ln(2)-\frac{417968750}{189}\ln(2) \ln(5) $\\
$b_5^{\ln{}}$ & $-\frac{208984375}{189}  \ln(5)+\frac{35096602624}{7875}\ln(2) +\frac{1825955498}{86625} -\frac{1061386821}{875} \ln(3) $\\
$b_{5.5}^{\rm c}$ & $ -\frac{375333092211461}{905313024000}\pi$\\
$b_6^{\rm c}$ & $-\frac{54126285229417}{73573500}\gamma-\frac{315130937024}{2025}\gamma\ln(2)+\frac{7843492521}{2450}\ln(2)\ln(3)+\frac{39285904041}{2450}\gamma\ln(3)+\frac{448936953125}{7938}\ln(2)\ln(5)$\\
&$+\frac{448936953125}{7938}\gamma\ln(5)-\frac{11892972284088646293}{31391360000}\ln(3)-\frac{1686162964063105097}{12770257500}\ln(2)-\frac{2314158285520063375}{36614882304}\ln(5)$\\
&$-\frac{192}{7}\zeta(3)+\frac{1878836255027762051}{6065280000}\ln(7)+\frac{4253856}{1225}\gamma^2-\frac{262462223346649}{10737418240}\pi^4-\frac{375306539275861}{23121100800}\pi^2$\\
&$+\frac{39285904041}{4900}\ln(3)^2+\frac{448936953125}{15876}\ln(5)^2-\frac{21523313234464}{70875}\ln(2)^2$\\
&$+\frac{384973167765003181159}{58736373696000} $\\
$b_6^{\ln{}}$ & $-\frac{54126285229417}{147147000} +\frac{4253856}{1225}\gamma -\frac{157565468512}{2025}\ln(2) +\frac{39285904041}{4900} \ln(3)+\frac{448936953125}{15876} \ln(5) $\\
$b_6^{\ln^2{}}$ & $+\frac{1063464}{1225}$\\
$b_{6.5}^{\rm c} $ & $-\frac{102893846003}{19845000}\pi\gamma+\frac{431653923653437}{19845000}\pi\ln(2)$\\
&$\frac{30475181893883804796413}{144994933923840000}\pi-\frac{2758233739833}{490000}\pi\ln(3)+\frac{961624729}{567000}\pi^3-\frac{22361328125}{3969}\pi\ln(5) $\\
$b_{6.5}^{\ln{}}$ & $-\frac{102893846003}{39690000}\pi $\\
\end{tabular}
\end{ruledtabular}
\end{table*}


\section{Transforming the  energy-gauge 2PM $Q$ term, $q_{2 \rm EG}(H_S) u^2$,
into its (closed-form) $p_r$-gauge version via an Abel transform}
\label{Abel_tr}

In the energy-gauge, the 2PM EOB $Q$ potential reads  $\widehat Q^{\rm 2PM}_{\rm EG}=q_{2 \rm EG}(\gamma) u^2$
where $\gamma=H_S$. We want to transform it in a $p_r$-dependent one, say $\widehat Q^{\rm 2PM}_{p_r}=q_{2}^{ (p_r)}(p_r) u^2$,
that leads to the same scattering angle. Using Eq. (4.22) of \cite{Damour:2017zjx}, this means that the two
functions must yield the same integral $\int_{- \infty}^{+ \infty} d\sigma Q$, where $d\sigma= dR/P^R$.  
Writing this condition at the 2PM level (neglecting any $O(G^3)$ correction) is easily seen to lead to the condition
\beq
q_{2 \rm EG}(\gamma)= \frac2{\pi} \int_0^{\sqrt{\gamma^2-1}} dp_r  \frac{q_{2}^{ (p_r)}(p_r)}{\sqrt{\gamma^2-1-p_r^2}}\,.
\eeq
Reexpressing this condition (and the two functions) in terms of the variables $c \equiv\gamma^2-1$ and $x\equiv p_r^2$ yields
\beq \label{abel1}
q_{2 \rm EG}(c)=  \frac1{\pi} \int_0^{c} dx \frac{q_{2}^{ (p_r)}(x)/\sqrt{x}}{\sqrt{c-x}}\,.
\eeq
The latter condition expresses the fact that the function $q_{2 \rm EG}(c)$ is the (usual) Abel transform of the function 
$q_{2}^{ (p_r)}(x)/\sqrt{x}$. But the Abel transform (with inverse square root kernel) is just (in the sense of Marcel Riesz'
integral operators) a derivative of order $ -\frac12$. Therefore the inverse transform (a derivative of order $ +\frac12$)
can simply be written as the composition of a derivative and an Abel transform. Hence, the following formula for the
inverse of Eq. \eqref{abel1}
\beq
\label{abel}
q_2^{ (p_r)}(x)=\sqrt{x}\frac{d}{d x}\, \int_0^x \frac{q_{2 \rm  EG }(c)}{\sqrt{x-c}} dc \equiv \sqrt{x}\frac{d}{d x}\,
I(x)\,.
\eeq
The function $q_{2 \rm  EG }(c)$ to be inserted in this formula is (after expressing $\gamma$ in terms of $c \equiv\gamma^2-1$
in  $q_{2\rm  EG }(\gamma)$, Eq. \eqref{q_energy_gauge_fun})
\beq
\label{q_2_di_c}
q_{2\rm  EG }(c)=\frac32 (4+5c)\left(1-\frac{1}{h(c)}  \right)\,,
\eeq
with $h(c)=\sqrt{1-2\nu+2\nu (1+c)^{1/2}}$.

One can first easily obtain the all-order PN expansion of the function $q_2^{ (p_r)}(x)$ (where we recall that $x=p_r^2$)
by expanding $q_{2\rm  EG }(c)$, Eq. \eqref{q_2_di_c}, in powers of $c$, and then inserting this expansion in Eq. \eqref{abel}. 
The result reads
\begin{widetext}
\begin{eqnarray}
q_2^{ (p_r)}(x)&=& 6\nu x+(8\nu-6\nu^2) x^2+\left(-\frac{9}{5}\nu-\frac{27}{5}\nu^2+6\nu^3\right) x^3
+\left(\frac{6}{7}\nu+\frac{18}{7}\nu^2+\frac{24}{7}\nu^3-6\nu^4\right) x^4\nonumber\\
&&+\left(-\frac{11}{21}\nu-\frac{11}{7}\nu^2-\frac{20}{7}\nu^3-\frac{5}{3}\nu^4+6\nu^5\right) x^5
+\left(\frac{4}{11}\nu+\frac{12}{11}\nu^2+\frac{170}{77}\nu^3+\frac{30}{11}\nu^4-6\nu^6\right) x^6\nonumber\\
&& +\left(-\frac{3}{11}\nu-\frac{9}{11}\nu^2-\frac{250}{143}\nu^3-\frac{35}{13}\nu^4-\frac{315}{143}\nu^5+\frac{21}{13}\nu^6+6\nu^7\right) x^7+O(x^8)\,.
\end{eqnarray}
 \end{widetext}
However, it is also possible to obtain a closed-form expression for the function $q_2^{ (p_r)}(x)$
by computing the integral $I(x)$ entering the inverse Abel transform Eq. \eqref{abel}.

To compute the integral $I(x)$, Eq. \eqref{abel}, we change the variable $c = \gamma^2-1$ back into $\gamma$. This yields
\beq
I(x)=\frac32 \int_{1}^{\sqrt{1+x}} \frac{(5\gamma^2-1) 2 \gamma d\gamma}{\sqrt{1+x-\gamma^2}}\left(1-\frac{1}{\sqrt{1+2\nu(\gamma-1) }}  \right)\,.
\eeq
We introduce then the notation
\beq
\gamma_r\equiv\sqrt{1+x}\,,\qquad \gamma_\nu\equiv\frac{1}{2\nu}-1\ge 1\,,
\eeq
so that
\begin{eqnarray}
I(x)&=&3\int_{1}^{\gamma_r} d\gamma\frac{(5\gamma^2-1)  \gamma }{\sqrt{\gamma_r^2-\gamma^2}}\left(1-\frac{1}{\sqrt{2\nu}\sqrt{\gamma_\nu+\gamma }}  \right)\nonumber\\
&=&
2(1+5\gamma_r^2)\sqrt{\gamma_r^2-1} \nonumber\\
&& -\frac{3}{\sqrt{2\nu}}  \int_{1}^{\gamma_r} d\gamma\frac{(5\gamma^2-1)  \gamma }{\sqrt{(\gamma_r^2-\gamma^2)(\gamma_\nu+\gamma)}} \nonumber\\
&\equiv & 2(1+5\gamma_r^2)\sqrt{\gamma_r^2-1} -\frac{3}{\sqrt{2\nu}}  J\,.
\end{eqnarray}
where we introduced
\begin{eqnarray}
J&\equiv&\int_{1}^{\gamma_r}  d\gamma\frac{(5\gamma^2-1)  \gamma }{\sqrt{(\gamma_r^2-\gamma^2)(\gamma_\nu+\gamma)}}\nonumber\\
 &\equiv &\int_{1}^{\gamma_r} d\gamma \frac{Q_3(\gamma)}{\sqrt{P_3(\gamma)}}\,.
\end{eqnarray}
Here $P_3$ and $Q_3$ denote the cubic polynomials in $\gamma$ entering the integrand of the integral $J$. 

At this stage it is already clear that the original integral $I(x)$ is the sum of an elementary term and of an elliptic integral given by $J$.
To get an explicit form of the elliptic integral $J$, we need to perform the Legendre reduction of $J$. This means writing the identity
\begin{eqnarray}
\label{identity_Leg_red}
&&[2(d_0+d_1\gamma)\sqrt{P_3}]'=2 d_1 \sqrt{P_3}+\frac{(d_0+d_1\gamma)P_3'}{\sqrt{P_3}}\nonumber\\
&&\qquad\qquad\qquad=\frac{2 d_1  P_3 +(d_0+d_1\gamma)P_3'}{\sqrt{P_3}}
\end{eqnarray}
and determining the coefficients $d_0$ and $d_1$ so as to reduce the integral 
$\int d\gamma Q_3(\gamma)/\sqrt{P_3(\gamma)}$ to an integral whose numerator is a polynomial
of degree 1.  Indeed, the choice
\beq
d_0=\frac{4}{3}\gamma_\nu \,,\qquad d_1=-1\,,
\eeq
implies
\begin{eqnarray}
2 d_1  P_3 +(d_0+d_1\gamma)P_3'=&&\nonumber\\
=Q_3+\left(  -3\gamma_r^2-\frac{8}{3}\gamma_\nu^2+1 \right)\gamma 
-\frac23 \gamma_\nu\gamma_r^2 && \nonumber\\
=Q_3+
\left(-3\gamma_r^2-\frac{8}{3}\gamma_\nu^2+1\right)(\gamma+\gamma_\nu)+ \frac{7}{3}\gamma_\nu\gamma_r^2&&\nonumber\\
+\frac{8}{3}\gamma_\nu^3-\gamma_\nu\,,\qquad\qquad\qquad &&
\end{eqnarray}
that is
\beq
2 d_1  P_3 +(d_0+d_1\gamma)P_3'\equiv Q_3+C_1 (\gamma+\gamma_\nu)+C_2\,,
\eeq
where
\begin{eqnarray}
C_1&=& -3\gamma_r^2-\frac{8}{3}\gamma_\nu^2+1\,,\nonumber\\
C_2&=& \frac{7}{3}\gamma_\nu\gamma_r^2+\frac{8}{3}\gamma_\nu^3-\gamma_\nu\,.
\end{eqnarray}
Therefore, the identity \eqref{identity_Leg_red} becomes
\beq
\label{identity_Leg_red_new}
[2(d_0+d_1\gamma)\sqrt{P_3}]'=\frac{Q_3+C_1 (\gamma+\gamma_\nu)+C_2}{\sqrt{P_3}}\,,
\eeq
so that integrating both sides  gives
\begin{eqnarray}
 -2\left(\frac{4}{3}\gamma_\nu -1\right)\sqrt{P_3(1)}&=&J\nonumber\\
&+& C_1\int_{1}^{\gamma_r} d\gamma \frac{\gamma+\gamma_\nu}{\sqrt{P_3}}\nonumber\\
&+& C_2\int_{1}^{\gamma_r} \frac{d\gamma}{\sqrt{P_3}}\,,
\end{eqnarray}
where
\beq
P_3(1)=(\gamma_r^2-1)(1+\gamma_\nu)\,.
\eeq
This yields the following expression for $J$:
\begin{eqnarray}
J&=& -2\left(\frac{4}{3}\gamma_\nu -1\right)\sqrt{(\gamma_r^2-1)(1+\gamma_\nu)}\nonumber\\
&-&
C_1\int_{1}^{\gamma_r} d\gamma \frac{\gamma+\gamma_\nu}{\sqrt{P_3(\gamma)}} \nonumber\\
&-&C_2 \int_{1}^{\gamma_r} \frac{d\gamma}{\sqrt{P_3(\gamma)}}\,.
\end{eqnarray}
The remaining integrals are then explicitly expressible in terms of complete Legendre elliptic integrals, namely
\begin{eqnarray}
{\mathbb I}_1&=& \int_{1}^{\gamma_r} \frac{d\gamma}{\sqrt{P_3(\gamma)}}\nonumber\\
&=&  \frac{2}{\sqrt{a - c} }\,\,
  {\rm EllipticF}\, \left({\rm arcsin} \sqrt{\frac{a-1}{a-b}}, \sqrt{\frac{ a - b }{ a - c }}\right)\,,\nonumber\\
{\mathbb I}_2&=&\int_{1}^{\gamma_r} d\gamma \frac{\gamma+\gamma_\nu}{\sqrt{P_3(\gamma)}}\nonumber\\
&=& -
  2\sqrt{a - c}\,\,
  {\rm EllipticE}\, \left({\rm arcsin} \sqrt{\frac{a-1}{a-b}}, \sqrt{\frac{ a - b }{ a - c }}\right)\,,\nonumber\\
\end{eqnarray}
where $a=\gamma_r$, $b=-\gamma_r$, $c=-\gamma_\nu$ and
\begin{eqnarray}
&&\sqrt{a-c}=\gamma_r+\gamma_\nu\,,\qquad \frac{a-1}{a-b}=\frac{\gamma_r-1}{2\gamma_r}\,,\nonumber\\
&& \frac{ a - b }{ a - c }=\frac{2\gamma_r}{\gamma_r+\gamma_\nu}\,.
\end{eqnarray}
Here, we got ${\mathbb I}_1$ from \cite{GR}, Eq. 6 pag. 254 sec. 3.131 and ${\mathbb I}_2$ from  \cite{GR}, Eq. 5 pag. 255 sec. 3.132, using in the latter case $(x-c)$ in the numerator of the integrand and simplifying the final result. The minus sign in ${\mathbb I}_2$ corresponds to a general prefactor $a/b$ which is $-1$ in the present case.   

Inserting the latter elliptic-integral representation of $J$ in the above expression of $I(x)$, and then inserting $I(x)$ in
Eq. \eqref{abel}, finally gives a closed-form expression for the 2PM-level $p_r$-gauge function $q_2^{ (p_r)}(x)$ (with $x=p_r^2$).
This exercise shows, however, that the energy-gauge expression  of the
2PM dynamics, involving the algebraic function $q_{2\rm  EG }(\gamma)$, Eq. \eqref{q_energy_gauge_fun},
is drastically simpler than its $p_r$-gauge retranscription.

\section{Radial action and periastron advance in the test-mass limit}
\label{C}

We recall the notations $\gamma=\e$,
\beq
 \gamma^2-1 \equiv p_{\infty}^2  \equiv - |p|^2 \,,
\eeq
and
\beq
e^2 \equiv 1+ p_{\infty}^2 j^2\,.
\eeq

The exact radial action in the test-mass (or Schwarzschild, or $\nu \to 0$)  limit reads
\beq \label{IrS1}
I_r^S(\gamma, j)=\frac{1}{2\pi}\oint du \frac{\sqrt{P_3(u)}} {u^2(1-2u)}\,,
\eeq
where $P_3(u)$ is the following cubic polynomial in $u=\frac1r$
\bea
P_3(u)&=&\gamma^2-(1-2u)(1+j^2u^2) \nonumber\\
&=& \gamma^2- \left(1-2u+j^2 u^2 - 2 j^2 u^3 \right) \nonumber\\
&=& \gamma^2 -1+2 u-j^2 u^2 + 2 j^2 u^3 \,.
\eea
Here, we are  interested in ellipticlike motions with $0<e^2<1$, {\it i.e.}, with $-1 < p_{\infty}^2 j^2<0$.
The  inequality $ |p|j<1$ does not {\it a priori} allow us  (contrary to the scattering-motion case)  to straightforwardly
use a PM expansion in powers of $\frac1j\propto G$ at a fixed value of $\gamma$ (or $p_{\infty}$).  The standard
expansion technique for ellipticlike motions is the PN expansion. A useful way to formalize the PN expansion
is to introduce a PN scaling, say with the bookkeeping parameter
$\eta$ introduced in the scaling relations \eqref{etascaling}. The main geometrical effect of this scaling is to introduce
a parametric separation between the two roots of the cubic polynomial $P_3(u)$ that are close to the roots,
\beq \label{upm}
u_{\pm }=\frac{ 1\pm e }{j^2}\,,
\eeq
of
\beq
P_2(u)= \gamma^2 -1 +2 u-j^2 u^2 =p_{\infty}^2+2 u- j^2 u^2\,,
\eeq
and the third root of $P_3(u)$. It is easily seen that this is formally equivalent to introducing a related
PN bookkeeping parameter, say $\epsilon$, and to write $P_3(u)$ as
\beq
P_3(u)= p_{\infty}^2 + 2u-j^2 u^2 + \epsilon \,( 2 j^2 u^3)\,.
\eeq
One can then expand the radial integral  \eqref{IrS1} in powers of $\epsilon$, using the technique explained in Ref. \cite{Damour:1988mr}.

From the general result given in Eqs. (3.8), (3.9) of Ref. \cite{Damour:1988mr} one can see that the PN expansion
of the sum $I^S_r+j$ defines,
when considered at a fixed (negative) value of $p_{\infty}^2$, an analytic function of the variable $\frac1j$ having an
expansion in powers of $\frac1j$ of the form
\beq \label{Laurent}
I^S_r(\gamma,j)+j= I^S_0(\gamma) + \sum_{n\geq0} \frac{I^S_{2n+1}(\gamma)}{j^{2n+1}}\,.
\eeq
We wish to algorithmically compute the coefficients $I^S_{0}(\gamma)$, $ I^S_{2n+1}(\gamma)$
entering the Laurent expansion \eqref{Laurent}.This expansion shows that, when keeping fixed 
$p_{\infty}^2$ (with $p_{\infty}^2 <0$), one can analytically continue
$I^S_r(\gamma,j)$ down to  $\frac1j \to 0$. 
In order to be able to use the integral definition \eqref{IrS1}
of $I^S_r(\gamma,j)$ in the limit $\frac1j \to 0$, one must (following the method of Sommerfeld used in Ref. \cite{Damour:1988mr})
interpret the integral $\oint du $ as a contour integral in the complex $u$-plane, along a closed contour $\cal C$
circling around the two roots of $P_3(u)$ close to \eqref{upm}. When $\frac1j \to 0$
the latter two roots become complex (because $e \approx \pm i \sqrt{- p_{\infty}^2} j$), and tend towards
$\pm i \frac{\sqrt{- p_{\infty}^2}}{j}$. The important point is that, in this limit, these two roots tend towards zero, and therefore
remain well separated from the third root which tends towards $\frac12$ (indeed the sum of the three roots of $P_3(u)$
is equal to $\frac12$). One can technically see  the possibility of expanding the contour integral defining $I^S_r(\gamma,j)$ in
this limit by introducing the scaled integration variable $y$ such that $u = \frac{y}{j}$. In terms of this variable we have the contour integral
\beq
I_r^S(\gamma, j)=\frac{j}{2\pi}\oint_{\cal C} dy \frac{\sqrt{p_{\infty}^2- y^2 + \frac{2}{j}(y+y^3)}} {y^2(1-\frac{2}{j} y)}\,.
\eeq
As the contour ${\cal C}$ circles around the roots $ \pm i \frac{\sqrt{- p_{\infty}^2}}{j}$ of $P'_2(y)=p_{\infty}^2- y^2$
(while avoiding them), it is allowed to expand the integrand in powers of $\frac1j$. The latter expansion leads to well-defined
complex-contour integral expressions for the looked-for coefficients $I^S_{0}(\gamma)$, $ I^S_{2n+1}(\gamma)$.
One can then  contract the complex contour ${\cal C}$ down to the (doubled) interval $[- i \sqrt{- p_{\infty}^2},
+ i \sqrt{- p_{\infty}^2}]$ along the imaginary axis, and thereby reduce the integrals to real integrals in the
variable $x=y/(i \sqrt{- p_{\infty}^2})$. The latter real integrals on the interval $x\in [-1,+1]$ can then be evaluated by using
Hadamard's Partie finie \cite{Damour:1988mr}. Using this technique we computed the exact expressions
of the test-mass coefficients $I^S_{0}(\gamma)$, $ I^S_{2n+1}(\gamma)$ given in the text.

Let us also note  that Ref. \cite{Damour:1988mr} (see Eq. A.8 there) has explicitly computed the 
(simpler) complete elliptic integral giving the test-particle periastron advance $K_{\rm Sch}= \Phi_{\rm Sch}/(2 \pi)$,
 i.e., the $j$ derivative of $I_{r}^{\rm Sch}(\e,j)$. They expressed the result in the simplified form
\beq
\label{K_schw_exact}
K_{\rm Sch}(\e,j)
=K_{\rm Sch,circ}(j)(1+\xi)^{1/4}F\left[\frac14, \frac34, 1, \frac{\xi}{3}\right]\,,
\eeq
where the prefactor
\beq
K_{\rm Sch,circ}(j)=\left(1-\frac{12}{j^2}\eta^2 \right)^{-1/4} \,,
\eeq
corresponds to the circular-orbit limit, and where the argument $\xi$ is defined as
\beq
\label{xi_def}
\xi=\tan^2 \left(\frac13 {\rm arcsin}(\sqrt{x}) \right)\,,
\eeq
in terms of
\begin{eqnarray}
x&\equiv&\frac{108}{j^2}\eta^4 \left(1-\frac{12}{j^2}\eta^2 \right)^{-3}\left(2\tilde E +\frac{1}{j^2}\right. \nonumber\\
&& \left.-36 \frac{\tilde E}{j^2}\eta^2 (1+3\tilde E\eta^2)-\frac{16}{j^4}\eta^2 \right)\,.
\end{eqnarray}
Here $\tilde E$ denotes
\beq
\tilde E \equiv \frac{\e^2-1}{2} \equiv \bar E_{\rm eff} \left(1+\frac12 \bar E_{\rm eff} \eta^2 \right)  \,, 
\eeq
where we introduced the further notation
\beq 
\gamma=\e 
\equiv1+\bar E_{\rm eff} \eta^2\,.
\eeq
Inserting the PN expansion of $\xi$ in terms of $x$, i.e.,
\beq
\xi =  \frac19 x+\frac{11}{243} x^2+\frac{169}{6561} x^3+\frac{1009}{59049}x^4+O(x^5)\,,
\eeq
with
\begin{eqnarray}
x &=& 108 \frac{(2\tilde E j^2+1)}{j^4} \eta^4+432\frac{ (9\tilde E j^2+5)}{ j^6}\eta^6\\
&& -3888\frac{ (-12\tilde E j^2-8+3\tilde E^2 j^4)}{ j^8}\eta^8\nonumber\\
&&-46656 \frac{(-8\tilde E j^2-8+9\tilde E^2 j^4)}{ j^{10}}\eta^{10}+O(\eta^{12})\,,\nonumber
\end{eqnarray}
in the expression of $K_{\rm Sch}(\e,j)$, then yields
\begin{eqnarray}
K_{\rm Sch}(\bar E_{\rm eff},j)&=& 1+\frac{3}{j^2}\eta^2+\left(\frac{105}{4 j^4}+\frac{15}{ 2 j^2}\bar E_{\rm eff}\right)\eta^4\nonumber\\
&+&\left(\frac{15}{4j^2}\bar E_{\rm eff}^2
+\frac{315}{ 2 j^4}\bar E_{\rm eff}+\frac{1155}{ 4 j^6}\right)\eta^6\nonumber\\
&+& \left( \frac{225225}{ 64 j^8}+\frac{4725}{16j^4}\bar E_{\rm eff}^2 +\frac{45045}{16j^6}\bar E_{\rm eff}\right)\eta^8\nonumber\\
&+& \left( \frac{765765}{ 16 j^8}\bar E_{\rm eff}+\frac{2909907}{ 64 j^{10}}+\frac{3465}{ 16 j^4}\bar E_{\rm eff}^3\right. \nonumber\\
&& \left.+\frac{315315}{32j^6} \bar E_{\rm eff}^2 \right)\eta^{10}+ O(\eta^{12})\,,
\end{eqnarray}
where we used the energy variable $\bar E_{\rm eff}=(\gamma-1)/\eta^2$. The latter expression is easily checked to agree
with (minus) the $j$ derivative of the $\nu \to 0$ limit of our 5PN-expanded radial action above, as given in
 Table \ref{table_barE}.

\section*{Acknowledgments}

DB thanks the IHES for warm hospitality at various stages during the development of the present project.

\end{document}